%% file: thesis.tex
\documentclass[11pt,Chicago]{uuthesis2e}
\usepackage[papersize={8.5in,11in}]{geometry}

\usepackage {amsthm}

\usepackage{mathpazo}

\fourlevels

\usepackage[numbers]{natbib}
\usepackage {amsmath,esint}
\usepackage {amssymb}
\usepackage {bm}
\usepackage {bibnames}
\usepackage {citesort}
\usepackage {graphicx}
\usepackage {graphpap}
\usepackage {longtable}
\usepackage {multirow}
\usepackage {tikz}
\usepackage {pdfpages}
\usepackage{amsmath}
\usetikzlibrary {decorations.markings}
\usetikzlibrary{patterns}
\usetikzlibrary{decorations.pathreplacing,decorations.markings}
\usetikzlibrary{plotmarks}
\usetikzlibrary{decorations.shapes}
\usepackage {varioref}
\tikzset{
  on each segment/.style={
    decorate,
    decoration={
      show path construction,
      moveto code={},
      lineto code={
        \path [#1]
        (\tikzinputsegmentfirst) -- (\tikzinputsegmentlast);
      },
      curveto code={
        \path [#1] (\tikzinputsegmentfirst)
        .. controls
        (\tikzinputsegmentsupporta) and (\tikzinputsegmentsupportb)
        ..
        (\tikzinputsegmentlast);
      },
      closepath code={
        \path [#1]
        (\tikzinputsegmentfirst) -- (\tikzinputsegmentlast);
      },
    },
  },
  mid arrow/.style={postaction={decorate,decoration={
        markings,
        mark=at position .5 with {\arrow[#1]{stealth}}
      }}},
}

\tikzset{decorate sep/.style 2 args=
{decorate,decoration={shape backgrounds,shape=circle,shape size=#1,shape sep=#2}}}

\usepackage {uuthesis-2016-h}  

\newcommand{\m}{\textbf{m}}
\def\be{\begin{equation}}
\def\ee{\end{equation}}
\def\ba{\begin{eqnarray}}
\def\ea{\end{eqnarray}}
\def\rr{{\bf r}}

\def\kk{{\bf k}}
\def\w{\omega}

\def\e{\epsilon}

\def\p{\partial}
\def\jj{\textbf{j}}
\def\pp{\textbf{p}}
\def\rr{\textbf{r}}
\def\qq{\textbf{q}}
\def\vv{\textbf{v}}

\def\EE{\textbf{E}}
\def\AA{\textbf{A}}
\def\BB{\textbf{B}}

\def\d{\delta}

\def\bra{\langle}
\def\ket{\rangle}

\usepackage {uuthesis-chapterbib}
\usepackage {mythesis}
\usepackage{float}

\author                 {Jing Ma}
\title                  {Geometric theory of the natural optical activity in noncentrosymmetric metals} 
\thesistype             {dissertation}

\dedication             {  I would like to dedicate this thesis to my parents, who always give me absolute understanding and support during my life. I also want to dedicate this thesis to all my friends. Life would be miserable without you guys.}

\degree                 {Doctor of Philosophy}

\approvaldepartment     {Physics and Astronomy}
\department             {Department of Physics and Astronomy}
\graduatedean           {David Kieda}
\departmentchair        {Ben Bromley}

\committeechair         {Dmytro Pesin}
\firstreader            {Yongshi Wu}
\secondreader           {Oleg Starykh}
\thirdreader            {Andrey Rogachev}
\fourthreader           {Yekaterina Epshteyn}
\chairtitle             {Professor}

\submitdate             {Dec 2017}
\copyrightyear          {2017}

\chairdateapproved      {20 Sep 2017}
\firstdateapproved      {20 Sep 2017}
\seconddateapproved     {20 Sep 2017}
\thirddateapproved      {20 Sep 2017}
\fourthdateapproved     {20 Sep 2017}

\begin{document}

\frontmatterformat
\titlepage
\copyrightpage
\dissertationapproval
\setcounter {page}     {2}             
\preface    {abstract} {Abstract}
\dedicationpage
\tableofcontents
\listoffigures
%

\optionalfront {Notation and Symbols} {\input{notation}}


\preface{acknowledge}{Acknowledgements}


\maintext       

\pagestyle{headings} 

\include {chap1}
\include {chap2}

\include {chap3}
\include {chap5}
\include {chap4}
\include {conclusion}

\numberofappendices = 3
\appendix       

\include {appa}

\include {appb}

\include {appc}

\end {document}

%% file: notation.tex

\begin{tabular}{ll}
    \hline
    $\alpha$      & fine-structure (dimensionless) constant, approximately $1 / 137$ \\
    $c$      &speed of light\\
    $C$     & Chern number\\
    $\delta(x)$   & Dirac's famous function \\
    $\nabla$     &nabla operator, a vector differential operator\\
    $e$    &the charge of an electron, $e\approx -1.6\times10^{-19}$ coulombs\\
    $\epsilon_0$       &vacuum permittivity, $\epsilon_0\approx 8.85\times 10^{-12}$ F/m\\
    $f(\rr,\kk,t)$     &electron distribution function\\
    $g_{ab}$     &gyrotropic tensor\\
    $\gamma(\omega)$ or $\eta(\omega)$      &chiral magnetic conductivity\\
    $\gamma_n$   &Berry phase\\
     $\mathcal{A}_n$   &Berry connection\\
    $\bm \Omega_n$  &Berry curvature\\
    $\hbar$    &reduced Planck's constant, with a value of $\sim6.58\times10^{−16}\rm eV\cdot s/rad$\\
    $h$        &Planck's constant, equals $2\pi \hbar$\\
    $\mathcal{I}$  &inversion symmetry\\
    $\mathcal{T}$    &time reversal symmetry\\
    $\bf m$     &magnetic moment\\
    $\mu$       &chemical potential\\
    $\mu_0$       &vacuum permeability, $\mu_0=4\pi\times 10^{-7}$ H/m\\
    $\sigma_{ij}$     &conductivity tensor\\
    $\bm \sigma$    &a triplet of Pauli matrices:$\bm{\sigma}=(\sigma_x, \sigma_y, \sigma_z)$\\
    $\omega_c$   & cyclotron frequency, $\omega_c=qB/m$\\
    $\Phi_0$    &the quantum of flux, $\Phi_0=h/e$\\
    $Q$      &chirality\\
    $\tau$       &relaxation time\\
    Tr            &trace\\
    $v_F$       &Fermi velocity\\
    $p_F$      &Fermi momentum\\
    \hline
\end{tabular}

%% file: chap1.tex

\chapter{Introduction: basic concepts in geometric band theory}

\newpage
The last decade has witnessed the flourishing in condensed matter field, especially in the study of systems with unconventional band structures, which are topologically protected, like graphene and topological insulator, or the recent Weyl semimetal. To study the transport properties of these materials, the most common way is applying some electromagnetic field on them. Theoretically, the most common method used is the semi-classical transport theory based on the Boltzmann transport equations. This chapter will start with Berry phase, the most basic and important concept of geometric band theory. After that, I will introduce the basic idea of semi-classical transport theory. In general, this chapter is a detailed introduction of some basic concepts and methods used in studying those topological band systems.

\section{Berry phase, Berry connection and Berry curvature}
Berry phase or Pancharatnam-Berry phase (named after S. Pancharatnam and Sir Michael Berry, as it was first discovered by S. Pancharatnam in 1956 \cite{Pancharatnam}, then rediscovered by M. V. Berry in 1984\cite{Berry1984}) is the most important concept in geometric band theory. However, its discovery was not specifically related to Bloch-periodic system, but to the general idea of quantum adiabatic transport. It is the geometric phase difference gained through a path of a cycle when a system is under a cyclic adiabatic process, which results from the geometrical properties of the parameter space of the Hamiltonian of the system. The so- called ``Berry connection" and ``Berry curvature" are local gauge potential and gauge field associated with the Berry phase.

\subsection{General Formalism of Berry Phase}
In quantum mechanics, the Berry phase emerges in a cyclic adiabatic evolution. The adiabatic theorem, introduced by Max Born and Vladimir Fock in 1928\cite{Born1928}, states that a physical system remains in its instantaneous eigenstate if a given perturbation is acting on it slowly enough and if there is a gap between the eigenvalue and the rest of the Hamiltonian's spectrum. We consider such system with a Hamiltonian $H(\mathbf{R})$ that depends on time through a vector parameter $\mathbf{R}=( R_1, R_2, R_3......)$. Here $R_i=R_i(t)$ are slowly varying parameters, which can be anything, such as electric field, magnetic field, strains and so on. By slowly varying, we mean that $\mathbf{R}$ changes slightly during the period of motion $T$: $T\frac{d \mathbf{R}}{dt}\ll\mathbf{R}$\cite{landau1}. Denote an instantaneous orthonormal basis of the instantaneous eigenstates as $|n(\mathbf{R})\ket$:
\be
H(\mathbf{R})|n(\mathbf{R})\ket=E_{n}(\mathbf{R}))|n(\mathbf{R})\ket.
\ee
This equation determines the eigenfunction $|n(\mathbf{R})\ket$ up to a phase. We want to study the phase of the wave function of a system that starts from an initial pure state $|n(\mathbf{R}(0))\ket$ as we move $\mathbf{R}(t)$ along the path $\mathcal{C}$.

Assuming that the eigenstates $E_{n}(\mathbf{R})$ is non-degenerate everywhere along $\mathcal{C}$, then according to the adiabatic theorem, as $\mathbf{R}(t)$ varies slowly along $\mathcal{C}$, $|n(\mathbf{R}(0))\ket$ evolves with $H(\mathbf{R})$, and hence $|n(\mathbf{R}(t))\ket$  stays as an instantaneous eigenstate of  $H(\mathbf{R(t)})$ during the whole process. Still, there can be arbitrary phases which may evolve with $\mathbf{R}$ as well. Let's assume an eigenstate $|\psi_{n}(t)\ket$ which differs from $|n(\mathbf{R}(t))\ket$ only by a phase factor $\theta(t)$: $|\psi_{n} (t)\ket=\mathrm{e}^{-i\theta(t)}|n(\mathbf{R}(t))\ket$. Applying $H(\mathbf{R}(t))$ on it, we have
\be
H(\mathbf{R(t)})|\psi_n(t)\ket=i\frac{\p}{\p t}|\psi_n (t)\ket,
\ee
which gives
\be
E_{n}(\mathbf{R}(t))\mathrm{e}^{-i\theta(t)}|n(\mathbf{R}(t))\ket=i\frac{\p}{\p t}(\mathrm{e}^{-i\theta(t)}|n(\mathbf{R}(t))\ket)=\mathrm{e}^{-i\theta(t)}\frac{\p \theta(t)}{\p t}|n(\mathbf{R}(t))\ket +i\mathrm{e}^{-i\theta(t)}\frac{\p}{\p t}|n(\mathbf{R}(t))\ket.
\ee
Here one may have noticed that I use the so-called natural unit system, and hence take $\hbar\to 1$.

Multiplying $\bra \psi_{n}(t)|=\bra n(\mathbf{R}(t)|\mathrm{e}^{i\theta(t)}$ on both sides, we end up with
\be
\frac{\p \theta(t)}{\p t}=E_{n}(\mathbf{R}(t))-i\bra n(\mathbf{R}(t))|\frac{\p}{\p t}|n(\mathbf{R}(t))\ket,
\ee
since $|n(\mathbf{R}(t))\ket$ is normalized: $\bra n(\mathbf{R}(t))|n(\mathbf{R}(t))\ket=1$.
Integrate from $0$ to $t$, we get
\be
\theta(t)=\int_0^t E_n(\mathbf{R}(t'))dt'-i\int_0^t \bra n(\mathbf{R}(t'))|\frac{\p}{\p t'}|n(\mathbf{R}(t'))\ket dt'.
\ee
Therefore, regarding the phase, the state at time t can be written as
\be
|\psi_n(t)\ket=\mathrm{e}^{i\gamma_n}\mathrm{e}^{-i\int_0^t E_n(\mathbf{R}(t'))dt'}|n(\mathbf{R}(t))\ket.
\ee
The second exponential term $\mathrm{e}^{-i\int_0^t E_n(\mathbf{R}(t'))dt'}$ is the ``dynamic phase factor", while the first phase term is the famous Berry phase:
\be
\gamma_n=i\int_0^t \bra n(\mathbf{R}(t'))|\frac{\p}{\p t'}|n(\mathbf{R}(t'))\ket dt'.
\ee
Time can be removed explicitly from this formula:
\be
\gamma_n=i\int_0^{t_{\mathrm{end\; cycle}}} \bra n(\mathbf{R}(t'))|\frac{\p}{\p \mathbf{R}}|n(\mathbf{R}(t'))\ket \frac{\p \mathbf{R}}{\p t'} dt'=i\int_{\mathcal{C}} \bra n(\mathbf{R})|\nabla_{\mathbf{R}}|n(\mathbf{R})\ket d\mathbf{R},
\ee
indicating that the Berry phase only depends on the path in the parameter space, and has nothing to do with the rate at which the path is traversed. It is a geometric phase. We also notice that the Berry phase $\gamma_n$ is real (it's Berry phase, not Berry decay), since $ \bra n(\mathbf{R})|\nabla_{\mathbf{R}}|n(\mathbf{R})\ket$ is purely imaginary: $ \bra n(\mathbf{R})|n(\mathbf{R})\ket=1$  
$\Rightarrow \bra n(\mathbf{R})|\nabla_{\mathbf{R}}|n(\mathbf{R})\ket=- \bra n(\mathbf{R})|\nabla_{\mathbf{R}}|n(\mathbf{R})\ket^*.$ Therefore, the Berry phase can also be written as
\be
\gamma_n=-Im\int_{\mathcal{C}} \bra n(\mathbf{R})|\nabla_{\mathbf{R}}|n(\mathbf{R})\ket d\mathbf{R}.
\ee

\subsection{From Aharonov-Bohm effect to Berry connection and Berry curvature}
The first geometric phase studied and observed was Aharonov-Bohm effect, which is an effect happened in real space, unlike the Berry phase, which is a geometric phase in momentum space. Aharonov-Bohm effect is a phenomenon in which a charged particle is affected by vector potential $\bf{A}$, despite being confined to a region in which both the magnetic field $\BB$ and electric field $\EE$ are zero. Before the discovery of Aharonov-Bohm effect, people believe that the vector potential $\bf{A}$ was introduced into physics only as a mathematical crutch, as there was no magnetic monopole discovered: $\nabla\cdot\BB=\rho_M=0\Rightarrow\BB=\nabla\times\bf{A}$. It was widely believed that vector potential $\bf{A}$ should not have any physical meaning or any physical effect, as said by the distinguished nineteenth-century physicist Heaviside: ``Physics should be purged of such rubbish as the scalar and vector potentials; only the fields $\EE$ and $\BB$ are physical."

\begin{figure}[H]
\begin{tikzpicture}
 \draw [draw=red, very thick, postaction={on each segment={mid arrow=red}},pattern=dots, pattern color=red]
  (8,1) circle(0.8);
\node[below][red] at(8, 1.3) {\Large$\mathbf{B}$};
\draw[thick,blue] (4,1) .. controls (7,2.5) and (9,2.5) ..
node[near start,sloped,above] {\large path 1}(12,1);
\node[below][red] at(8, 1.3) {\Large$\mathbf{B}$};
\draw[thick,green] (4,1) .. controls (7,-0.5) and (9,-0.5) ..
node[near start,sloped,below] {\large path 2}(12,1);
\node[left] at(4,1) {\large electrons};
\node[right] at (12,1) {\Large$\Phi_{\mathbf{B}}$};
\end{tikzpicture}
\caption{Magnetic solenoid effect. Confined magnetic field in the infinite solenoid affects the phase of the electrons traveling beside it.}
\label{fig:ab}
\end{figure}
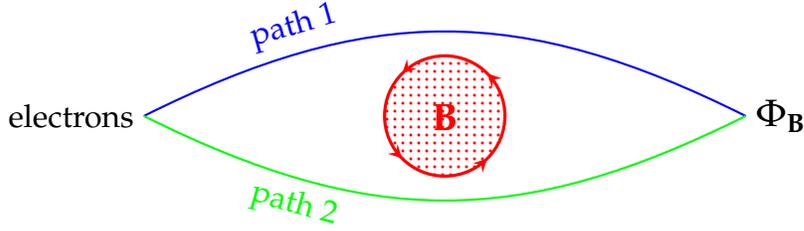

The Abaronov-Bohm effect was brought out in 1959\cite{PhysRev.115.485}. The idea is: Considering an electron passing by an infinite solenoid which has a magnetic field $\BB$ confined in it, see Fig.\ref{fig:ab}. The electromagnetic theory implies that by traveling along a path, the electron acquires a phase $\varphi=e\int_P\mathbf{A}\cdot d\vec{x}$. Thus, when calculate the probability for the propagation, there will be an interference between the contributions from path 1 and path 2
\be
(\mathrm{e}^{ie\int_{P_1}\mathbf{A}\cdot d\vec{x}})(\mathrm{e}^{ie\int_{P_2}\mathbf{A}\cdot d\vec{x}})^*=\mathrm{e}^{ie\oint\mathbf{A}\cdot d\vec{x}}=\mathrm{e}^{ie\int\mathbf{B}\cdot d\vec{S}}=\mathrm{e}^{ie\Phi_B}.
\ee
The electron feels the magnetic potential in the region where the magnetic field is zero. Therefore, the vector potential is essential in physics.

Similarly, we can define a ``vector potential" for Berry phase, which is called Berry connection:
\be
\mathcal{A}_n(\mathbf{R})=i\bra n(\mathbf{R})|\nabla_{\mathbf{R}}|n(\mathbf{R})\ket ,\qquad \gamma_n=\int_{\mathcal{C}}d\mathbf{R}\cdot \mathcal{A}_n(\mathbf{R}).
\ee
Just like the vector potential $\rm{A}$, the Berry connection $\mathcal{A}_n$ is gauge dependent. Under a gauge transformation $|n(\mathbf{R})\ket\to\mathrm{e}^{i\zeta(\mathbf{R})}|n(\mathbf{R})\ket$, where $\zeta(\mathbf{R})$ is a smooth, single-valued function. The Berry connection transforms in the usual way:
\be
\mathcal{A}_n(\mathbf{R})\to\mathcal{A}_n(\mathbf{R})-\frac{\p}{\p \mathbf{R}}\zeta(\mathbf{R}),
\ee
so that the Berry phase is changed by
\be
\gamma_n \to \gamma_n -\int_{\mathcal{C}}\frac{\p}{\p \mathbf{R}}\zeta(\mathbf{R})=\gamma_n-\zeta(\mathbf{R}_f)+\zeta(\mathbf{R}_i).
\ee
Here $\mathbf{R}_i$ and $\mathbf{R}_f$ stand for the initial and final $\bf{R}$ respectively when moving along path $\mathcal{C}$. One may doubt whether we will always be able to cancel the Berry phase by a smart choice of the gauge factor $\zeta(\bf{R})$. The answer is no. We can consider a closed path $\mathcal{C}$, $\mathbf{R}_f=\mathbf{R}_i$, hence $|n(\mathbf{R}_f)\ket=|n(\mathbf{R}_i)\ket$. Gauge transformation must maintain this property $\mathrm{e}^{i\zeta(\mathbf{R}_i)}|n(\mathbf{R}_i)\ket=\mathrm{e}^{i\zeta(\mathbf{R}_f)}|n(\mathbf{R}_f)\ket=\mathrm{e}^{i\zeta(\mathbf{R}_f)}|n(\mathbf{R}_i)\ket$. Therefore, $\zeta(\mathbf{R}_f)-\zeta(\mathbf{R}_i)=2\pi \rm{m}$, where $\rm{m}$ stands for any integer. That is to say, under a closed path, the Berry phase cannot be canceled unless it is equal to $2\pi$ times an integer. Therefore, Berry phase is not trivial.\cite{TITS}

Considering only closed paths, with Stoke's theorem, the ``field", Berry curvature is obtained as:
\be
\bm{\Omega}_n(\mathbf{R})=\nabla_{\mathbf{R}}\times \mathcal{A}_n=i\bra \nabla_{\mathbf{R}} n(\mathbf{R})|\times|\nabla_{\mathbf{R}}n(\mathbf{R})\ket,
\ee
\ba
\gamma_n&=&\int_{\mathcal{C}}d\mathbf{R}\cdot \mathcal{A}_n(\mathbf{R})=i\int_{\mathcal{S}}d\mathbf{S}\cdot(\nabla\times\bra n(\mathbf{R})|\nabla_{\mathbf{R}}|n(\mathbf{R})\ket )\nonumber\\
&=&i\int_{\mathcal{S}}dS_i \e_{ijk}\nabla_j\bra n(\mathbf{R})|\nabla_k|n(\mathbf{R})\ket )
=\int_{\mathcal{S}}d\mathbf{S}\cdot\boldsymbol{\bm{\Omega}}_n(\mathbf{R}).
\ea


Here I need to add a comment: The discovery of Aharonov-Bohm effect does not confirm that the magnetic monopole should not exist.\cite{PhysRevD.12.3845} In other words, the existence of magnetic monopole does not conflict the reality of vector potential $\bf{A}$, even though the magnetic monopole can be written as the divergency of the magnetic field, $\nabla\cdot \mathbf{B}=\nabla\cdot(\nabla\times \bf{A})$, and the divergence of the curl of a vector field is zero. $\nabla\cdot(\nabla\times \bf{A})=0$ is actually the ``Bianchi identity"($dd=0$) in electromagnetism: acting differential operation $d$ on any form twice gives zero\cite{gauge}. ($A=A_{\mu}dx^{\mu}$ is the potential 1-form, with $A_{\mu}$ denoting the electromagnetic 4-potential. $F=(1/2!)F_{\mu\nu}dx^{\mu}\wedge dx^{\nu}$ is the field 2-form, with $F_{\mu\nu}$ denoting the electromagnetic tensor. Of course $F=dA$). The Poincar\'e lemma states that a closed form is locally exact. (A $p$-form $\alpha$ is said to be closed when $d\alpha=0$, and it is exact if there exists a ($p$-1)-form $\beta$ such that $\alpha=d\beta$). Thus, if the curl of a vector field vanishes, the vector field is locally the gradient of some scalar field; if the divergence of a vector field vanishes, the vector field is locally the curl of some vector field. However, a closed form does not need to be globally exact, while only globally exact form ensure a trivial result after integrating over the whole manifold. Just think about a sphere surrounding a magnetic monopole with magnetic charge g.  The magnetic 2-form is $F=(g/2\pi)d \cos\theta d \phi$. Then we find that the potential 1-form $A$ is well-defined everywhere except for that on the south pole or north pole of the sphere, and that is essentially how we end up with a possible nonzero magnetic monopole. Similar to this, we may have Berry monopoles in reciprocal space which gives nonzero Chern number.

\subsection{Berry phase in Bloch band theory}
As mentioned at the very beginning, the Berry phase was studied since 1959, but those works around it were mainly on quantum adiabatic transport. It had not been related to the Bloch-periodic system until 1984, a Japanese guy Mahito Kohmoto\cite{kohmoto1985topological}, as well as a group of people from University of Washington, Qian Liu, D. J. Thouless and Yong-Shi Wu\cite{PhysRevB.31.3372}, build a relation between the Berry curvature and the Hall conductivity, making the Berry phase a significant concept in topological band theory. In Bloch band theory, because of the periodic structure of lattices, the Hamiltonian eigenstates are expressed as
\be
\psi_{n\kk}=\mathrm{e}^{i\mathbf{k}\cdot \rr}u_{n\mathbf{k}}(\rr),
\ee
where $n$ is a band index, $\mathbf{k}$  is a wave vector in the reciprocal-space (Brillouin zone), and $u_{n{\mathbf{k}}}({\mathbf  r})$ is a periodic function of $\mathbf{r}$. Letting $\mathbf {k}$  play the role of the parameter $\mathbf {R}$, one can define Berry connections, and Berry curvatures in the reciprocal space:
\be
\mathcal{A}_n(\mathbf{k})=i\bra n(\mathbf{k})|\nabla_{\mathbf{k}}|n(\mathbf{k})\ket ,
\ee
\be
\bm{\Omega}_n(\mathbf{k})=i\bra \nabla_{\mathbf{k}} n(\mathbf{k})|\times|\nabla_{\mathbf{k}}n(\mathbf{k})\ket.
\ee
The Berry phase across the Brillouin zone is called Zak's phase
\be
 \gamma_n=\oint d\mathbf{k}\cdot \bra n(\mathbf{k})|i\nabla_{\mathbf{k}}|n(\mathbf{k})\ket=\oint d\kk\cdot\mathcal{A}_n(\kk)=\oiint_{BZ}d\kk\cdot\bm{\Omega}_n(\kk)=2\pi C,
\ee
where $C$ is the Chern number.

For a two-dimensional band insulator, the Hall conductivity of it is given by
\be
\sigma_{xy}=\frac{e^2}{\hbar}\int_{BZ}\frac{d^2k}{(2\pi)^2}\bm{\Omega}_{k_x,k_y}.
\ee
Here, the $\hbar$ is restored, and this expression is in CGS.
\subsubsection{Position operator}
We know that without lattice, the momentum operator $\pp=-i\hbar\nabla_\rr$ and the position operator $\rr=i\hbar\nabla_\pp$, which can be straightly forward obtained by plane wave expansion:
\ba
\Psi(\kk)&=&\int d\rr \Psi(\rr)\rm{e}^{-i\kk\cdot\rr},\\
\Psi(\rr)&=&\int d\kk  \Psi(\kk)\rm{e}^{i\kk\cdot\rr}.
\ea
\ba
\kk\Psi(\rr)&=&\int d\kk \kk\Psi(\kk)\mathrm{e}^{i\kk\cdot\rr}=\int d\kk\Psi(\kk)(-i\p_\rr)\mathrm{e}^{i\kk\cdot\rr}=-i\nabla_\rr\Psi(\rr)\\
\rr\Psi(\kk)&=&\int d\rr \rr\Psi(\rr)\mathrm{e}^{-i\kk\cdot\rr}=\int d\rr\Psi(\rr)(i\p_\kk)\mathrm{e}^{-i\kk\cdot\rr}=i\nabla_\kk\Psi(\kk).
\ea
 By restoring $\hbar$ (or $\pp=\hbar\kk$), we get the operators which we are familiar with. Also, but less straight forward we can get the position operator in momentum space by
\be
\rr\Psi(\rr)=\int d\kk (\tilde{\rr}\Psi(\kk))\mathrm{e}^{i\kk\cdot\rr}=\int d\kk \rr\Psi(\kk)\mathrm{e}^{i\kk\cdot\rr}=\int d\kk \Psi(\kk)(-i\p_\kk)\mathrm{e}^{i\kk\cdot\rr}=\int d\kk (i\nabla_\kk \Psi(\kk))\mathrm{e}^{i\kk\cdot\rr}.\label{lsf}
\ee
Here, we use $\tilde{\rr}$ to denote the operator and distinguish with the vector $\rr$. What happens if we have a lattice?

Similarly, any wave function can be written as a superposition of the Bloch waves, in the case of lattice:
\be
\Psi(\rr)=\sum_n \int d\kk \Psi_n(\kk)u_{n\mathbf{k}}(\rr)\mathrm{e}^{i\mathbf{k}\cdot \rr}.
\ee
To obtain the position operator $\rr$ in momentum space, we need to do same thing as what we did in Eq.(\ref{lsf}):
\ba
\rr \Psi(\rr)&=&\sum_n \int d\kk (\tilde{\rr}\Psi_n(\kk))u_{n\mathbf{k}}(\rr)\mathrm{e}^{i\mathbf{k}\cdot \rr}=\sum_n \int d\kk \Psi_n(\kk)u_{n\mathbf{k}}(\rr)(-i\p_\kk)\mathrm{e}^{i\mathbf{k}\cdot \rr}\nonumber\\
&=&\sum_n \int d\kk\{ (i\p_\kk \Psi_n(\kk))u_{n\mathbf{k}}(\rr)\mathrm{e}^{i\mathbf{k}\cdot \rr}+ \Psi_n(\kk)(i\p_\kk u_{n\mathbf{k}}(\rr))\mathrm{e}^{i\mathbf{k}\cdot \rr}\}\nonumber\\
&=&\sum_n \int d\kk\{ (i\nabla_\kk \Psi_n(\kk))u_{n\mathbf{k}}(\rr)\mathrm{e}^{i\mathbf{k}\cdot \rr}+ \Psi_n(\kk)[\int d \rr' \delta(\rr-\rr')i\p_\kk u_{n\mathbf{k}}(\rr')]\mathrm{e}^{i\mathbf{k}\cdot \rr}\}\nonumber\\
&=&\sum_n \int d\kk\{ (i\nabla_\kk \Psi_n(\kk))u_{n\mathbf{k}}(\rr)\mathrm{e}^{i\mathbf{k}\cdot \rr}+ \Psi_n(\kk)[\int d \rr' \sum_m u_{m\kk}^*(\rr')u_{m\kk}(\rr)i\p_\kk u_{n\mathbf{k}}(\rr')]\mathrm{e}^{i\mathbf{k}\cdot \rr}\}\nonumber\\
&=&\sum_n \int d\kk\{ (i\nabla_\kk \Psi_n(\kk))u_{n\mathbf{k}}(\rr)\mathrm{e}^{i\mathbf{k}\cdot \rr}+ \Psi_n(\kk) \sum_m[\int d \rr' u_{m\kk}^*(\rr')i\p_\kk u_{n\mathbf{k}}(\rr')]u_{m\kk}(\rr)\mathrm{e}^{i\mathbf{k}\cdot \rr}\}\nonumber\\
&=&\sum_n \int d\kk\{ (i\nabla_\kk \Psi_n(\kk))u_{n\mathbf{k}}(\rr)\mathrm{e}^{i\mathbf{k}\cdot \rr}+ \Psi_n(\kk) \sum_m \mathcal{A}_{mn}(\kk)u_{m\kk}(\rr)\mathrm{e}^{i\mathbf{k}\cdot \rr}\}\nonumber\\
&=&\sum_n \int d\kk\{[\sum_m (i\nabla_\kk\delta_{mn}+ \mathcal{A}_{mn}(\kk)) \Psi_n(\kk)]u_{m\mathbf{k}}(\rr)\mathrm{e}^{i\mathbf{k}\cdot \rr}\nonumber\\
&=&\sum_n \int d\kk\{[\sum_m (i\nabla_\kk\delta_{mn}+ \mathcal{A}_{mn}(\kk)) \Psi_m(\kk)]u_{n\mathbf{k}}(\rr)\mathrm{e}^{i\mathbf{k}\cdot \rr}.
\ea
Therefore, we have
\be
\tilde{\rr}\Psi_n(\kk)=\sum_m (i\nabla_\kk\delta_{mn}+ \mathcal{A}(\kk)) \Psi_m(\kk),
\ee
thus, the position operator in reciprocal lattice (here I took the tilde off, since there is no need to make distinguishment) is
\be
\rr_{mn}=\sum_m i\nabla_\kk\delta_{mn}+ \mathcal{A}_{mn}.\label{eqrmnB}
\ee
For single band $n$, we have
\be
\rr=i\nabla_{\kk}+\mathcal{A}_n.\label{eqrnB}
\ee
This result can be compared to the momentum operator for a charged particle:
\be
\pp=-i\nabla_\rr+e\AA.
\ee


\subsubsection{Equations of motion}
It is well known that the equation of motion for a particle with charge $e$ under the electromagnetic field is
\be
\dot{\pp}=\mathbf{F}=e\EE+e\dot{\rr}\times\BB=-e\nabla \phi+e\dot{\rr}\times(\nabla\times\AA),
\ee
which is just Lorentz force.
The dual equation is easily guessed out, as shown previously that $\pp$ and $\rr$ are conjugate variables, and $\mathcal{A}$ acts just like $\AA$, $\bm{\Omega}_n$ acts just like $\BB$ in the momentum space: (Also, this equation can be carefully derived, see the reference \cite{Xiao2010})
\be
\dot{\rr}=\frac{\e_{n,\pp}}{\p \pp}-\dot{\pp}\times(\nabla_{\pp}\times\mathcal{A}_n)=\vv_{n\pp}-\dot{\pp}\times \bm{\Omega}_{n\pp}.
\ee

These two equations are tangled with each other. We want to separate $\dot{\rr}$ and $\dot{\pp}$ out. Taking the Lorentz force equation into the velocity equation, we have
\be
\dot{\rr}=\vv_{n\pp}-e\EE\times \bm{\Omega}_{n\pp}-e\dot{\rr}\times\BB\times \bm{\Omega}_{n\pp}=\vv_{n\pp}-e\EE\times \bm{\Omega}_{n\pp}+e\dot{\rr}(\bm{\Omega}_{n\pp}\cdot\BB)-e\BB(\bm{\Omega}_{n\pp}\cdot\dot{\rr}).
\ee
Take the third term on the right hand side to the left hand side:
\be
(1-e\BB\cdot\bm{\Omega}_{n\pp})\dot{\rr}=\vv_{n\pp}-e\EE\times \bm{\Omega}_{n\pp}-e\BB(\bm{\Omega}_{n\pp}\cdot\dot{\rr}).\label{eq34kin}
\ee
Take dot product with $\bm{\Omega}_{n\pp}$ on both sides of the equation:
\be
(1-e\BB\cdot\bm{\Omega}_{n\pp})\dot{\rr}\cdot\bm{\Omega}_{n\pp}=\vv_{n\pp}\cdot\bm{\Omega}_{n\pp}-e(\BB\cdot\bm{\Omega}_{n\pp})(\bm{\Omega}_{n\pp}\cdot\dot{\rr}).
\ee
\be
\Rightarrow \qquad \dot{\rr}\cdot\bm{\Omega}_{n\pp}=\vv_{n\pp}\cdot\bm{\Omega}_{n\pp}.
\ee
Substitute it into Eq.(\ref{eq34kin}). Finally, we get
\be
\dot{\rr}=\frac{1}{1-e\BB\cdot\bm{\Omega}_{n\pp}}[\vv_{n\pp}-e\EE\times \bm{\Omega}_{n\pp}-e\BB(\vv_{n\pp}\cdot\bm{\Omega}_{n\pp})].\label{eqrd}
\ee
The first term is the normal group velocity; the second term is the anomalous velocity appeared due to the interband coherence effects induced by the electric part of Lorentz force; the third term is interband coherence effects induced by the magnetic part of Lorentz force.

Same way, we obtain
\be
\dot{\pp}=\frac{1}{1-e\BB\cdot\bm{\Omega}_{n\pp}}[e\EE+e\vv_{n\pp}\times\BB-e^2(\EE\cdot\BB)\bm{\Omega}_{n\pp}].\label{eqpd}
\ee
The first and second terms represent Lorentz force, and the third term is the origin of chiral anomaly.


\subsection{Time reversal symmetry  and inversion symmetry}
In this section, I am going to talk about the time-reversal symmetric $\mathcal{T}$ and inversion symmetric $\mathcal{I}$ properties of Berry curvature and Chern number.

Time reversal symmetry $\mathcal{T}$ can be represented by
\be
T=UK,
\ee
where $U$ is a unitary matrix and $K$ is complex conjugation. It changes $\kk$ to $-\kk$, $i$ to $-i$, and the periodic part of the Bloch wave function this way:
\be
\mathcal{T}u_{n\kk}(\rr)=T u_{n\kk}(\rr) T^{-1}=u_{n,-\kk}^*(\rr).
\ee
Act time reversal symmetry on Berry connection $\mathcal{A}_n(\kk)=i\bra u_{n\kk}|\nabla_\kk|u_{n\kk}\ket$:
\ba
\mathcal{T A}_n(\kk)&=&-i\bra\mathcal{T}u_{n\kk}|-\nabla_\kk|\mathcal{T}u_{n\kk}\ket=i\int d\rr u_{n,-\kk}(\rr)\p_\kk u_{n,-\kk}^*(\rr)=-i\int d\rr u_{n,-\kk}^*(\rr)\p_\kk u_{n,-\kk}(\rr)\nonumber\\
&=&\mathcal{A}_n(-\kk).
\ea

If a system has time reversal symmetry,
\be
\mathcal{T A}_n(\kk)=\mathcal{A}_n(\kk)+\nabla_\kk \zeta(\kk)\quad \Rightarrow \quad \mathcal{A}_n(-\kk)=\mathcal{A}_n(\kk)+\nabla_\kk \zeta(\kk),
\ee
$\mathcal{A}_n(-\kk)$ and $ \mathcal{A}_n(\kk)$ differ only by a gauge transformation $\nabla_\kk \zeta(\kk)$.

Under time reversal symmetry, the Berry curvature $\bm{\Omega}_{n,i}(\kk)=i\e_{ijl}\bra\p_{k_j}u_{n\kk}|\p_{k_l}u_{n\kk}\ket$ changes:  (here $\e_{ijl}$ is the Levi-Civita symbol.)
\ba
\mathcal{T}\bm{\Omega}_{n,i}(\kk)&=&i\e_{ijl}\bra\p_{k_j}\mathcal{T}u_{n\kk}|\p_{k_l}\mathcal{T}u_{n\kk}\ket=i\e_{ijl}\int d\rr \p_{k_j}u_{n,-\kk}(\rr)\p_{k_l}u_{n,-\kk}^*(\rr)\nonumber\\
&=&-i\e_{ijl}\int d\rr \p_{k_l}u_{n,-\kk}(\rr)\p_{k_j}u_{n,-\kk}^*(\rr)=-i\e_{ijl}\int d\rr \p_{-k_j}u_{n,-\kk}^*(\rr) \p_{-k_l}u_{n,-\kk}(\rr)\nonumber\\
&=&-\bm{\Omega}_{n,i}(-\kk).
\ea

For a system with time-reversal symmetry,
\be
\mathcal{T}\bm{\Omega}_n(\kk)=\bm{\Omega}_n(\kk), \quad \Rightarrow \quad \bm{\Omega}_n(-\kk)=-\bm{\Omega}_n(\kk),
\ee
thus, the Berry curvature $\bm{\Omega}_n(\kk)$ is an odd function of $\kk$. Since the Chern number is the total integral of Berry curvature over the whole Brillouin Zone, it must vanish when Berry curvature is an odd function. Therefore, a system with time reversal symmetry always has zero Chern number.

The inversion $\mathcal{I}$ changes $\kk$ to $-\kk$, and the periodic part of Bloch wave function this way:
\be
\mathcal{I}u_{n\kk}(\rr)=u_{n,-\kk}(-\rr).
\ee
Act on the Berry connection:
\be
\mathcal{IA}_n(\kk)=i\bra\mathcal{I}u_{n\kk}|-\nabla_\kk|\mathcal{I}u_{n\kk}\ket=\mathcal{A}_n(-\kk).
\ee
Therefore, if a system has inversion symmetry,
\be
\mathcal{I A}_n(\kk)=\mathcal{A}_n(\kk)+\nabla_\kk \zeta(\kk)\quad \Rightarrow \quad \mathcal{A}_n(-\kk)=\mathcal{A}_n(\kk)+\nabla_\kk \zeta(\kk),
\ee
$\mathcal{A}_n(-\kk)$ and $ \mathcal{A}_n(\kk)$ differs only by a gauge transformation $\nabla_\kk \zeta(\kk)$.

Act inversion on Berry curvature:
\be
\mathcal{I}\bm{\Omega}_n(\kk)=\bm{\Omega}_n(-\kk).
\ee
If a system is invariant under $\mathcal{I}$:
\be
\bm{\Omega}_n(-\kk)=\bm{\Omega}_n(\kk).
\ee
Berry curvature should be an even function over $\kk$.

If a system has both time reversal symmetry and inversion symmetry, the Berry curvature is both odd and even over $\kk$, thus vanish everywhere in the Brillouin zone. That is essentially the reason why we need to break one of these two symmetries in order to get a topological nontrivial Weyl semi-metal.


\section{Semi-classical transport:  Wave packet and  Boltzmann equation}
Semi-classical theory is a theory in which one part is described with quantum mechanism while the other part is described classically. Transport properties of electrons in crystals are usually studied by semi-classical transport theory.  In the free electron theory, electrons move between two collisions according to the classical equations of motion, while the collisions obey the quantum mechanical Fermi-Golden rule. Therefore, semi-classical models are natural choices. Taking the periodic structure of crystal structure into account, the free electron models have to be extended by the Bloch's theory, in which the electrons are described by Bloch's wave functions. How could the motion between two collisions being classical? Thus, we need the concept of wave packet.

Essentially, electron transport in the crystal is a complicated quantum mechanical many-body problem: we've taken the periodic structure of the crystal into the Hamiltonian to get the Bloch's wave function, but we still need to consider the impurities, crystal defects, and thermal vibrations of the irons, which is electron-phonon interaction. Even if we are lucky enough to find a solution, which must be rather complicated, it would be hard to extract transport properties from the solution. Therefore, a semi-classical theory is preferred, and a statistical treatment is required. Since we are working with statistical mechanism, we need to get the  distribution function of the electrons, and that is why Boltzmann equation lies at the heart of the transport theory.

In this part, I will give a short introduction to the wave packet construction, which is the basis for the semi-classical  theories, and the Boltzmann equation, which plays a key role in the transport mechanism.

\subsection{Wave packet construction and its orbital moment}
From the quantum mechanical point of view, the equations of motion of electrons in periodic potential describe the behavior of wave packets constructed by the superposition of single free electron eigenstates. Therefore, a semi-classical theory works only when the electron position is measured with an accuracy of the wave packet width. Just like what we did to obtain the position operator in the reciprocal lattice in the Berry phase part of this paper, the plane waves are  replaced by Bloch's wave, considering the electrons in crystal. The wave packet we construct with the Bloch functions $\psi_{n\kk}=\mathrm{e}^{i\kk\cdot\rr}u_{n\kk}(\rr)$ from the $n$th band:
\be
|W_0\ket=\int d\kk\; \omega(\kk,t)|\psi_{n\kk} \ket
\ee
$\omega(\kk,t)$ must have sharp distribution, such that the wave vector of $\kk_0$ of the the wave packet makes sense:
\be
\kk_0=\int d\kk\; \kk|\omega(\kk,t)|^2,
\ee
\be
f(\kk_0)=\int d\kk\; f(\kk)|\omega(\kk,t)|^2\label{eqfk}
\ee
What is the criterion for the ``sharpness"? Since we are working in crystal, a natural choice is the Brillouin zone dimension: The width of the wave packet $\delta k$ should be much smaller than the Brillouin zone dimensions, which are of the order of the inverse lattice constant $1/a$. It follows that $\Delta R=1/\Delta k$ must be larger than $a$. Thus, a wave packet of Bloch levels with  a wave vector that is well-defined on the scale of the Brillouin zone must be spread in the real space over many primitive cells. The semi-classical model describes the response of the electrons to externally applied electron and magnetic fields that vary slowly over the dimension of such a wave packet (a few primitive cells).

Unlike a classical point particle, a wave packet has a finite spread around its center of mass, denoted by $\rr_c$, in real space:
\ba
\rr_c&=&\bra W_0|\rr|W_0\ket\nonumber\\
&=&\int d\kk' d\kk \; \omega^*(\kk')\omega(\kk)\bra \psi_{n \kk'}|(-i\frac{\p}{\p \kk}\mathrm{e}^{i\kk\cdot\rr})|u_{n \kk}\ket\nonumber\\
&=&\int d\kk' d\kk \; \omega^*(\kk')\omega(\kk)[(-i\frac{\p}{\p \kk})\delta(\kk-\kk')+\delta(\kk-\kk')\bra u_{n \kk'}|i\frac{\p}{\p \kk}|u_{n \kk}\ket]\nonumber\\
&=&\int d\kk[i(\frac{\p}{\p \kk} {\omega}^*(\kk)){\omega}(\kk)+|{\omega}(\kk)|^2\mathcal{A}_n (\kk)],\label{eqrc}
\ea
where $\mathcal{A}_n (\kk)=i\bra u_{n \kk'}|\frac{\p}{\p \kk}|u_{n \kk}\ket$ is the Berry connection. We have already known that the Berry connection involved with the position operator in the reciprocal lattice from Eq.(\ref{eqrmnB}) and Eq.(\ref{eqrnB}), so this $\rr_c$ formula we get here looks alright. Because of its finite spread in the real space, it may possess a self-rotation around its center of mass, which leads to an orbital magnetic moment \cite{Chang1996}:
\ba
\mathbf{m}(\kk_0)&=&\frac{1}{2}\bra W_0|(\rr-\rr_c)\times\mathbf{j}|W_0\ket\nonumber\\
&=&\frac{e}{2m}\bra W_0|(\rr-\rr_c)\times\mathbf{P}|W_0\ket\nonumber\\
&=&\frac{e}{2m}\int d\kk'\int d \kk\; \omega^*(\kk')\omega(\kk)\bra \psi_{n \kk'}|(\rr-\rr_c)\times\mathbf{P}|\psi_{n \kk}\ket\nonumber\\
&=&\frac{e}{2m}\int d\kk'\int d \kk\; \omega^*(\kk')\omega(\kk)\bra u_{n \kk'}|\mathrm{e}^{i\rr\cdot(\kk-\kk')}(\rr-\rr_c)\times\mathbf{P}|u_{n \kk}\ket\nonumber\\
&=&\frac{e}{2m}\int d\kk'\int d \kk\; \tilde{\omega}^*(\kk')\tilde{\omega}(\kk)\bra u_{n \kk'}|\mathrm{e}^{i(\kk-\kk')\cdot(\rr-\rr_c)}(\rr-\rr_c)\times\mathbf{P}|u_{n \kk}\ket\nonumber\\
&=&\frac{e}{2m}\sum_{n'}\int d\kk'\int d \kk\; \tilde{\omega}^*(\kk')\tilde{\omega}(\kk)\bra u_{n \kk'}|\mathrm{e}^{i(\kk-\kk')\cdot(\rr-\rr_c)}(\rr-\rr_c)|u_{n'\kk}\ket\bra u_{n'\kk}|\times\mathbf{P}|u_{n \kk}\ket\nonumber\\
&=&\frac{e}{2m}\sum_{n'}\int d\kk'\int d \kk\; \tilde{\omega}^*(\kk')\tilde{\omega}(\kk)\bra u_{n \kk'}|i\frac{\p}{\p \kk'}\mathrm{e}^{i(\kk-\kk')\cdot(\rr-\rr_c)}|u_{n'\kk}\ket\bra u_{n'\kk}|\times\mathbf{P}|u_{n \kk}\ket\nonumber\\
&=&\frac{e}{2m}\sum_{n'}\int d\kk'\int d \kk\; \tilde{\omega}^*(\kk')\tilde{\omega}(\kk)[i\frac{\p}{\p \kk'}\bra u_{n \kk'}|\mathrm{e}^{i(\kk-\kk')\cdot(\rr-\rr_c)}|u_{n'\kk}\ket\nonumber\\
& &-\bra i\frac{\p}{\p \kk'} u_{n \kk'}|\mathrm{e}^{i(\kk-\kk')\cdot(\rr-\rr_c)}|u_{n'\kk}\ket]\bra u_{n'\kk}|\times\mathbf{P}|u_{n \kk}\ket\nonumber\\
&=&\frac{e}{2m}\sum_{n'}\int d\kk'\int d \kk\; \tilde{\omega}^*(\kk')\tilde{\omega}(\kk)[i\frac{\p}{\p \kk'}\delta_{n,n'}\delta(\kk-\kk')\nonumber\\
& &-\delta(\kk-\kk')\bra i\frac{\p}{\p \kk'} u_{n \kk'}|u_{n'\kk}\ket]\bra u_{n'\kk}|\times\mathbf{P}|u_{n \kk}\ket\nonumber\\
&=&-i\frac{e}{2m}\int d \kk\; [\frac{\p}{\p \kk} \tilde{\omega}^*(\kk)]\tilde{\omega}(\kk)\bra u_{n\kk}|\times\mathbf{P}|u_{n \kk}\ket-i\frac{e}{2m}\int d \kk\;|\tilde{\omega}(\kk)|^2 \bra \frac{\p}{\p \kk} u_{n \kk}|\times\mathbf{P}|u_{n \kk}\ket\nonumber\\
&=&-i\frac{e}{2m}\int d \kk\; [\frac{\p}{\p \kk} \tilde{\omega}^*(\kk)]\tilde{\omega}(\kk)\times\bra\mathbf{P}\ket_n-i\frac{e}{2m}\int d \kk\;|\tilde{\omega}(\kk)|^2 \bra \frac{\p}{\p \kk} u_{n \kk}|\times\mathbf{P}|u_{n \kk}\ket.
\ea
 $\mathbf{P}=m\frac{\p H(\kk)}{\hbar \p \kk}$ is the mechanical momentum operator, hence $(\rr-\rr_c)\times \mathbf{P}$ is the mechanical angular momentum operator. Also we defined $\tilde{\omega}(\kk)=\mathrm{e}^{i\kk\cdot\rr_c}\omega(\kk)$ to meet the requirement of calculation. The integrant part of the second term can be written as:
\ba
 \bra \frac{\p}{\p \kk} u_{n \kk}|\times\mathbf{P}|u_{n \kk}\ket&=&\frac{m}{\hbar} \bra \frac{\p}{\p k_1} u_{n \kk}|\frac{\p H}{\p k_2}|u_{n \kk}\ket-(k_1\leftrightarrow k_2)\nonumber\\
&=&\frac{m}{\hbar}\frac{\p }{\p k_2} \bra\p_{k_1} u_{n \kk}|H|u_{n \kk}\ket-\frac{m}{\hbar} \bra\p_{k_1} u_{n \kk}|H|\p_{k_2}u_{n \kk}\ket-(k_1\leftrightarrow k_2)\nonumber\\
&=&\frac{m}{\hbar} \bra\p_{k_1} u_{n \kk}|u_{n \kk}\ket\frac{\p \e_n}{\p k_2}+\frac{m}{\hbar} \bra\p_{k_1} u_{n \kk}|\p_{k_2}u_{n \kk}\ket\ \e_n-\frac{m}{\hbar} \bra\p_{k_1} u_{n \kk}|H|\p_{k_2}u_{n \kk}\ket\nonumber\\
& &-(k_1\leftrightarrow k_2)\nonumber\\
&=& \bra\p_{\kk} u_{n \kk}|u_{n \kk}\ket\times\bra\mathbf{P}\ket_n+\frac{m}{\hbar} \bra\p_{\kk} u_{n \kk}|\times (\e_n-H)|\p_{\kk}u_{n \kk}\ket.
\ea
Therefore, the formula of orbital magnetic moment becomes
\ba
\mathbf{m}(\kk_0)&=&-i\frac{e}{2m}\int d \kk\; [\frac{\p}{\p \kk} \tilde{\omega}^*(\kk)]\tilde{\omega}(\kk)\times\bra\mathbf{P}\ket_n-i\frac{e}{2m}\int d \kk\;|\tilde{\omega}(\kk)|^2  \bra\p_{\kk} u_{n \kk}|u_{n \kk}\ket\times\bra\mathbf{P}\ket_n\nonumber\\
& &-i\frac{e}{2\hbar}\int d \kk\;|\tilde{\omega}(\kk)|^2 \bra\p_{\kk} u_{n \kk}|\times (\e_n-H)|\p_{\kk}u_{n \kk}\ket\nonumber\\
&=&-\frac{e}{2m}\int d \kk\;\{ [i\frac{\p}{\p \kk} \tilde{\omega}^*(\kk)]\tilde{\omega}(\kk)+|\tilde{\omega}(\kk)|^2\mathcal{A}_n (\kk)\}\times\bra\mathbf{P}\ket_n\nonumber\\
& &-i\frac{e}{2\hbar}\int d \kk\;|\tilde{\omega}(\kk)|^2 \bra\p_{\kk} u_{n \kk}|\times (\e_n-H)|\p_{\kk}u_{n \kk}\ket\nonumber\\
&=&-\frac{e}{2\hbar}\int d \kk\;|{\omega}(\kk)|^2\rr_c \times\frac{\p\e_{n\kk}}{\p\kk}-i\frac{e}{2\hbar}\int d \kk\;|\tilde{\omega}(\kk)|^2 \bra\p_{\kk} u_{n \kk}|\times (\e_n-H)|\p_{\kk}u_{n \kk}\ket\nonumber\\
&=&i\frac{e}{2\hbar}\bra \nabla_{\kk_0} u|\times[H(\kk_0)-\epsilon(\kk_0)]|\nabla_{\kk_0} u\ket.
\ea
Here I used Eq.(\ref{eqrc}) and Eq.(\ref{eqfk}) to get the first two terms vanish.

Finally we obtain the obital magnetic moment
\be
\mathbf{m}(\kk)=i\frac{e}{2\hbar}\bra \nabla_{\kk} u|\times[H(\kk)-\epsilon(\kk)]|\nabla_{\kk} u\ket,
\ee
which does not depend on the actual shape and size of the wave packet but only on the Bloch functions. It does not depend on the way the wave packet was constructed as well, since there is no $\omega(\kk)$ dependence in the final formula. The orbital moment transforms exactly like the Berry curvature under discrete symmetry operations. Therefore, it vanishes if both time reversal symmetry and inversion symmetry are protected. This intrinsic orbital moment act exactly like electron spin: By applying a magnetic field, it couples to the field through a Zeeman  term $-\mathbf{m}(\kk)\cdot\BB$.

\subsection{Boltzmann equation}
The Boltzmann equation is used to describe the statistical behavior of a thermodynamic system, which is not in a state of equilibrium. To be exact, it  describes the time evolution of the electron distribution function $f(\rr,\kk,t)$. Its physical interpretation is that $f(\rr,\kk, t) d\rr d\kk$ is the number of electrons at point $\rr$ with wave vector $\kk$ (or wave packets with mean position $\rr$ and mean momentum $\kk$), in the phase space volume $d\rr d\kk$. Thus, the total integral of  $f(\rr,\kk,t)$ over the whole phase space gives the number of electrons. The time variation of it comes from three effects: diffusion, drift and collision. Diffusion is caused by any nontrivial gradient in the electron concentration,
$\nabla_\rr f(\rr,\kk,t)$ (or $\p_\rr  f$) . Drift is caused by external forces, which can be deemed as diffusion in $k$ space, $\nabla_\kk f(\rr,\kk,t)$ (or $\p_\kk f$), since
\be
\frac{d}{d t}f(\rr,\kk,t)=\frac{\p}{\p t}f(\rr,\kk,t)+\dot{\rr}\frac{\p}{\p \rr}f(\rr,\kk,t)+\dot{\kk}\frac{\p}{\p \kk}f(\rr,\kk,t),
\ee
and $\dot{k}$ is apparently a force term given by Newton's second law. The third term (drift term) can be deemed as the dual of the second term (diffusion term) in $k$ space.

If without collision, $\frac{d}{d t}f(\rr,\kk,t)=0$, given by Liouville's theorem, which can be proven by continuity equation:
\be
\frac{\p}{\p t}f+\frac{\p (f\dot{\rr})}{\p \rr}+\frac{\p (f\dot{\kk})}{\p \kk}=0,
\ee
where $(f, f\dot{\rr},f\dot{\kk})$ is a conserved current in $r$-$k$ space.\cite{Mermin}

According to Bloch's theory, an electron in a perfect lattice should experience no collision at all. However, real lattices are not prefect: they have impurities and crystal defects, which scatter the electrons. What is more, the ions that form the lattices are not fixed: they have thermal vibrations, which is usually described as phonons in quantum mechanics. The phonon-electron interaction usually dominates the collision term in room temperature, while at low temperature, the impurity and defect scatterings dominate, since the vibration of the ions declined with the temperature drop.

The distribution function changed by all kinds of collisions is denoted as $(\frac{d f}{d t})_{coll}$. If it is positive, it means that the collisions lead to an increasing number of electrons at $(\rr,\kk,t)$, and negative means decrease. Obviously, we get
\be
\bigg(\frac{d f(\rr,\kk,t)}{d t}\bigg)_{coll}=\frac{\p}{\p t}f(\rr,\kk,t)+\dot{\rr}\frac{\p}{\p \rr}f(\rr,\kk,t)+\dot{\kk}\frac{\p}{\p \kk}f(\rr,\kk,t).
\ee
This is the famous Boltzmann equation. $\dot{\rr}$ and $\dot{\kk}$ are obtained by solving equation of motion, which is done in the previous Berry phase part.

Now, l am going to deal with the collision part. First,  let me introduce the scattering probability $W_{\kk,\kk'}$, which is defined this way: assume an electron with wave vector $\kk$ is scattered into any one of the group of levels  (with the same spin) contained in the infinitesimal $k$-space volume $d\kk'$ around $\kk'$. (Suppose all these levels are unoccupied, hence not forbidden by the exclusion principle). The probability for this scattering to happen, in an infinitesimal time interval $dt$, is
\be
\frac{W_{\kk,\kk'}dtd\kk'}{(2\pi)^3}.
\ee
$W_{\kk,\kk'}$ is usually obtained by Fermi's Golden Rule:
\be
W_{\kk,\kk'}=\frac{2\pi}{\hbar}n_i\delta(\e(\kk)-\e(\kk'))|\bra\kk'|U|\kk\ket|^2,
\ee
where $n_i$ is the impurity density, and $U$ describes the interaction between the impurity and the electron. Here, we assume a low temperature case with sufficiently dilute impurities, and the interaction is sufficiently weak. More general cases are studied in Many Body Quantum Field Theory, where loop graphs and Green's functions based on different approximations for different circumstances were used to get the scattering amplitude. We use $U(\rr)=U_0\delta(\rr)$ to represent short range disorders, while for long range disorders we use correlation functions such as $\bra U(\rr_1)U(\rr_2)\ket=W^2$ with $\bra U(\rr)\ket=0$.

Because of the exclusive principle, only unoccupied $\kk'$ states are not forbidden, so the total probability for any electron with wave vector $\kk$ scattered into states with wave vector $\kk'$ per unit time is given by
\be
\frac{1}{\tau(\kk)}=\int d\kk' \;W_{\kk,\kk'}[1-f(\kk')].
\ee
$\tau(\kk)$ is the relaxation time. By integrate over $d\kk$, I mean integrating over $\frac{d\kk}{(2\pi)^3}$.  Usually, I neglect $(2\pi)^3$ when performing the formulas, but during the calculations, we know that it has to be there. To make the collision happen, we need not only $\kk'$ states unoccupied but also  $\kk$ states occupied. Therefore, the change of the distribution function per unit time, because of electrons that have wave vector $\kk$ scattering out, is
\be
\bigg(\frac{d f(\kk)}{d t}\bigg)_{out}=-\frac{f(\kk)}{\tau(\kk)}=-f(\kk)\int d\kk' \;W_{\kk,\kk'}[1-f(\kk')].
\ee
Similarly, we can get the change of distribution function per unit time, that comes from``scattering in" collisions (electrons that was not in $\kk$ states get their wave vector changed to $\kk$ after scattering):
\be
\bigg(\frac{d f(\kk)}{d t}\bigg)_{in}=[1-f(\kk)]\int d\kk'\;W_{\kk',\kk}f(\kk').
\ee
Therefore, the total contribution from collision is
\be
\bigg(\frac{d f(\kk)}{d t}\bigg)_{coll}=-\int d\kk'\{W_{\kk,\kk'}f(\kk)[1-f(\kk')]-W_{\kk',\kk}f(\kk')[1-f(\kk)]\}.
\ee
In the relaxation-time approximation this is simplified to
\be
\bigg(\frac{d f(\kk)}{d t}\bigg)_{coll}=-\frac{f(\kk)-f_0(\kk)}{\tau(\kk)},\label{relaxt}
\ee
where $f_0(\kk)$ is the equilibrium distribution function, which is just Fermi-Dirac distribution function for electrons.



\bibliographystyle{apsrev}
\bibliography{chap1} 

%% file: chap2.tex

\chapter{An overview of Weyl semimetal phase}

\newpage
In the last century, or even in a more distant past, physics is mostly centered on symmetries, which are intimately related to conservation laws by Noether's theorem\cite{Noether1971}. One of the most important examples of symmetry in physics is that the speed of light is the same in any coordinate system, which is indicated by mathematicians as ``Poincar\'e group'', the symmetry group of the special relativity. Another important example is the invariance of the form of physical laws under arbitrary differentiable coordinate transformations, which is a representative idea in general relativity. Also last century witnessed the prosperity of high energy physics not only the discovery of all those fundamental particles but also when quantum electrodynamics, Glashow-Weinberg-Salam theory of electroweak processes, quantum chromodynamics come together, forming the so called standard model, while from the symmetry point of view, an $\bf SU(3)\times SU(2)\times U(1)$ group.

Our current century is believed to be a century of topology. In mathematics, topology is a concept concerned with the properties of space that are preserved under continuous deformations, such as stretching, bending and crumpling, but not tearing or gluing. Since the study into the topological insulators at very the beginning of our century\cite{Haldane, Hasan2010, Xiaoliang2011}, the study of electron structure topology of crystalline material has been an extremely important subject in  the modern condensed matter physics. These topological phases have some property to which an integer can be assigned (Chern number, as introduced in chapter 1), which is robust and only depends on global properties: they cannot be destroyed by local perturbations such as disorder and scattering, as long as the bulk gap is not closed, and  that is why we call them ``topological". In 2016, David J. Thouless, F. Duncan M. Haldane, and J. Michael Kosterlitz won Nobel Prize in Physics for ``theoretical discoveries of topological phase transitions and topological phases of matter". Recently, the enthusiasm on this theme shifted to topological semimetals or metals, due to the theoretical prediction of topological Weyl semimetals\cite{Burkov2011, Burkov20112, Wan2011p, Weng2015, huang2015weyl} and Dirac semimetals\cite{wang2012, Zhijun2013, Young2012}, followed by experimental realizations of them: Dirac semimetal\cite{Xu1256742, PhysRevLett.118.146402, neupane2013observation} and topological Weyl semimetal\cite{Lv2015, Lu622, Xu613}.

\section{From Graphene to Weyl semimetal}
The Nobel prize in physics for 2010 was awarded to Andre Geim and Konstantin Novoselov ``for groundbreaking experiments regarding the two-dimensional material gra- -phene". Graphene is a single sheet of carbon atoms arranged in a honeycomb lattice. Electrons moving around the carbon atoms interact with the periodic potential of the honeycomb lattice,  which give rise to a Fermi surface with six double cones where the valence and conductance bands touch each other. The dispersion relation is linear near the band touching point, which brings about relativistic Dirac fermions as quasi-particles, where the speed of light is replaced by the Fermi velocity and the spins are replaced by pseudo-spins associated with the sublattices. The general Hamiltonian is usually expressed as
\be
H=v_F(\sigma_x p_x+\sigma_y p_y).
\ee
Clearly, it can be easily gapped out by perturbations (mass terms) proportional to σ$\sigma_z$. The stability of graphene comes from the extra symmetries under time-reversal and spatial inversion, which enforce the vanishing of terms proportional to σ$\sigma_z$, ensuring that no gap is induced, when perturbations do not break time-reversal and inversion symmetry.

It is possible to generalize this model into 3D, and we can have Hamiltonian looks like this:
\be
H=
\begin{bmatrix}
0 &&v_F \bm{\sigma}\cdot\pp\\
v_F \bm{\sigma}\cdot\pp &&0
\end{bmatrix}.
\ee
Here $\bm{\sigma}=(\sigma_x, \sigma_y, \sigma_z)$ is a triplet of Pauli matrices. This is a simplest model for Dirac semimetal. It is still easily gapped out by $4\times 4$ diagonal matrices, and need crystal symmetries to protect it. However, it is possible to split the degeneracy of the Dirac node in momentum space by breaking either time-reversal or inversion symmetry. This generates two Weyl nodes with opposite chiralities (or helicities), of which dispersion is given by the massless Weyl Hamiltonian.

The robustness Weyl nodes is obvious: the Hamiltonian of Weyl semimetal is written with $2\times 2$ Pauli matrices,  and we have used all three of them, so it is hard to gap it out. From topological point of view, Weyl points with opposite nonzero Chern number, associated with the 2D Fermi surface sheet, are separated, so they are topologically protected: a Gauss surface surround a Weyl node detect the chirality of it, preventing it from disappear unless  another Weyl node with opposite chirality enter in the surface. One might be familiar with the Gauss-Bonnet theorem, which is a theorem on 2D manifold. The origin for 2D topological insulators to be ``topological'' is that their completely filled bands leave the first Brillouin Zone a perfect torus in momentum space, where Gauss-Bonnet theorem applies. Weyl semimetal has a 3D Brillouin Zone, so we must choose a  closed 2D manifold around the Weyl points, within the 3D bulk Brillouin Zone. The Chern number is defined as
\be
C=\frac{1}{2\pi}\int_{\mathcal{S}}\bm{\Omega}(\kk)d\bf{S},
\ee
where $\mathcal{S}$ denotes a closed 2D manifold in the Brillouin Zone, compared to the 2D topological insulator case, which is integral over the whole Brillouin Zone. Apparently, the Chern number only depends on the number of Weyl nodes or monopoles contained in the closed manifold, and that is why it is a topological property. Since for metals, everything interesting happens around the Fermi surface, we require Weyl nodes being located near the Fermi surface, to make it a Weyl (semi-)metal. Those with Weyl nodes located exactly at the Fermi level are called ``Weyl semimetals'', and those with  Weyl nodes around the Fermi energy are named as ``Weyl metals''. I may not distinguish them if not necessary to.

In the vicinity of Weyl points, the Berry curvature takes a general form (similar to the assumed magnetic monopole formula):
\be
\bm{\Omega}(\kk)=\pm\frac{\kk}{2k^3},
\ee
which leads to a Chern number $C=\pm 1$, indicating a sink or source enclosed by the 2D surface we chose. A nonzero Chern number, or topological number, is the reason why Weyl semimetal is called ``topological Weyl semimetal'', compared to Dirac semimetal, where all bands are doubly-degenerated due to the Kramers theorem, and Berry curvature is zero everywhere.

Near the Weyl points, the Hamiltonian takes a universal form as well:
\be
H=\pm v_F \bm{\sigma}\cdot\kk.
\ee
This Hamiltonian is identical to the Hamiltonian of free Weyl fermions proposed by Hermann Weyl in 1929, if one replace $v_F$ with the speed of light $c$. The $\pm$ sign, which indicates the opposite Chern numbers, now corresponds to Left- and Right- handed Weyl fermions. In 1929, Hermann Weyl had noticed that the Dirac equation could be separated into two Weyl spinors with opposite chirality. However, as the lightest neutrinos turned out to have nonzero mass, it seemed  hopeless to find a natural particle which is a Weyl fermion. Now we know that, in condensed matter field, it is accomplishable to get quasi-particles with every property a Weyl fermion supposed to have. Weyl semimetals might be able to provide us a platform for testing theories about Weyl fermions, with a lower ``speed of light''.

$Remark:$ How to gap a Weyl (semi-)metal out, as it is topologically stable? Since the total Chern number in the entire Brillouin zone has to be zero, known as the ``Nielsen-Ninomiya theorem''\cite{nielsen1981no}, we cannot gap out a single Weyl point. Instead, we have to move two Weyl points with opposite  chirality on top of each other to annihilate them. Then further perturbation might gap out the system. If we go beyond band theory, superconductivity might be able to gap the system by breaking U(1) symmetry. In addition, charge density wave (CDW) and disorders may gap the Weyl points out by breaking translational symmetries. Weyl (semi-)metals are protected by U(1) charge symmetry and translational symmetries\cite{hasan2017discovery}.

\section{Theoretical model and special surface states}
To provide a more intuitive idea about Weyl semimetal, let me at least adopt a theoretical model\cite{turner2013beyond}:
\ba
\hat{h}(\kk)&=&a(\kk)\sigma_x+b(\kk)\sigma_y+c(\kk)\sigma_z,\quad H=\sum_\kk \hat{h}(\kk)\nonumber\\
a(\kk)&=&-2t_x(\cos k_x-\cos k_0)+m(2-\cos k_y-\cos k_z),\nonumber\\
b(\kk)&=&2t_y\sin k_y,\quad c(\kk)=2t_z\sin k_z.
\ea
Obviously, it has two nodes at $\kk=\pm k_0 \hat{x}$. Denote $\pp^\pm=(\pm k_x\mp k_0, k_y, k_z)$, and expand the Hamiltonian around the nodes, we get:
\be
H^\pm=2t_x\sin k_0 p^\pm_x\sigma_x+2t_y p^\pm_y\sigma_y+2t_z p^\pm_z\sigma_z,
\ee
which is an anisotropic version of $H=\pm v_F \bm{\sigma}\cdot\kk.$ It is okay to have different Fermi velocity in different directions, since we are in the lattice, where Lorentz invariance is not required. The fact that the Lorentz invariance need not exist is also the origin of a new type of Weyl semimetal: the Type \uppercase\expandafter{\romannumeral2} Weyl semimetal\cite{Soluyanov2015}. We are not going to give a further introduction to the Type \uppercase\expandafter{\romannumeral2} Weyl semimetal, since it is not the points of focus of this thesis.

With this model, we can calculate the eigenstates:
\ba
\psi_+&=&\frac{1}{\sqrt{2d(d+2t_z p_z)}}\begin{bmatrix} 2t_z p_z+d \\ 2t_x p_x+2i t_y p_y \end{bmatrix},\nonumber\\
\psi_-&=&\frac{1}{\sqrt{2d(d-2t_z p_z)}}\begin{bmatrix} 2t_z p_z-d \\ 2t_x p_x+2i t_y p_y \end{bmatrix},\nonumber\\
d&=&\sqrt{(2t_x p_x)^2+(2t_y p_y)^2+(2t_z p_z)^2},
\ea
and the Berry connection
\be
\mathcal{A}_i(\kk)=i\bra \psi_-|\p_{k_i}|\psi_-\ket=\frac{1}{d(d-2t_z p_z)}[2t_y p_y\p_i (t_xp_x)-2t_x p_x\p_i (t_y p_y)],
\ee
and the Berry curvature
\be
\Omega_{i}=\frac{1}{2d^3}\e_{ijk}\e_{abc}(2t_a p_a)\p_j (2t_b p_b)\p_k (2t_c p_c).
\ee
When integrating over the base manifold, we can find that the Chern number is always an integer, $\pm 1$.

Now we look into the surface states of this model. The first nontrivial question to ask  is: how can surface states be protected from hybridizing with bulk states in a gapless system? The answer lies in the translational symmetry:  the translational symmetry in the lattice guarantees the conservation of momentum of surface states, so the surface states at the Fermi level are stable at any momenta where there are no bulk states at the same energy, since they cannot decay into bulk states. For Weyl semimetals, which have Fermi energy strictly located at the Weyl nodes, the bulk states at Fermi level are just at the projections of the Weyl nodes to the surface Brillouin zone, and all other surface states at the Fermi energy are stable and well-defined. This leads to unique, non-closed surface states at the Fermi energy: the Fermi arcs.

\begin{figure}[H]
 \centering
  \includegraphics[width=0.9\textwidth]{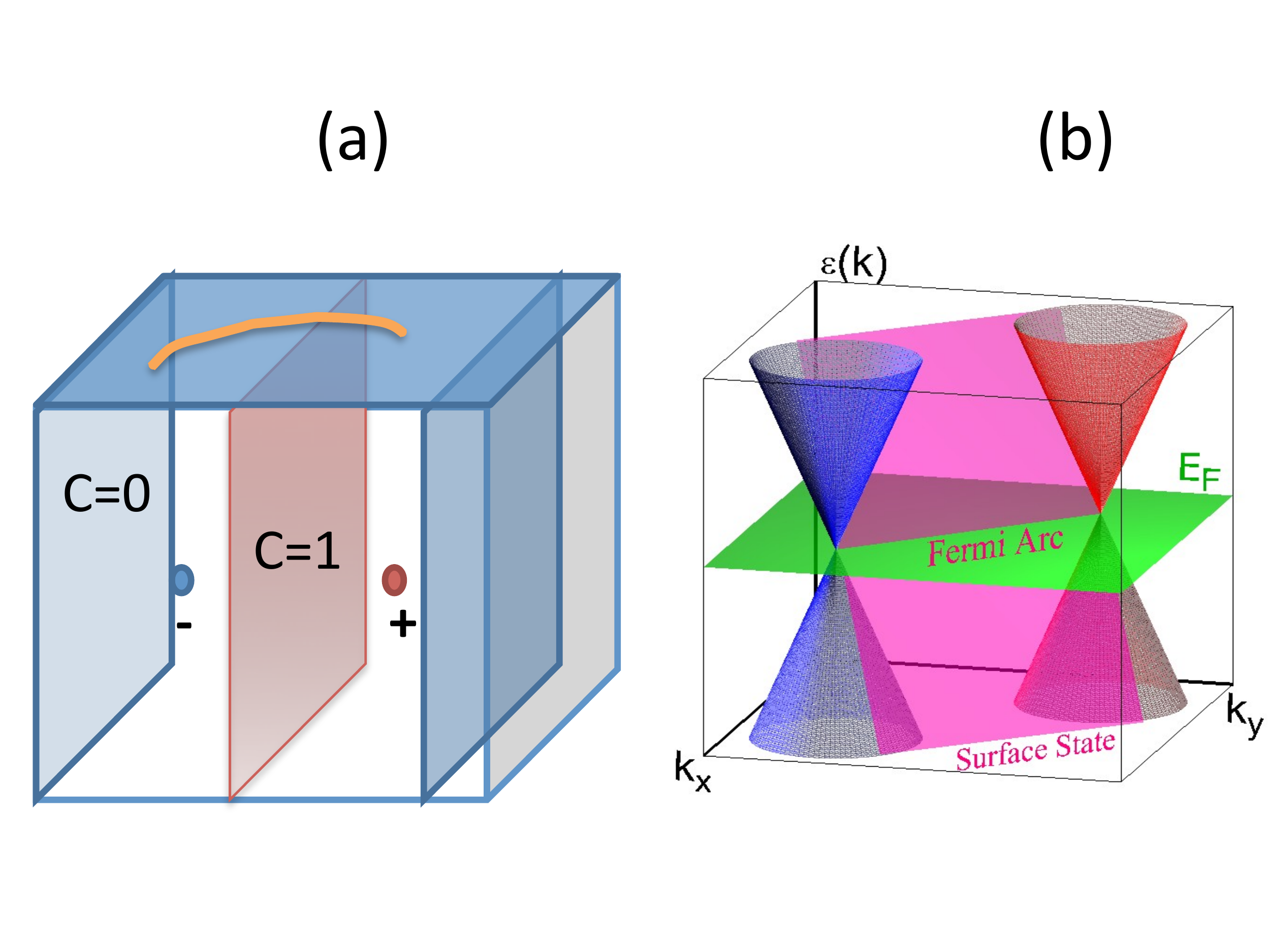}
 \caption{Surface states of a Weyl semimetal. (a) C=0 slices correspond to 2D normal insulators (NI), and C=1 slices correspond to Chern insulators with Chern number C=1. Fermi arc is formed by adding all edge states of 2D Chern insulators. (b) A graph of the dispersion relation of surface states (the pink plane) and bulk states (blue and red cone). We can see how the surface state joins to the bulk states here. These two graphs are from the paper ``Beyond Band Insulators: Topology of Semi-metals and Interacting Phases'' by Ari M. Turner and Ashvin Vishwanath\cite{turner2013beyond}. Permission has been received from the authors.}
   \label{Fermiarcs}
 \end{figure}

Think the momentum space of the 3D Weyl system as a stack of 2D slices as shown in Fig.(\ref{Fermiarcs}). Certainly, for slices that do not contain a Weyl node, they are totally gapped 2D systems, and we can calculate the Chern number by integrating the Berry curvature over the 2D surface. Those slices with 0 Chern number are just 2D normal insulators. The slices between two Weyl points are Chern insulators with Chern number C=1. We know that Chern insulators have a chiral edge state. The Fermi arc is nothing special, but a stacking of chiral edge states of Chern insulators.

Fermi arcs have been observed by ARPES measurements in monophosphides TaAs class Weyl semimetals \cite{Lv2015, Xu613, XuSY2015, Xue1501092}. However, to be exact the Fermi arcs detected in TaAs is not disjoint arcs but closed contours, dubbed as Fermi kinks. What is a Fermi kink? Why can it be a proof of the existence of the Weyl nodes? We know that the two ends of a Fermi arc are the projection of a pair of Weyl points on the surface Brillouin zone. What if there are more than one Weyl point projecting on the same point of the surface Brillouin zone? Yes, we will have more than one Fermi arcs start and end at the same Weyl points projection. If the number of pairs of Weyl points sharing same projection points on the surface is an even number, we have closed contours. That is the case for TaAs. However, closed contours does not mean that we will miss the signal of Weyl points, because generally speaking, those Fermi arcs attach to same Weyl projection points with different slopes, which  leads to a kink in the contour. Therefore, either odd number of Fermi arcs or Fermi kinks being observed can stand as the evidence of the existence of Weyl points.

\section{Where to find Weyl things?}
In this section, I will first try  to recover the most famous attempts that physicists took to search possible materials and realize Weyl (semi-)metals in the past decade. Followed by this, I will talk about experimental breakthroughs in this area.

\subsection{Search for possible materials}
Let us give a guess first: where to start if one is seeking for Weyl material? There is a hint lying in the last section where we introduce Fermi arcs by viewing Weyl system as a stack of 2D normal insulators and Chern insulators. We might be able to find Weyl (semi-)metals at some quantum phase transition point. That is an idea proposed by Shuichi Murakami in 2007\cite{murakami2007phase}. Murakami suggests that a gapless phase (Weyl points) appears between the quantum spin Hall (QSH) phase and the insulator phase in 3D inversion symmetry broken systems. In order to get Weyl (semi-)metals, three necessary conditions are required: either breaking inversion symmetry or time reversal symmetry to avoid double degeneracy, gapless points as Weyl nodes which are sinks and sources of Berry flux and Fermi energy located at/near the Weyl points. We might be able to find Weyl nodes at some phase transition point in a magnetic system (time reversal symmetry broken) or a inversion-symmetry-break system. How to make the Fermi energy locate at the Weyl points? Luttinger's theorem, which is obtained directly from Pauli exclusion principle, states that the volume enclosed by a Fermi surface of a material is directly proportional to its electron density. Therefore, the Fermi level is only determined by the electron density. This suggests that, at least theoretically, one way to get Weyl (semi-)metals is to find an insulator with time reversal symmetry or inversion symmetry broken, and then close a gap. This will lead the Fermi level being pinned to the band touching point.
Followed by this intuition, in 2011, A. A. Burkov and Leon Balents proposed the idea of realizing Weyl semimetal in a magnetic topological insulator(TI) multilayer, which consists of identical thin films of a magnetically dopped (to break time reversal symmetry of course) 3D topological insulator, separated by ordinary insulator layers\cite{PhysRevLett.107.127205}. They used a simple tight-binding model which showed that their multilayer exhibit a quantum Hall plateau transition from $\sigma_{xy} = 0$ (normal insulator) to $\sigma_{xy} = e^2/h$ (quantum Hall insulator) when varying the tunneling between the top and bottom surface of the TI layer, or the exchange spin-splitting (raised by the magnetic impurities). Weyl nodes appear as an intermediate phase in this quantum phase transition. This model is still to be realized in experiments.

Historically, somehow, nearly all attempts of seeking and prediction of Weyl materials were concentrated on time reversal symmetry breaking Weyl semimetals at the very beginning. The first material candidate for Weyl semimetals was a family of magnetic pyrochlore iridates, $R_2\rm{Ir}_2 O_7$,  where $R$ is a rare earth element. It was suggested by Wan $et$ $al.$ in 2011\cite{Wan2011p}. Theoretical study shows that  $R_2\rm{Ir}_2 O_7$ exhibits a transition from an ordinary magnetic metal to a Mott insulator, with the increasing of the on-site Coulomb interaction, and a Weyl semimetal sits in between. They also studied the possible Fermi arcs in this system. Their calculation results brought a fever in Weyl material seeking projects. However, attempts to realize a Weyl semimetal in $R_2\rm{Ir}_2 O_7$ encountered unprecedented difficulties. The metal-insulator phase transition was observed and transport behaviors were roughly the way it should be for a semimetal, but the overall results were not persuasive enough, not to say that ARPES measurements were lacking. (ARPES stands for angle-resolved photoemission spectroscopy. It is a technique provide direct observation of bulk and surface states. Intuitively, it works by shedding light on material and  measuring energy, momentum, and spin of the photoelectrons.) After $R_2\rm{Ir}_2 O_7$, there are also some other proposals about time reversal symmetry breaking Weyl material, like $\mathrm{Hg}_{1-x-y}\mathrm{Cd}_x\mathrm{Mn}_y\rm{Te}$\cite{PhysRevB.89.081106}, and magnetic doping of Dirac semimetals, but none of them got a successful ARPES measurement. They all faced same difficulties such, that it is hard to grow high quality magnetically dopped crystals with a useful magnetic order, as disorder from doping degrades the sample quality, while the spin splitting from magnetic doping might be too small to produce Weyl points.

Since there are many problems involved with magnetic doping for time reversal symmetry broken Weyl semimetals, people started to think, that it might be easier to realize inversion symmetry broken Weyl material. Since inversion symmetry broken is a property of crystal structure, which can be detected by X-ray diffraction, the research into the inversion symmetry broken Weyl material benefits a lot from those databases provided by X-ray diffraction experiments in the past century. Inversion symmetry breaking adds no complications to $ab$ $initio$ calculation, compared to those magnetic materials. In 2012, Halasz and Balents proposed $\rm{HgTe}/CdTe$ heterostructure as possible inversion symmetry broken Weyl material \cite{PhysRevB.85.035103}. They state a Weyl semimetal phase between the normal insulator and the topological insulator phases in this system, which is similar to the time reversal symmetry broken topological multilayer. This proposal is still not realized, as it requires topological insulator and normal insulator with matching lattice structures that can both be grown by a thin film technique. In 2014, with first principle calculation, Jianpeng Liu and David Vanderbilt  suggested there was a robust Weyl-semimetal phase exists in the solid solutions $\rm LaBi_{1-x}Sb_{x}Te_3$ and $\rm LuBi_{1-x}Sb_{x}Te_3$ for $x\sim38\%-41.9\%$ and $x \sim 40.5\%-45\%$, which remains unrealized\cite{LiuJP2014}. There are also a lot of other unrealized proposals for inversion symmetry broken Weyl materials, like tellurium or selenium crystals under pressure. For further reading, see the review paper by M. Zahid Hasan $et$ $al$\cite{ZahidHasan}.

In 2015, two groups of people independently predict Weyl semimetal phase in noncentrosymmetric transition-metal monophosphides,
 $\rm{TaAs}$ class with first principles calculations\cite{ Weng2015, Huang2015}. The $\rm{TaAs}$ family are natural Weyl semimetals, with 12 pairs of Weyl points for each of them. They are experimentally detected to have Weyl points in the bulk, as well as Fermi arcs on the surface shortly after the theoretical prediction\cite{Lv2015, Xu613}.
\subsection{Milestones in experiments}
In 2015, both Weyl points in bulk states and Fermi arcs  in surface states were directly observed by ARPES measurements in $\rm{TaAs}$ by Zahid Hasan's group in Princeton University\cite{Xu613} and Hong Ding's group in Beijing National Laboratory for Condensed Matter Physics\cite{Lv2015}. Soon after the identification of  $\rm{TaAs}$, experiments on other $\rm{TaAs}$ family Weyl semimetals, like $\rm NbAs$\cite{XuSY2015} and $\rm TaP$\cite{Xue1501092} also got prominent results.

Meanwhile, Marin $\rm Solja\check{c}i\acute{c}$'s group from MIT, and Lixin Ran's group from Zhejiang University observed Weyl states in a photonic crystal, which is a material with a periodic pattern of holes that only transmits light with certain frequencies\cite{Lu622}. They carved arrays of holes into ceramic layers with a computer controlled milling machine, and then stacked them to make a 3D inversion symmetry broken photonic crystal.

$\rm TaAs$ has a body-centered tetragonal lattice, with a space group $\rm I4_1md (\#109)$. Its structure lacks an inversion symmetry. A non-zero Chern number of Weyl points in TaAs was directly measured from an ARPES measurement\cite{PhysRevLett.116.066802}, which again verifies that it is a Weyl semimetal. As we mentioned in the previous section, that Fermi arcs do not always appear as disjoint
arcs in Weyl semimetals. In TaAs, the Fermi arcs appear in pairs which together form a closed contour, a surface state kink.

\section{Landau levels, Chiral anomaly and magnetotransport properties}
Perhaps the most significant consequence of a nontrivial electronic band topology is that it may bring unique transport phenomena or response to external probing, such as quantum hall effect. In this section, we will present the most famous consequence of having a nontrivial Weyl points topology -- chiral anomaly, followed by the negative longitudinal magnetoresistance induced by it, and an inverse process called chiral magnetic effect. To explain the mechanism of chiral anomaly in condensed matter, we have to start with Landau levels.

\subsection{Landau levels basics}
Landau levels in quantum mechanics are the result of quantization of the cyclotron orbits of charged particles in magnetic fields\cite{landau3}. You might be familiar with it,  but let's give a brief review before coming to the Landau levels of system with linear dispersion relationship.

Assume a 2D system of non-interacting electrons with charge denoted as $e$ (in my notation, $e\approx-1.6\times10^{-19}C$ is the charge of electron, not the positive elementary charge), confined in an area $L_xL_y$. Applying a uniform magnetic field $\BB=B\hat{e}_z$, the Hamiltonian of this system can be written as
\be
H=\frac{1}{2m}[\hat{p}_x^2+(\hat{p}_y-eB\hat{x})^2],
\ee
when we choose the Landau gauge: $\AA=(0, Bx, 0)$.

Since the Hamiltonian is independent of $y$, it commutes with $\hat{p}_y$: $[H,\hat{p}_y]=0$. Therefore, the eigenstates of $H$ are simultaneously eigenstates of $\hat{p}_y$:
\ba
H|\Psi\ket&=&\e|\Psi\ket,\\
\hat{p}_y|\Psi\ket&=&\hbar k_y|\Psi\ket.
\ea
We may simply replace $\hat{p}_y$ in the $H$ with $\hbar k_y$. With a cyclotron frequency defined as $\omega_c=qB/m$, we reach the simplified Hamiltonian:
\be
H=\frac{\hat{p}_x^2}{2m}+\frac{1}{2}m\omega_c^2(\hat{x}-\frac{\hbar k_y}{m\omega_c})^2,
\ee
which is exactly the Hamiltonian of the quantum harmonic oscillation. The energy of this system is
\be
\e_n=\hbar\omega_c(n+\frac{1}{2}),
\ee
$n\geq 0$ is the energy level. $\e_n$ is independent of $k_y$, so it is degenerate.

Since $\hat{p}_y$ commutes with $H$, the wavefunctions can be written as
\be
\Psi(x,y)=\mathrm{e}^{ik_y y}\psi_n(x-x_0),
\ee
with $x_0=\frac{\hbar k_y}{m\omega_c}$. Each set of wave functions with the same value of $n$ is known as a Landau level, where $k_y=\frac{2\pi N}{L_y}$, and $N$ is the degeneracy of the Landau level (without considering the spin degeneracy). Remember that the system is confined in an area $A=L_xL_y$, so $0\leq x_0\leq L_x$. Since $x_0=\frac{\hbar k_y}{m\omega_c}=\frac{2\pi\hbar N}{m\omega_c L_y}$, we can easily get
\be
0\leq N \leq \frac{m \omega_c L_x L_y}{2\pi \hbar}=\frac{\Phi}{\Phi_0},
\ee
where $\Phi=BA$ is the magnetic flux, and $\Phi_0=h/e$ is the quantum of flux. We also need to consider the spin degeneracy, so $N$ is twice in case of electron. For particle with charge $q=Ze$ and spin $s$, the upper limit is $Z(2s+1)\Phi/\Phi_0$.

\subsection{Landau quantization in Weyl metals}
Chiral anomaly is found as a consequence of the Landau quantization in system with linear dispersion relation\cite{Nielsen1983}. Therefore, lets study the Landau quantization of that kind of system first. Instead of the quadratic momentum Hamiltonian we used above, we need to use a simplest Weyl metal model:
\be
H=\pm v\boldsymbol{\sigma}\cdot\pp,
\ee
where $\pm$ represents different chirality of two Weyl points,  $v$ is the band speed, $\boldsymbol{\sigma}=(\sigma_x,\sigma_y,\sigma_z)$ are Pauli  matrices. Still, we choose  $\AA=(0, Bx, 0)$. Because of the magnetic field in z-direction, $\hat{p}_y$ in the original Hamiltonian should be replaced by $\hat{p}_y-eB\hat{x}$, and it commutes with the Hamiltonian, $[\hat{p}_y,H]=0$. Therefore, $\hat{p}_y$ in the Hamiltonian can be replaced by $\hbar k_y$. In the same way, $\hat{p}_z$ can be replaced by $\hbar k_z$.
\be
\pm v
\begin{pmatrix}
\hat{p}_z  &\hat{p}_x-i(\hat{p}_y-eB\hat{x})\\
\hat{p}_x+i(\hat{p}_y-eB\hat{x})  &-\hat{p}_z
\end{pmatrix}
\begin{pmatrix}
\psi_1\\
\psi_2
\end{pmatrix}
=\e
\begin{pmatrix}
\psi_1\\
\psi_2
\end{pmatrix}
\ee
\be
\Rightarrow \{
\begin{array}{ll}
(\hat{p}_z-\frac{\e}{\pm v})\psi_1+(\hat{p}_x-i(\hat{p}_y-eB\hat{x}))\psi_2=0\\
(\hat{p}_x+i(\hat{p}_y-eB\hat{x}))\psi_1-(\hat{p}_z+\frac{\e}{\pm v})\psi_2=0
\end{array}
\ee
\be
\Rightarrow \{\hat{p}_z^2-(\frac{\e}{\pm v})^2+\hat{p}_x^2+(\hat{p}_y-eBx)^2+eBi[\hat{x},\hat{p}_x]\}\psi_{1,2}=0
\ee
\be
\Rightarrow \frac{1}{2m} [\hat{p}_x^2+(\hbar k_y-eBx)^2+(\hbar k_z)^2-\hbar eB]\psi_{1,2}=\frac{1}{2m}(\frac{\e}{v})^2\psi_{1,2}
\ee
We may define a new ``Hamiltonian" $H'=\frac{\hat{p}_x^2}{2m}+\frac{1}{2}m\w_c(x-\frac{\hbar k_y}{m\w_c})^2+\frac{(\hbar k_z)^2}{2m}-\frac{1}{2}\hbar \w_c$, which is again a harmonic oscillator, but with a free motion in z-direction. Therefore, its eigenvalue is:
\be
\frac{1}{2m}(\frac{\e_n}{v})^2=\hbar \w_c(n+\frac{1}{2})+\frac{(\hbar k_z)^2}{2m}-\frac{1}{2}\hbar\w_c=n\hbar\w_c+\frac{(\hbar k_z)^2}{2m},
\ee
where $n\geq 0$ is the energy level of harmonic oscillation, but since $H'$ is a fake Hamiltonian, the real one $H$ is not a harmonic oscillator, so there is no reason for $n$ to be positive for $H$. Thus, we replace $n$ here with $|n|$, and allow the new $n$ to be negative. Actually, here the positive $n$ stands for conducting bands, while negative $n$ stands for valence bands.

Then we get the eigenvalue of the previous Hamiltonian $H$:
\be
\e_n=\pm v \sqrt{2eB\hbar |n|+\hbar^2 k_z^2}.
\ee
Since we want positive $n$ to be conducting bands and negative $n$ to be valence bands, we can write the above result for $n\neq 0$ this way:
\be
\e_n=\mathrm{sgn}(n) v \sqrt{2eB\hbar |n|+\hbar^2 k_z^2}.
\ee

The zeroth Landau level has a linear dispersion relation:
\be
\e_0=\pm v \hbar k_z=\pm vp_z.
\ee
Here the $\pm$ stand for right and left moving Weyl fermions. 

\subsection{Chiral anomaly}
\begin{figure}[H]
\centering
\begin{tikzpicture}
\draw[->](0,0)--(5,0) node[right]{$k_z$};
\draw[->](6,0)--(11,0) node[right]{$k_z$};
\draw[->](2.5,-3.5)node[below]{L}--(2.5,3.5) node[above]{\Large $\epsilon$};
\draw[->](8.5,-3.5)node[below]{R}--(8.5,3.5) node[above]{\Large $\epsilon$};
\draw (5,-2.5)--(0,2.5);
\draw[decorate sep={2mm}{5mm},fill=white] (5,-2.5)--(0,2.5);
\draw[decorate sep={2mm}{5mm},fill=blue] (5,-2.5)--(0.5,2);
\draw (6,-2.5)--(11,2.5);
\draw[decorate sep={2mm}{5mm},fill=white] (6,-2.5)--(11,2.5);
\draw[decorate sep={2mm}{5mm},fill=green] (6,-2.5)--(10.8,2.3);
\draw (0,-3.5) parabola bend (2.5,-2.5)  (5,-3.5);
\draw[decorate sep={2mm}{5mm},fill=blue] (0.19,-3.36) parabola bend (2.5,-2.5)  (5,-3.5);
\draw (0,-3) parabola bend (2.5,-1.7)  (5,-3);
\draw[decorate sep={2mm}{5mm},fill=blue] (0.29,-2.73) parabola bend (2.5,-1.7)  (5,-3);
\draw (0,3.5) parabola bend (2.5,2.5)  (5,3.5);
\draw[decorate sep={2mm}{5mm},fill=white] (0.19,3.36) parabola bend (2.5,2.5)  (5,3.5);
\draw (0,3) parabola bend (2.5,1.7)  (5,3);
\draw[decorate sep={2mm}{5mm},fill=white] (0.29,2.73) parabola bend (2.5,1.7)  (5,3);
\draw[decorate sep={2mm}{5mm},fill=blue] (2,1.75)--(2.5,1.7)--(3.2,1.75);
\draw (6,-3.5) parabola bend (8.5,-2.5)  (11,-3.5);
\draw[decorate sep={2mm}{5mm},fill=green] (6.19,-3.36) parabola bend (8.5,-2.5)  (11,-3.5);
\draw (6,-3) parabola bend (8.5,-1.7)  (11,-3);
\draw[decorate sep={2mm}{5mm},fill=green] (6.29,-2.73) parabola bend (8.5,-1.7)  (11,-3);
\draw (6,3.5) parabola bend (8.5,2.5)  (11,3.5);
\draw[decorate sep={2mm}{5mm},fill=white] (6.19,3.36) parabola bend (8.5,2.5)  (11,3.5);
\draw (6,3) parabola bend (8.5,1.7)  (11,3);
\draw[decorate sep={2mm}{5mm},fill=white] (6.29,2.73) parabola bend (8.5,1.7)  (11,3);
\draw[decorate sep={2mm}{5mm},fill=green] (7.08, 2.12) parabola bend (8.5, 1.7) (10,2.17);
\draw[red,thick] (0,1.83)node[left] {$\mu_L$}--(5,1.83);
\draw[red,thick] (6,2.3)--(11,2.3)node[right] {$\mu_R$};
\draw[black, line width=1pt][->] (4,-0.5) .. controls (5,-2) and (6,-2) .. (7.5,-0.5);
\draw[red,line width=1pt][->] (5.35,3.5)--(5.35,4.5) node[left]{\large$\bf{E}$};
\draw[red,line width=1pt][->] (5.65,3.5)--(5.65,4.5) node[right]{\large$\bf{B}$};
\end{tikzpicture}
\caption{Chiral Anomaly. Applying $\EE$ and $\BB$ both in z-direction, electron in left valley are bumped into electrons in right valley, which lead to an imbalance of the chemical potential which can be and has been detected.}
\label{fig:chiralanomaly}
\end{figure}
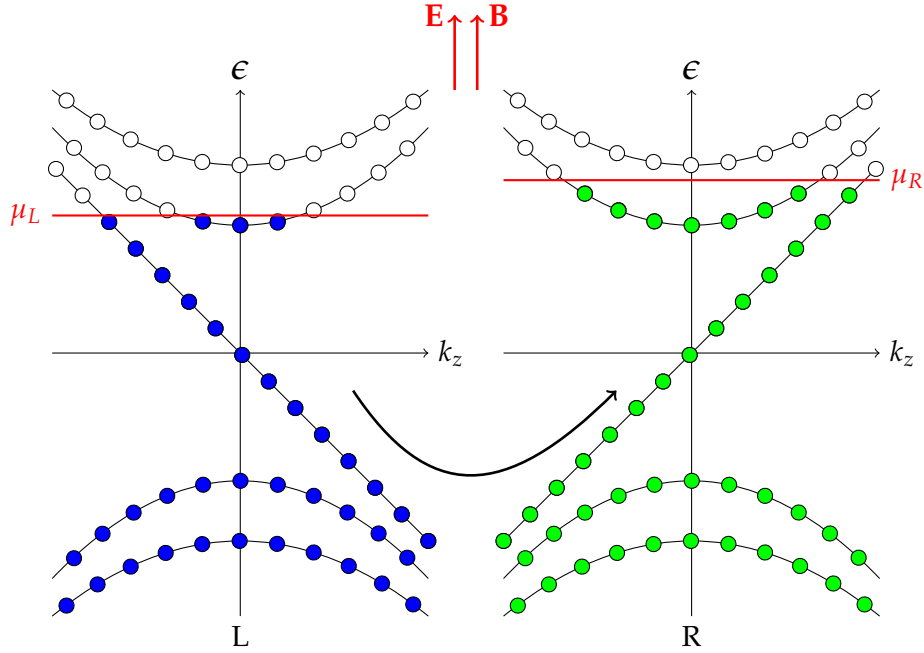

The Adler-Bell-Jackiw (ABJ) axial anomaly \cite{Adler1969}\cite{Bell1969} or chiral anomaly is the anomalous non-conservation of the chiral current. It was introduced to condensed matter field by Nielson and Ninomiya\cite{Nielsen1983} in 1983. The idea is: Followed by the above section, we have strong magnetic field applied on the Weyl metal system to get Landau levels, and at the zeroth Landau level we get right and left moving Weyl fermions. Now apply a uniform electric field $\EE=E\hat{e}_z$ parallel to the magnetic field. The equation of motion has been obtained as Eq.(\ref{eqrd}) and Eq.(\ref{eqpd}) (we consider the zeroth band only, therefore no $n$ is needed here):
\ba
\dot{\rr}&=&\frac{1}{1-e\BB\cdot\bm{\Omega}_{\pp}}[\vv-e\EE\times \bm{\Omega}_{\pp}-e\BB(\vv\cdot\bm{\Omega}_{\pp})],\\
\dot{\pp}&=&\frac{1}{1-e\BB\cdot\bm{\Omega}_{\pp}}[e\EE+e\vv\times\BB-e^2(\EE\cdot\BB)\bm{\Omega}_{\pp}].
\ea
Substituting them into the Boltzmann equation
\be
\bigg(\frac{d f(\rr,\pp,t)}{d t}\bigg)_{coll}=\frac{\p}{\p t}f(\rr,\pp,t)+\dot{\rr}\frac{\p}{\p \rr}f(\rr,\pp,t)+\dot{\pp}\frac{\p}{\p \pp}f(\rr,\pp,t),
\ee
with relaxation time approximation Eq.(\ref{relaxt}), we get
\ba
-\frac{\delta f^i(\rr,\pp,t)}{\tau}&=&\frac{\p}{\p t}f^i(\rr,\pp,t)+\frac{1}{1-e\BB\cdot\bm{\Omega}^i_{\pp}}\{[\vv-e\EE\times \bm{\Omega}^i_{\pp}-e\BB(\vv\cdot\bm{\Omega}^i_{\pp})]\frac{\p}{\p \rr}f^i(\rr,\pp,t)\nonumber\\
& &+[e\EE+e\vv\times\BB-e^2(\EE\cdot\BB)\bm{\Omega}^i_{\pp}]\frac{\p}{\p \pp}f^i(\rr,\pp,t)\}
\ea
where $i$ stands for different valleys. $\tau$ is the relaxation time, which contain both intravalley scattering contribution $\tau_{intr}$ and intervalley scattering $\tau_v$ contribution. We assume $\tau_{intr}\ll \tau_v$, the anisotropy of the distribution function within each valley can be neglected, and the latter depends only on the energy $f^{i}(\pp)=f^i(\e)$. Denote the density of states\cite{Son2013}
\be
\rho^i(\e)=\int \frac{d\pp}{(2\pi\hbar)^3}(1-e\BB\cdot\bm{\Omega}^i_{\pp})\delta(\e_\pp-\e).
\ee
In the homogeneous case ($\frac{\p}{\p \rr}f^i(\rr,\pp,t)=0$, average of the direction of $\pp$ makes first two force terms vanish), and we get the Boltzmann equation:
\be
\frac{\p}{\p t}f^i(\e)-\frac{k^i}{\rho^i(\e)}\frac{e^2}{4\pi^2\hbar^2}(\EE\cdot\BB)\frac{\p f^i(\e)}{\p\e}=-\frac{\delta f^i(\e)}{\tau_v}\label{homoB}
\ee
where
\be
k^i=\frac{1}{2\pi\hbar}\oint d\mathbf{S}\cdot \bm{\Omega}_\pp^i=0,\pm 1,...
\ee
for right and left valley, its just $\pm 1$.
The electron density in each valley is given by
\be
N^i=\int d\e \rho^i(\e)f^i(\e).
\ee
Integrating Eq.(\ref{homoB}) over $\rho^i(\e)d\e$ we get
\be
\frac{\p N^i}{\p t}=k^i\frac{e^2}{4\pi^2\hbar^2}(\EE\cdot\BB)-\frac{\delta(N^i)}{\tau_v}.
\ee
The number of electrons in each valley is not conserved even when $\tau_v\to \infty$.
Finally, we get
\be
\frac{\p N^R}{\p t}-\frac{\p N^L}{\p t}=\frac{e^2}{2\pi^2\hbar^2}(\EE\cdot\BB)-\frac{N^R-N^L}{\tau_v}.
\ee
Electrons are bumped from left valley to the right one, thus a deviation from the thermodynamic equilibrium appears, which can be expressed by different chemical potential $\mu_R$ and $\mu_L$, see Fig.(\ref{fig:chiralanomaly})\cite{Behrends2016}.

\subsection{Negative longitudinal magnetoresistance} The  existence of chiral anomaly in Weyl semimetals results in a negative longitudinal magnetoresistance, which has been detected in TaAs family Weyl semimetals\cite{PhysRevX.5.031023, zhang2016signatures, ghimire2015magnetotransport, PhysRevB.92.205134, moll2016magnetic, zhang2015large, arnold2016negative}. It does not seem unique to Weyl semimetals, since this phenomenon is also observed in Dirac semimetals\cite{kim2013dirac, xiong2015evidence, zhang2015detection}. However, in Dirac semimetals,  the magnetic field not only produces a chiral anomaly but also works in splitting a Dirac cone into a pair of Weyl cones of opposite chirality by time reversal symmetry break. Therefore, the mechanism is deemed as similar to the chiral anomaly induced negative magnetoresistance in Weyl semimetals. ($Remark$: A Dirac cone split into two Weyl cones is one explanation for the observation of the negative magnetoresistance phenomenon in Dirac semimetals, but it is not a fully convincing explanation. For instance, it cannot explain the negative magnetoresistance in $\rm ZrTe_5$\cite{Li2016}, when photoemission and STM experiments\cite{wu2016evidence, zhang2017electronic} show that the robust electronic ground state of ZrTe5 has a small gap($\sim$ 50 meV), not a Dirac cone to be split into Weyl cones. The theoretical explanation of it is still under investigation.)

Now let's talk about the mechanism of chiral anomaly induced negative longitudinal magnetoresistance. Magnetoresistance refers to the increase of resistance (decrease of conductivity) when applying a magnetic field in the same direction of the current. The ordinary magnetoresistance was first discovered by William Thomson in 1856\cite{thomson1856electro}. After that, other related effects, such as negative magnetoresistance, giant magnetoresistance, tunnel magnetoresistance, colossal magnetoresistance, and extraordinary magnetoresistance are studied. Negative magnetoresistance means an increasing of conductivity with parallel magnetic field applied. In order to verify that it is the case for Weyl semimetal because of the chiral anomaly, we need to get the conductivity for our previous model of a pair of Weyl nodes in parallel electric and magnetic field, shown in Fig.(\ref{fig:chiralanomaly}). To calculate the chiral anomaly related distribution of the conductivity, we can use the Boltzmann equations\cite{PhysRevB.89.195137}, and keep the second order term of electromagnetic field, but the calculation will be very complicated. Son and Spivak used an easier way to calculate it in their paper\cite{Son2013}, by estimating the rate of entropy production $\dot{S}$ in the presence of an electric field, and use the relation $\dot{S}=\sigma E^2/T$ to get $\sigma$ tensor. The chiral anomaly term enters the equation of conductivity through the electron density difference between the right valley and left valley, which affects the rate of entropy $\dot{S}$. By keeping only the chiral anomaly $\EE\cdot\BB$ related term, the conductivity tensor has only one nonzero component, which is
\be
\sigma_{zz}=\frac{e^2}{4\pi^2\hbar c}\frac{u}{c}\frac{(eB)^2v^2}{\mu^2}\tau,
\ee
where $z$ is the direction of electric field as well as the magnetic field. This result indicates that the longitudinal conductivity  increases with an increasing magnetic field, which is dubbed as negative magnetoresistance.

There are a number of  other effects which also lead to a negative magnetoresistance in metals, and some of them appear just like chiral anomaly induced negative magnetoresistance: they are prominent only when electric and magnetic fields are parallel\cite{PhysRev.104.900} and the presence of magnetoresistance does not depend on the direction of the electric field with respect to the crystalline axis. The uniqueness of the  chiral anomaly induced negative magnetoresistance is that the magnitude of it have an inverse dependence on the square of the chemical potential $1/\mu^2$\cite{zhang2016signatures}.
\subsection{The chiral magnetic effect: the inverse of the chiral anomaly}
Previously, we had parallel magnetic and electric fields applied to Weyl semimetal, which bump electrons in the zeroth Landau level from one valley to the other, leading to an imbalance of the chemical potential. Think about an inverse process: If we have an imbalance of the chemical potential in two valleys of a Weyl metal, say $\mu_R>\mu_L$, and then we apply a (static) magnetic field on it, what will happen? There will be a current produced along the magnetic field:\cite{PhysRevD.22.3080}
\be
\jj^{CME}_{\w=0}=\frac{e^2(\mu_R-\mu_L)}{4\pi^2}\BB,\label{staticnonequi}
\ee
which can be easily obtained by calculating conductivity tensor (or gyrotropic tensor) and keeping up to the linear order of the electromagnetic field. Details about calculations and theoretical analysis can be found in chapter 3 (part \uppercase\expandafter{\romannumeral3}.B).

 This effect is dubbed as chiral magnetic effect (CME). Just like chiral anomaly, chiral magnetic effect was firstly studied in QED and QCD\cite{PhysRevD.22.3080, PhysRevD.80.034028, PhysRevLett.81.3503}, then because of the recent study of chiral anomaly in Weyl metals, it was brought into condensed matter field, to be exact, into the study of Weyl metals\cite{li2016chiral, PhysRevB.85.165110, PhysRevB.86.115133, PhysRevB.88.245107, PhysRevLett.111.027201, PhysRevB.88.125105}.

\begin{figure}[H]
\centering
\includegraphics[width=0.9\textwidth]{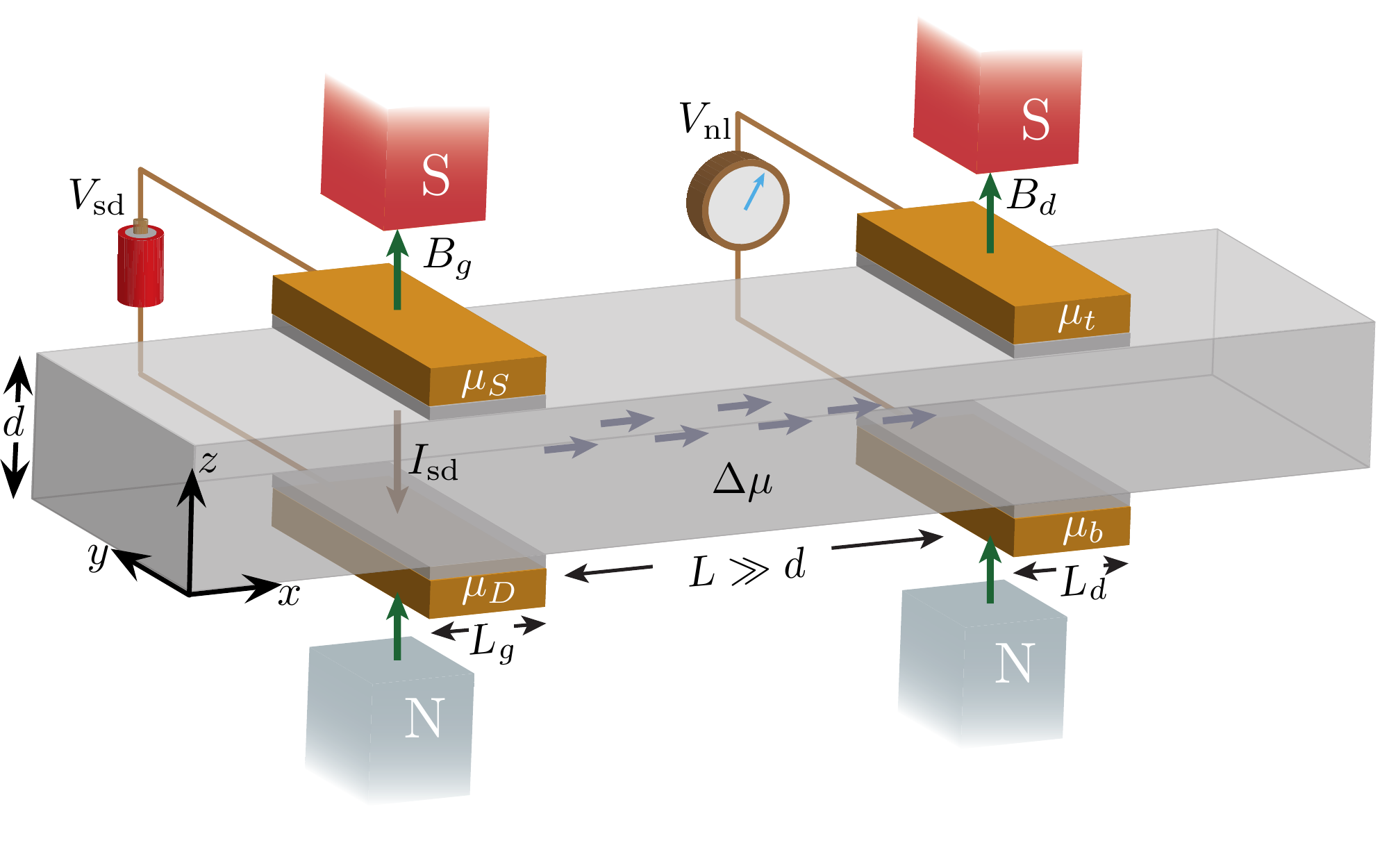}
\caption{ Nonlocal transport experiment. A source-drain current
$I_{SD}$ is injected into a Weyl metal slab with thickness $d$, by the potential difference $V_{sd}$. With a local generation magnetic field $B_g$, a chemical imbalance $\delta\mu\sim|\mu_R-\mu_L|$ is created because of
the chiral anomaly and that imbalance diffuses a distance $L\ll d$ away. If a
probe magnetic field $B_d$ is applied, potential difference $V_{NL}$ between top and bottom
will be detected.  This graph is published by S. A. Parameswaran $et$ $al$\cite{dima2014}. Permission has been received from the authors.}
\label{fig:anomalybasefig}
\end{figure}

Chiral magnetic effect can be detected via nonlocal transport experiments as suggested by my advisor Dima Pesin and his co-workers in 2013\cite{dima2014}, when the chiral anomaly stimulates the imbalance of the chemical potential,  $\mu_R\neq\mu_L$, and then a probe magnetic field converts this imbalance into a measurable
voltage drop far from source and drain, indicated by the mechanism of  CME. The basic idea is shown in Fig.(\ref{fig:anomalybasefig}).

From Eq.(\ref{staticnonequi}), we see that in equilibrium case, where $\mu_R=\mu_L$, there is no chiral magnetic current at all, which is quite reasonable: nobody would expect a current induced by a static magnetic field in equilibrium system. However, if we have a slowly vibrating magnetic field, that is $\w\neq 0$, there might be a current, even when system is in equilibrium $\mu_R=\mu_L$. The phenomenon of nontrival current response to slowly oscillating magnetic field is named as dynamic chiral magnetic effect (dCME). We will show that it happens in equilibrium Weyl metal system when a pair of Weyl nodes have different energies, $E_L\neq E_R$, which could happen in nature. Further discussions are in chapter 3.
\section{Our motivation: to learn the optical and transport properties of Weyl metals}
For high energy physicists, the exciting part of the new phase of Weyl semimetal is that it provides us a plateau to test theories about 3D massless Dirac fermions, although these Weyl fermions are quasi-particles, not real particles. For condensed matter physicists, the most interesting part is the new phase itself: will it bring any new phenomena, or will it response differently to external probe? Therefore, our concentration would be on the topic of revealing possibly unique optical and transport properties.

In the last section, we have already had a quick glance about the chiral anomaly related negative longitudinal magnetoresistance and the chiral magnetic effect, which were thought to be candidates of unique phenomena associated with Weyl (semi-)metal. Chapter3 will be a deep research into the mechanism and properties of chiral magnetic effect (CME) and optical activity in (Weyl) metals, in which we will find that the dynamic CME-like response, originated from local geometry of electronic bands (not topology), is actually not unique to Weyl metals only: it might happen even without Berry monopoles. Followed by this is a natural question: how to measure the dynamic chiral magnetic conductivity? The answer is Faraday rotation, as the Faraday rotation angle is directly proportional to the chiral magnetic conductivity, which gives a most direct way to measuring it. Based on that insight, chapter 4 gives a prediction about the chiral magnetic conductivity measured with Faraday rotation experiment. However, when studying optical and transport phenomena in a Weyl metal, or more generally speaking, a gapless topological system, the challenge is: in principle, with a gapless bulk, it manifests all responses similar to a normal metal with the same symmetries. In order to distinguish them, in chapter 4, we studied the omnipresent disorder effects: how macroscopic sample inhomogeneities affect the dynamic chiral magnetic conductivity to be measured by Faraday rotation experiment. We pushed our study a bit further to the current induced magnetization in chapter 5. One thing has to be pointed out is that, our research was ignited by Weyl semimetals, but all our results have a wider application.

\setcounter{NAT@ctr}{0}
\bibliographystyle{apsrev}
\bibliography{chap2}

%% file: chap3.tex

\chapter{Chiral magnetic effect and natural optical activity in (Weyl) metals}

The article in this chapter was originally published in
\emph{PHYSICAL REVIEW B} {\bf 92}, 235205 (2015).  It is reproduced 
here with permission of the publisher.

\uudummysection {Introduction}                            {1}
\uudummysection {Semiclassical theory of natural optical activity in metals}       {2}
\uudummysubsection {Gyrotropic current}                      {3}
\uudummysubsection {Gyrotropic tensor: Relation to previous works}       {4}
\uudummysection {Natural optical activity and chiral magnetic effect in simple models} {6}
\uudummysubsection {Isotropic noncentrosymmetric metal}           {6}
\uudummysubsection {Weyl metal with particle-hole symmetric Weyl points}    {7}
\uudummysubsection {Chiral magnetic effect without Berry monopoles}                   {8}
\uudummysection {Conclusion}       {9}
\uudummysection {Appendix A: Derivation of (18) from Kubo formula}     {9}
\uudummysubsection {Intraband part}      {9}
\uudummysubsection {Interband part}      {10}
\uudummysubsection {Current in the static limit}                 {13}
\uudummysubsection {Current in the dynamic limit}                 {13}
\uudummysection {Appendix B: Vanishing current in the static limit}                                 {13}
\uudummysection {References}                              {14}
\includepdf [
                pages = -,          
                scale = 0.91,       
                pagecommand = {\pagestyle{plain}} 
            ]
            {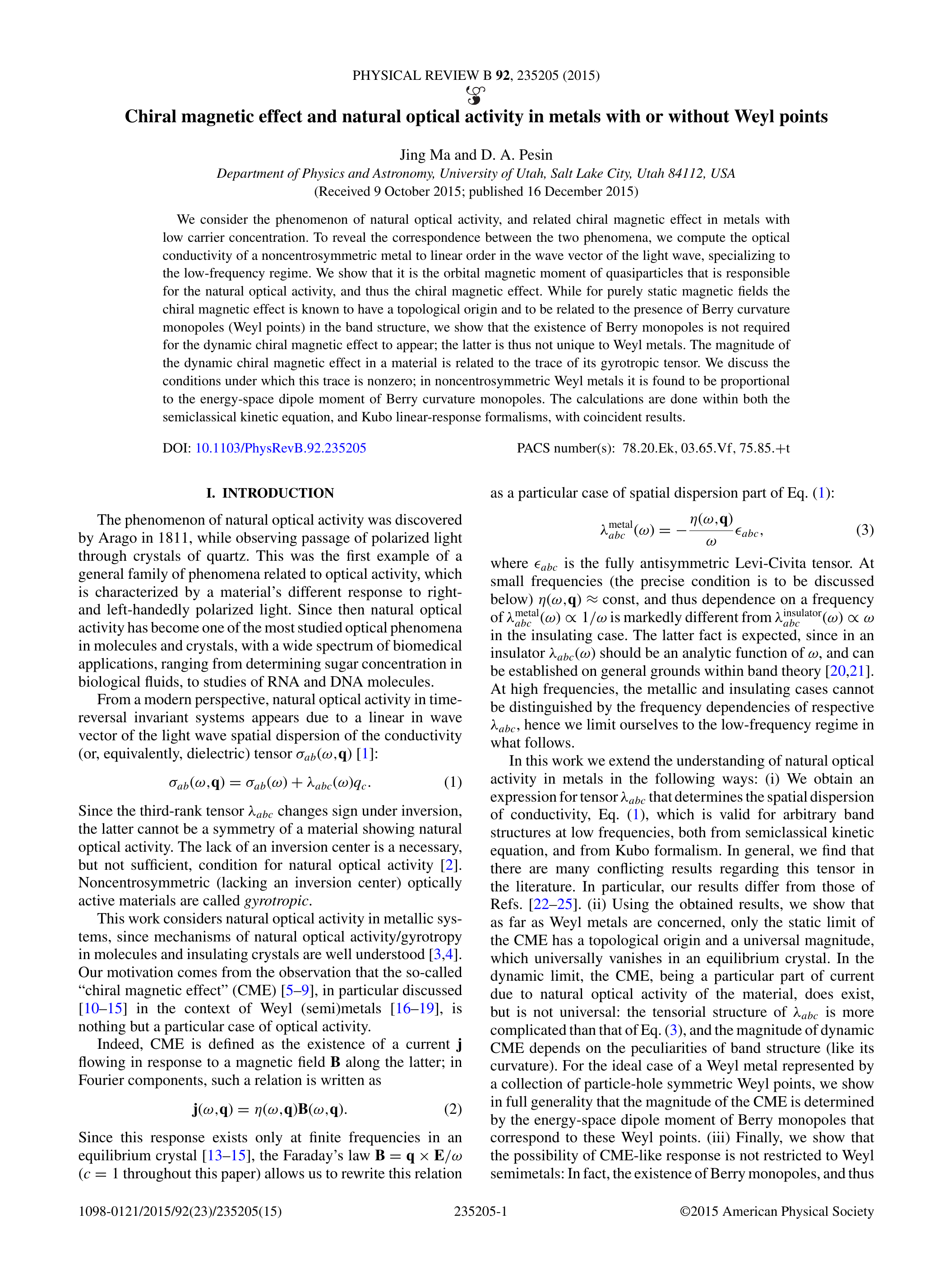}

\uudummyfigure        {(Color online) Schematic representation of a pair of Weyl points located at different energies close to the Fermi level of a material. $Q_{1,2}$ and $E_{1,2}$ are their chiralities and energies, respectively. The specific position of the Fermi energy EF with respect to the energies of the Weyl points (here chosen to be in between) is not
important.}           {6}

\uudummytable     {Chiral magnetic conductivity for a Weyl metal for different cases of linear-response theory. The rows correspond to static and dynamic limits of response. The columns pertain to the equilibrium response of a Weyl metal with Weyl points located at different energies, and to nonequilibrium response of a Weyl metal with Weyl points at the same energy.} {8}

%% file: chap5.tex
\chapter{Onsager relations and current-induced magnetization}

\newpage
In this Chapter, I discuss the role of Onsager relations for the conductivity tensor in establishing its general symmetry properties, and its physical content. Based on the Onsager relations, I discuss the implications of the time-reversal (TR) symmetry for the relation between bulk gyrotropic currents and an effective surface Hall effect, which is required by the TR-symmetry. In particular, such surface Hall effect, related to the spatial change of the gyrotropic tensor invariably present near sample boundaries, is responsible for canceling the polar Kerr effect due to the bulk polarization rotation.\cite{AgranovichYudson, HosurKerr2014} In this sense, there is “electromagnetic bulk-edge correspondence” in gyrotropic metals. Further, studying phenomenological magnetoelectric response of a noncentrosymmetric medium, and again relying on the Onsager relations, I will show that the gyrotropic tensor also determines the so-called kinetic mangnetoelectric effect, equivalent to the phenomenon of current-induced magnetization in noncentrosymmetric metals.
 
Let us start with a general discussion of the Onsager relations’ consequences for the conductivity tensor. Allowing for a moment for the possibility of time-reversal symmetry breaking in a noncentrosymmetric crystal, with the existence of magnetization $\bm M$, the optical conductivity tensor $\sigma_{ab}(\w,\qq;\bm M)$ can be written  as
\begin{equation}\label{eq:conductivity_general}
  \sigma_{ab}(\w,\qq;\bm M)\approx \sigma_{ab}(\w)+\chi_{abc}(\w)M_c+\lambda_{abc}(\w)q_c,
\end{equation}
for small $M$ and $q$.
In this expression, $\sigma_{ab}(\w)$ is the usual local optical conductivity, the pseudotensor $\chi_{abc}$ is the anomalous Hall effect term, and tensor $\lambda_{abc}$ describe the natural optical activity, as we have discussed in previous chapters. These tensors determine the antisymmetric part of the conductivity tensor.
 The Onsager relations\cite{Melrose} tell us:
\begin{equation}
  \sigma_{ab}(\w,\qq;\bm M)= \sigma_{ba}(\w,-\qq;-\bm M),
\end{equation}
implying that $\chi_{abc}$ and $\lambda_{abc}$ are antisymmetric with respect to the first pair of indices, which would be derived below:
\begin{align}
  \begin{bmatrix}
  \lambda_{abc}\\
  \chi_{abc}
  \end{bmatrix}=-
  \begin{bmatrix}
  \lambda_{bac}\\
  \chi_{bac}
  \end{bmatrix}.
  \label{lambdato}
\end{align}

 \section{Onsager relations, antisymmetry of $\lambda_{abc}$ tensor, and the bulk-surface correspondence}
 Let us derive Eq.(\ref{lambdato}) with Onsager relation\cite{ LL8}.  Since  $\chi_{abc}$ works similar to $\lambda_{abc}$, we will just derive the $\lambda_{abc}$ part of Eq.(\ref{lambdato}).

 Any component of current density can be written as
 \be
 j_a(\rr)=\int d \rr'\sigma_{ab}(\rr,\rr')E_b(\rr').
 \ee
 Then, we can get a relation about the dot product of electric field and current:
 \ba
 \int d\rr \EE'(\rr)\cdot\jj(\rr)&=&\int d\rr E'_a(\rr)j_a(\rr),\nonumber\\
 &=&\int d\rr E'_a(\rr)\int d \rr'\sigma_{ab}(\rr,\rr')E_b(\rr'),\nonumber\\
 &=&\int d\rr d\rr' E'_a(\rr)E_b(\rr')\sigma_{ab}(\rr,\rr'),\nonumber\\
 &=&\int d\rr d\rr' E'_a(\rr)E_b(\rr')\sigma_{ba}(\rr',\rr),\nonumber\\
 &=&\int d\rr [d\rr' E'_a(\rr')\sigma_{ba}(\rr,\rr')]E_b(\rr),\nonumber\\
 &=&\int d\rr \EE(\rr)\cdot\jj'(\rr).\label{relationEj}
 \ea

Current density in a sample with nontrival natural optic activity can be written as
\be
j_a=\sigma_{ab}E_b+\lambda_{abc}i\p_c E_b+i(\p_c \gamma_{abc})E_b,\label{currentsample}
\ee
 with $\gamma_{abc}$ term describing a boundary distribution to the total current density.

 Express currents in Eq.(\ref{relationEj}) with the form shown in Eq.(\ref{currentsample}). For the left hand side:
 \be
 \int d\rr E'_a(\rr)j_a(\rr)=\int d\rr E'_a(\rr)[\sigma_{ab}E_b+\lambda_{abc}i\p_c E_b+i(\p_c \gamma_{abc})E_b],
 \ee
 and for the right hand side:
 \ba
  \int d\rr E_a(\rr)j'_a(\rr)&=&\int d\rr E_a(\rr)[\sigma_{ab}E'_b+\lambda_{abc}i\p_c E'_b+i(\p_c \gamma_{abc})E'_b],\nonumber\\
  &=&\int d\rr E_a E'_b\sigma_{ab}-i\int d\rr E'_b\p_c E_a\lambda_{abc}-i\int d\rr E'_b E_a\p_c\lambda_{abc}+i\int d\rr E_a E'_b\p_c\gamma_{abc},\nonumber\\
   &=&\int d\rr E'_a E_b\sigma_{ab}-i\int d\rr E'_a\lambda_{bac}\p_c E_b-i\int d\rr E'_a E_b\p_c\lambda_{bac}+i\int d\rr E_b E'_a\p_c\gamma_{bac},\nonumber\\
 \ea
 where we replaced index $a\leftrightarrow b$ to get the final result.

 With $\int d\rr E'_a(\rr)j_a(\rr)= \int d\rr E_a(\rr)j'_a(\rr)$, we obtain Eq.(\ref{lambdato}):
 \ba
 & &\int d\rr E'_a(\rr)\lambda_{abc}i\p_c E_b=-i\int d\rr E'_a\lambda_{bac}\p_c E_b\nonumber\\
 &\Rightarrow& \lambda_{abc}=-\lambda_{bac},
 \ea
together with a relation between bulk contribution $\lambda_{abc}$ and the boundary contribution $\gamma_{abc}$:
\ba
& &\int d\rr E'_a(\rr)i(\p_c \gamma_{abc})E_b=-i\int d\rr E'_a E_b\p_c\lambda_{bac}+i\int d\rr E_b E'_a\p_c\gamma_{bac}\nonumber\\
&\Rightarrow&\p_c\gamma_{abc}=-\p_c \lambda_{bac}+\p_c\gamma_{bac}\nonumber\\
&\Rightarrow&\gamma_{abc}=\frac{1}{2}\lambda_{abc}.
\ea
\section{Onsager relations in the magnetoelectric effect}
 We take the dc limit, ``$q\to 0$ first, then $\w\to 0$'' of linear response to electric field. Therefore, the response of the crystal to electromagnetic field is fully determined by the electric polarization and the magnetization:
 \begin{subequations}\label{eq:magnetoelectricresponse}
  \begin{align}
    P_a(\w,\qq)&=\chi^\textrm{e}_{ab}(\w)E_b(\w,\qq)+i\chi^\textrm{em}_{ab}(\w)B_b(\w,\qq),\\
    M_a(\w,\qq)&=-i\chi^\textrm{me}_{ab}(\w)E_b(\w,\qq)+\chi^\textrm{m}_{ab}(\w)B_b(\w,\qq).
  \end{align}
\end{subequations}
 Here we only keep up to linear order of $q$, so we can neglect the $\qq$ dependence of the response tensors $\chi^{\textrm{e},\textrm{m},\textrm{em},\textrm{me}}$ as well. The magnetoelectric susceptibility $\chi^{\textrm{me}}_{ab}$ describes the magnetization response to a transport electric field, known as the kinetic magnetoelectric effect\cite{Levitov}.

 A macroscopic current density can be written as
 \be
 \jj=-i\w\mathbf{P}(\w,\qq)+i\qq\times\mathbf{M}(\w,\qq)\label{currPM}.
 \ee
 If we substitute Eq.(\ref{eq:magnetoelectricresponse}) into Eq.(\ref{currPM}), and use Faraday's law $\BB=\qq\times\EE/\w$, we get
 \be
 j_a=-i\w\chi_{ab}^\textrm{e} E_b-i\w\chi_{ab}^\textrm{em}\frac{\e_{bcd}q_c E_d}{\w}+i\e_{abc}q_b\chi_{cd}^\textrm{me}E_d+i\e_{abc}q_b\chi_{cd}^\textrm{m}\frac{\e_{drs}q_rE_s}{\w},
 \ee
 from which we write out the conductivity tensor
 \be
 \sigma_{ab}=-i\w\chi_{ab}^\textrm{e}-i\chi_{ad}^\textrm{em}\e_{bdc}q_c+i\e_{adc}q_d\chi^\textrm{me}_{cb}+i\e_{asc}q_s\chi_{cd}^\textrm{m}\e_{drb}q_r/\w.
 \ee

 From Onsager relation $\sigma_{ab}(\w,\qq)=\sigma_{ba}(\w,-\qq)$ we get
 \ba
 & &-i\w\chi_{ab}^\textrm{e}-i\e_{bdc}\chi_{ad}^\textrm{em}q_c+i\e_{adc}q_d\chi^\textrm{me}_{cb}+i\e_{asc}q_s\chi_{cd}^\textrm{m}\e_{drb}q_r/w\nonumber\\
 &=&-i\w\chi_{ba}^\textrm{e}+i\e_{adc}\chi_{bd}^\textrm{em}q_c-i\e_{bdc}q_d\chi^\textrm{me}_{ca}+i\e_{bsc}q_s\chi_{cd}^\textrm{m}\e_{dra}q_r/w.
 \ea
 By comparing terms in the left-hand side and those in the right-hand side, we get
 \be
 \chi_{ab}^\textrm{e}=\chi_{ba}^\textrm{e},\quad\chi_{cd}^\textrm{m}=\chi_{dc}^\textrm{m},\quad\chi_{ac}^\textrm{em}=\chi_{ca}^\textrm{me}.
 \ee
 \section{Current-induced magnetization}

 The gyrotropic part of current in Eq.(\ref{currPM}) can be written as
 \be
  j_{g,a}=(\chi^{\textrm{em}}_{ad}\epsilon_{dcb}+\e_{acd} \chi^{\textrm{me}}_{db})q_c E_b,
 \ee
 and from the previous section we've already know that $\quad\chi_{ad}^\textrm{em}=\chi_{da}^\textrm{me}$. Therefore we obtain
 \ba
& &\lambda_{abc}=(\e_{acd} \chi^{\textrm{me}}_{db}-\epsilon_{bcd}\chi^{\textrm{me}}_{da}).\\
&\Rightarrow&g_{ab}=\frac{1}{2}\e_{cda}\lambda_{cdb}=\chi^{\textrm{me}}_{ab}-\d_{ab}\mathrm{Tr}\chi^{\textrm{me}}.
 \ea
 Since $\mathrm{Tr} g=\mathrm{Tr} g-3\mathrm{Tr} g=-2\mathrm{Tr}g$, we get
 \be
  \chi^{\textrm{me}}_{ab}=g_{ab}-\frac{1}{2}\d_{ab}\mathrm{Tr} g.
 \ee

 The gyrotropic tensor has been obtained in chapter 3. For a disordered system, we write it as
 \be
 g_{ab}=\frac{e}{(\w+\frac{i}\tau)} \int(d\pp)
  (m^{\textrm{int}}_{\pp a} \p_bf^0_\pp -\d_{ab}\m^{\textrm{int}}_{\pp}\cdot\p_\pp f^0_\pp),
 \ee
 where $\m^{\textrm{int}}_{\pp}=\frac {i\hbar e}2 \bra \p_\pp u_{\pp}|\times (h_\pp-\e_{\pp})|\p_\pp u_{\pp}\ket$ is the intrinsic orbital magnetic moment.
 Therefore, we reach the formula for the magnetoelectric susceptibility:
 \be
   \chi^{\textrm{me}}_{ab}=\frac{e}{(\w+\frac{i}{\tau})}\int(d\pp)
  m^{\textrm{tot}}_{\pp a} \p_bf^0_\pp ,
  \ee
  which gives the following expression for the magnetization\cite{PhysRevB.96.035120}:
  \be
   \bm M^\textrm{int}=\int(d\pp)
 \m^{\textrm{int}}_\pp\frac{e\EE\cdot\p_\pp f^0_\pp}{(i\w-\frac{1}{\tau})}.\label{magnetization11}
  \ee
The reason we name this current as induced magnetization is that, the time-reversal parity of quantities in the left-hand side and right-hand side of the linear relationship $M\propto E$ indicates the existence of a dissipative process, in other words, a current flow in our time-reversal invariant noncentrosymmetric system\cite{Levitov, PhysRevLett.116.077201}.

 Those ``int'' superscripts stand for ``intrinsic'', indicating that they are just intrinsic part of contribution. To get total magnetization, we have to consider the ``extrinsic'' contributions: skew scattering and side-jump. The ``extrinsic'' contributions have a similar form as Eq.(\ref{magnetization11}), but the orbital magnetic moments are different. As I have mentioned at the beginning of this chapter, we are doing tight binding calculation of the current induced magnetization on Tellurium. We are calculating different contributions separately, so that we are able to compare them to find which one dominates at what kind of conditions.
 \setcounter{NAT@ctr}{0}
\bibliographystyle{apsrev}
\bibliography{chap5} 

%% file: chap4.tex

\chapter{Dynamic Chiral Magnetic Effect and Faraday Rotation
in Macroscopically Disordered Helical Metals}

The article in this chapter was originally published in
\emph{PHYSICAL REVIEW LETTERS} {\bf 118}, 107401 (2017).  It is reproduced 
here with permission of the publisher.
\setupuuchapterbib
\uudummysection {Dynamic chiral magnetic effect and Faraday rotation
in macroscopically disordered Helical metals}                            {1}
\uudummysection {References}                              {5}

\includepdf [
                pages = -,          
                scale = 0.91,       
                pagecommand = {\pagestyle{plain}} 
            ]
            {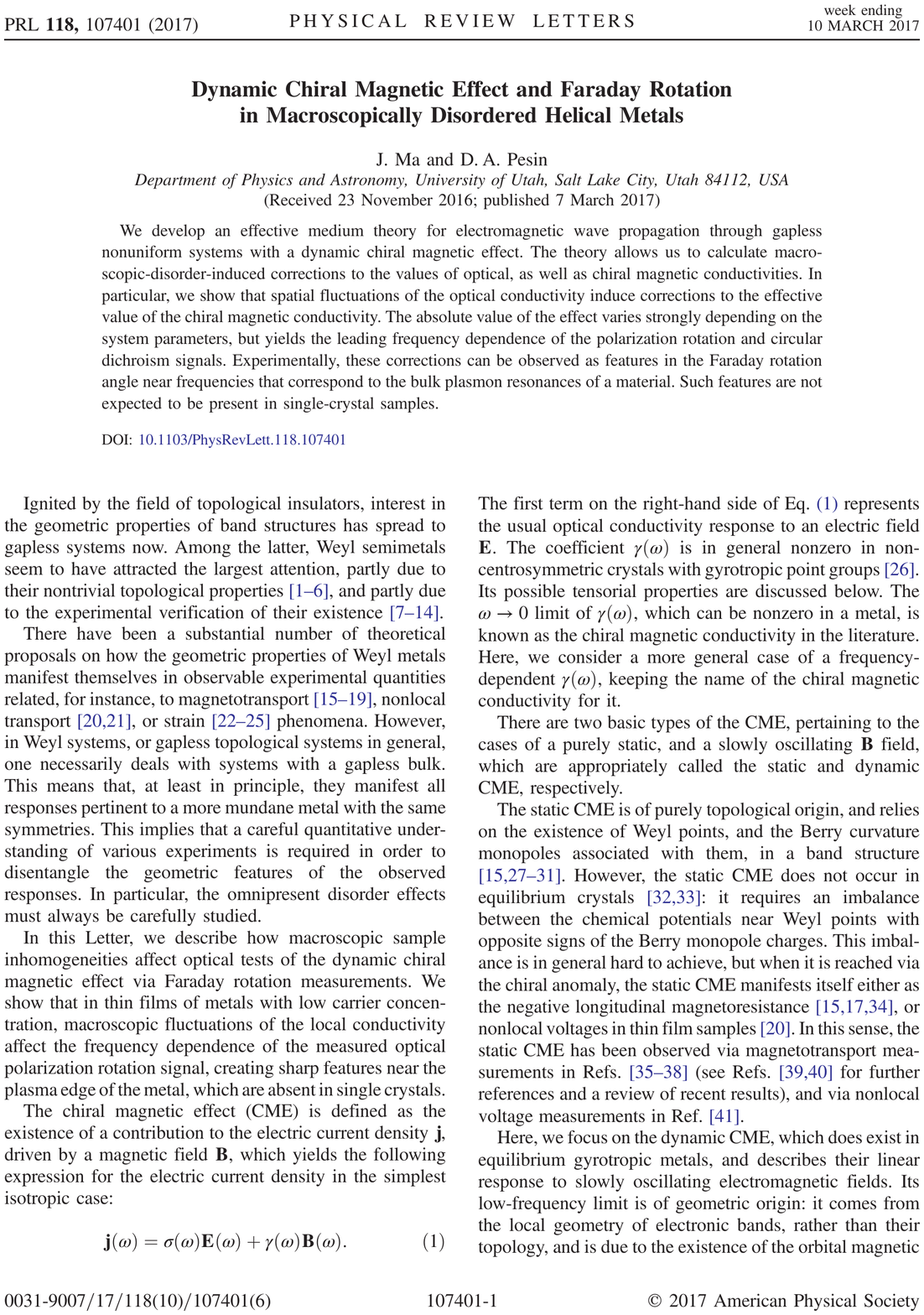}
            
\uudummyfigure        {The Feynman diagrams for the self-energy in the self-consistent Born approximation.}           {4}            
\uudummyfigure        {Relative change in the real (a), and imrarts
tive chiral magnetic conductivity for andThe latter valuthe applicability limit of the present theory. For smaller the curves retain their shape, but have to be scaled down appropriately, see the main text.}           {5}            

%% file: conclusion.tex
\chapter{Conclusions}
This work is focused on the study in the natural optical activity and chiral magnetic effect in noncentrosymmetric metals. The main part of the research is contained in Chapter 3, while the fundamental knowledge and relevant calculation methods introduced in Chapter 1 have paved the road to our work. Since this work is motivated by the recently discovery of Weyl semimetals, Chapter 2 has provided a brief review of the background of our research and the track of our reasoning: from chiral anomaly to chiral magnetic effect. Chapter 4 and Chapter 5 has pushed the study further, either by including the current induced magnetization, or by taking disorders into consideration.

We have argued that the chiral magnetic effect is essentially a specific case of the natural optical activity, and therefore, we need to study the antisymmetric part of the optical conductivity to understand both effects. In Chapter 3, a general expression (Eq.(17) in Chapter 3) for the leading contribution to the gyrotropic current in a metallic system at low frequencies and wave vectors has been derived, which is the central result of our work. This expression holds for low frequencies $\w$ and wave vectors $q$ compared to the energy scale of the chemical potential $\mu$ and the relevant energy gap $E_g$: $\omega, vq\ll\mu, E_g$, where $v$ is the relevant speed. The main physical conclusion is that the intrinsic orbital magnetic moment of quasiparticles is the source of the natural optical activity, the dynamic chiral magnetic effect, and the current induced magnetization in (semi)metals. Unlike the static chiral magnetic effect, the dynamic chiral magnetic effect does not have a topological origin. In general, the latter one exists in metallic systems with a gyrotropic point group, and can be observed in an inversion symmetry broken Weyl semimetal that is equipped with such point groups. However, these effects, the natural optical activity, the dynamic chiral magnetic effect and the current induced magnetization, are not limited to Weyl semimetals: the presence of the Weyl points, or the associated Berry monopoles is not required in general for the existence of these effects. Also, we have argued that the trace of the gyrotropic tensor determines the magnitude of the dynamic chiral magnetic effect in a system with a point group of relatively high symmetry, particularly in an isotropic system. The magnitude of the dynamic chiral magnetic effect can be measured by the Faraday rotation experiments. The rotatory angle $\theta$ is directly proportional to the dynamic chiral magnetic conductivity $\gamma$ and the depth of the measured material $d$: $\theta(\omega)=\frac{\mu_0}{2}\gamma(\w) d$. With an effective medium theory, we have shown that macroscopic inhomogeneities (the inhomogeneities occur on length scales large compared to the microscopic ones, like the Fermi wavelength, or elastic mean free path) make the effective observable chiral magnetic conductivity different from the chiral magnetic conductivity obtained by the band structure calculations. The effective one has sharp features around the plasma edge of the metal, which is not expected from the band structure calculations. The disorder-induced correction does not affect the magnitude of the effective chiral magnetic conductivity, but it is the reason for the observable sharp feature. This effective medium theory is pertinent at the circumstance of smoothly varying electromagnetic fields on the inhomogeneity scale. In particular, our work with effective medium theory is applicable to the case of Weyl semimetals with low electron density, near the terahertz frequency range. In Chapter 4, we have obtained a relation between the magnetoelectric susceptibility and the gyrotropic tensor. This relation allow us to calculate the current induced magnetization easily with the obtained gyrotropic tensor. Apparently, with all these results from theoretic derivations, the next thing to expect is the numerical calculations on models for specific materials, and then hopefully the experimental confirmations.

%% file: appa.tex
\chapter{General Expression for Optical Conductivity}

A general Hamiltonian 
\begin{equation}
H=\int d\mathbf{r} \; \psi_{\sigma}^{\dagger}[\frac{(\mathbf{p}-\frac{e}{c}\mathbf{A})^2}{2m}+\lambda\bm{\sigma} \cdot \mathbf{E}\times (\mathbf{p}-\frac{e}{c}\mathbf{A})+U(r)-g\mu\mathbf{B}\cdot\bm{\sigma}]_{\sigma \sigma^{\prime} }\psi_{\sigma^{\prime}}.
\end{equation}
give a current containing three part
\begin{equation}
\mathbf{j}=\mathbf{j}_{grad}+\mathbf{j}_{dia}+\mathbf{j}_s.
\end{equation}

Magnetic current from spin density is
\begin{equation}
\mathbf{j}_{spin}=\mu_{B}gc \nabla\times (\psi^{\dagger}\bm{\sigma}\psi).
\end{equation}

Diamagnetic current
\begin{equation}
\mathbf{j}_{dia}=-\frac{e^2}{mc}\bra \psi^{\dagger}_{\sigma}(\mathbf{r})\psi_{\sigma}(\mathbf{r})\ket \mathbf{A}(\mathbf{r},\tau).
\end{equation}
Averaging over a unit cell, the diamagnetic current become
\begin{equation}
\bar{\mathbf{j}}_{dia}=-\frac{e^2}{mc}\cdot\frac{N}{V}\cdot \mathbf{A}.
\end{equation}

Gradient part of the current with the form
\begin{equation}
\mathbf{j}_{grad}=\frac{e\hbar}{2mi}(\psi_{\sigma}^{\dagger}\nabla \psi_{\sigma}-\nabla \psi_{\sigma}^{\dagger}\cdot\psi_{\sigma})
\end{equation}
result in 
\begin{equation}
j^i_{grad}(\Omega, \mathbf{q})=Q^{ij}(\Omega,\mathbf{q})A_j(\mathbf{q},\Omega),
\end{equation}
with
\begin{align}
Q^{ij}(\Omega, \mathbf{q})=&-\frac{e^2N^2}{4m^2cV^3}\sum_{\mathbf{k},n,n^{\prime}}\frac{f_{n^{\prime}(\mathbf{k}-\mathbf{q})}-f_{n\mathbf{k}}}{i\Omega+\xi_{n^{\prime}(\mathbf{k}-\mathbf{q})}-\xi_{n\mathbf{k}}}\bra u_{n^{\prime}(\mathbf{k}-\mathbf{q})}|(2\frac{\hbar}{i}\partial_{r_i}+2\hbar k_i-\hbar q_i)|u_{n\mathbf{k}}\ket \\
&[\bra u_{n\mathbf{k}}|(2\frac{\hbar}{i}\partial_{\rho_j}+2\hbar k_j-\hbar q_j)|u_{n^{\prime}(\mathbf{k}-\mathbf{q})}\ket ].\label{eq8}
\end{align}

For tight binding model:
\begin{equation}
H=\sum_{ij}t_{ij}^{\alpha\beta}\e^{\frac{ie}{\hbar c}\int_{\mathbf{R}_j}^{\mathbf{R}_i}d \mathbf{r} \mathbf{A}(\mathbf{r})} a_{i\alpha}^{\dagger} a_{j\beta},
\end{equation}
our gradient part of current can be written as
\begin{equation}
\mathbf{j}_{grad}=\frac{e}{\hbar}\sum_{\mathbf{k}} a^{\dagger}_{\mathbf{k}-\mathbf{q}/2,\alpha}\partial_{\mathbf{k}}t^{\alpha\beta}_{\mathbf{k}}a_{\mathbf{k}+\mathbf{q}/2,\beta},
\end{equation}
while diamagnetic current have this form:
\begin{equation}
\mathbf{j}_{dia}=-\frac{e^2}{\hbar^2 c}\sum_{\mathbf{k}} a^{\dagger}_{\mathbf{k}-\mathbf{q}/2,\alpha}\partial_{\mathbf{k}}(\mathbf{A}\cdot\partial_{\mathbf{k}}t^{\alpha\beta}_{\mathbf{k}})a_{\mathbf{k}+\mathbf{q}/2,\beta}.
\end{equation}

With the help of green's function, after a litter bit calculation, we end up with a equation similar to Eq.(\ref{eq8}) for tight binding model
\begin{equation}
Q^{ij}(\Omega, \mathbf{q})_{grad}=-\frac{e^2}{\hbar^2 c}\sum_{\mathbf{k},n,n^{\prime}}\frac{f_{n^{\prime},\mathbf{k}-\mathbf{q}/2}-f_{n,\mathbf{k}+\mathbf{q}/2}}{i\Omega+\xi_{n^{\prime},\mathbf{k}-\mathbf{q}/2}-\xi_{n,\mathbf{k}+\mathbf{q}/2}}\bra u_{n^{\prime},\mathbf{k}-\mathbf{q}/2}|\frac{\partial t_{\mathbf{k}}}{\partial k_i}|u_{n,\mathbf{k}+\mathbf{q}/2}\ket \bra u_{n,\mathbf{k}+\mathbf{q}/2}|\frac{\partial t_{\mathbf{k}}}{\partial k_j}|u_{n^{\prime},\mathbf{k}-\mathbf{q}/2}\ket .
\end{equation}
This is a useful expression known as Kubo formula. Which is the foundation of our research as we are interested in the magnetization and optical activities, so the main part is calculating optical conductivity tensor.

Detailed derivations are given in appendix B and appendix C.

%% file: appb.tex
\chapter{Detailed Derivation for Optical Conductivity for system of a general Hamiltonian}
We can write a general Hamiltonian 
\begin{equation}
H=\int d\mathbf{r} \; \psi_{\sigma}^{\dagger}[\frac{(\mathbf{p}-\frac{e}{c}\mathbf{A})^2}{2m}+\lambda\bm{\sigma} \cdot \EE \times (\mathbf{p}-\frac{e}{c}\mathbf{A})+U(r)-g\mu\BB \cdot\bm{\sigma}]_{\sigma \sigma^{\prime} }\psi_{\sigma^{\prime}}.\label{H}
\end{equation}
Here we know that $\BB =\vec{\nabla}\times \mathbf{A}$.

Current density $\mathbf{j}$ can be derived from the Hamiltonian by
\begin{equation}
\mathbf{j}=-c\frac{\delta H}{\delta \mathbf{A}}.
\end{equation}

Therefore, we can get a general formula of current density $\mathbf{j}$ (we neglect spin orbit interaction$\lambda\bm{\sigma} \cdot \EE \times (\mathbf{p}-\frac{e}{c}\mathbf{A})$ in our calculation):
\begin{equation}
\mathbf{j}=\frac{e\hbar}{2mi}(\psi_{\sigma}^{\dagger}\nabla \psi_{\sigma}-\nabla \psi_{\sigma}^{\dagger}\cdot\psi_{\sigma})-\frac{e^2}{mc}\psi^{\dagger}_{\sigma}\psi_{\sigma}\mathbf{A}+magnetization\; current \; from \;spin\;density.\label{1}
\end{equation}

Naturally $\mathbf{j}$ has three parts, as shown in Eq.(\ref{1}):
\begin{equation}
\mathbf{j}=\mathbf{j}_{grad}+\mathbf{j}_{dia}+\mathbf{j}_s.
\end{equation}
$\mathbf{j}_s$ come from the spin- magnetic interaction part of the hamiltonian. $\mathbf{j}_{dia}$ is the diamagnetic current. $\mathbf{j}_{grad}$ is just the first part of Eq.(\ref{1}), involving gradient of $\psi_{\sigma}$. We are going to study them separately.
\newpage
\section{Magnetic current from spin density}
\begin{align}
\mathbf{j}_{s}^{l}=&c\frac{\delta}{\delta A_{l}}\int d\mathbf{r}\psi^{\dagger} g\mu_{B}\vec{\nabla}\times\mathbf{A}\cdot\bm{\sigma}\psi\\
=& \mu_{B}gc\frac{\delta}{\delta A_{l}} \int d\mathbf{r}(\psi^{\dagger}_{\sigma}\bm{\sigma}_{\sigma\sigma^{\prime}}\psi_{\sigma})\cdot\vec{\nabla}\times\mathbf{A}\\
=& \mu_{B}gc\frac{\delta}{\delta A_{l}} \int d\mathbf{r}\;s_i\epsilon_{ijk}\partial_j A_k\\
=& -\mu_{B}gc\frac{\delta}{\delta A_{l}} \int d\mathbf{r}\;(\partial_j s_i)\epsilon_{ijk} A_k\\
=& -\mu_{B}gc \int d\mathbf{r}\;(\partial_j s_i)\epsilon_{ijk} \delta_{lk}\\
=&-\mu_{B}gc (\partial_j s_i)\epsilon_{ijl}\\
=&\mu_{B}gc (\epsilon_{jil}\partial_j s_i).
\end{align}
Here, we denote $\vec{s}=\psi^{\dagger}_{\sigma}\bm{\sigma}_{\sigma\sigma^{\prime}}\psi_{\sigma}$.

Now we can write down our final result:
\begin{equation}
\mathbf{j}_{spin}=\mu_{B}gc \nabla\times (\psi^{\dagger}\bm{\sigma}\psi) .
\end{equation}
\section{Diamagnetic current}
\begin{equation}
\mathbf{j}_{dia}=-\frac{e^2}{mc}\bra \psi^{\dagger}_{\sigma}(\mathbf{r})\psi_{\sigma}(\mathbf{r})\ket \mathbf{A}(\mathbf{r},\tau).
\end{equation}
Expand $\psi_{\sigma}$:
\begin{equation}
\psi_{\sigma}(\mathbf{r})=\sum_{n\mathbf{k}} \varphi_{n\mathbf{k}\sigma}(\mathbf{r})a_{n\mathbf{k}}.
\end{equation}
Therefore, we have
\begin{align}
\mathbf{j}_{dia} =&-\frac{e^2}{mc}\sum_{n\mathbf{k}n^{\prime}\mathbf{k}^{\prime}} \varphi_{n\mathbf{k}\sigma}^{*}\varphi_{n^{\prime}\mathbf{k}^{\prime}\sigma}\bra a_{n\mathbf{k}}^{\dagger}a_{n^{\prime}\mathbf{k}^{\prime}}\ket \mathbf{A}(\mathbf{r},\tau)\\
=&-\frac{e^2}{mc}\sum_{n\mathbf{k}\sigma}\lvert \varphi_{n\mathbf{k}\sigma}\rvert^2 f_{n\mathbf{k}}\mathbf{A}(\mathbf{r},\tau)\\
=&-\frac{e^2}{mc}\sum_{n\mathbf{k}\sigma}\frac{1}{V}\lvert u_{n\mathbf{k}\sigma}(\mathbf{r})\rvert^2f_{n\mathbf{k}}\mathbf{A}(\mathbf{r},\tau).
\end{align}
$f_{n\mathbf{k}}$ is Pauli-Dirac distribution function. 
Average diamagnetic current over a unit cell:
\begin{equation}
\bar{\mathbf{j}}_{dia}=-\frac{e^2}{mc}\sum_{n\mathbf{k}\sigma}\frac{1}{Vv_0}\int_{v_0} d\rho \lvert u_{n\mathbf{k}\sigma}(\mathbf{r})\rvert^2f_{n\mathbf{k}}\mathbf{A}(\mathbf{r},\tau).
\end{equation}
Since particle number
\begin{align}
N =&\int d\mathbf{r}\bra \psi_{\sigma}^{\dagger}\psi_{\sigma}\ket \\
=&\int d\mathbf{r}\sum_{n\mathbf{k}}\frac{\lvert u_{n\mathbf{k}\rvert^2}}{V}f_{n\mathbf{k}}\\
=&\sum_{n\mathbf{k}}\frac{1}{V}\sum_{R}\int_{v_0} d\rho \lvert u_{n\mathbf{k}}\lvert^2f_{n\mathbf{k}}\\
=&\sum_{n\mathbf{k}}\frac{1}{v_0}\int_{v_0} d\rho \lvert u_{n\mathbf{k}}\lvert^2f_{n\mathbf{k}},
\end{align}
We can simplify our final result as
\begin{equation}
\bar{\mathbf{j}}_{dia}=-\frac{e^2}{mc}\cdot\frac{N}{V}\cdot \mathbf{A}.\label{dia}
\end{equation}
\section{Gradient part of the current}
The gradient part of the current has been given by Eq.(\ref{1}):
\begin{equation}
\mathbf{j}_{grad}=\frac{e\hbar}{2mi}(\psi_{\sigma}^{\dagger}\nabla \psi_{\sigma}-\nabla \psi_{\sigma}^{\dagger}\cdot\psi_{\sigma}).
\end{equation}
Obviously, the eigenvalue of $\mathbf{j}_{grad}$ can be achieved by
\begin{equation}
\bra \mathbf{j}_{grad}(\mathbf{r},\tau)\ket =\frac{e\hbar}{2mi}\lim_{\rr^{\prime}\to \rr}(\partial_{\rr}-\partial_{\rr^{\prime}})\sum_{\sigma}G_{\sigma\sigma}(\mathbf{r},\mathbf{r}^{\prime};\tau,\tau+\delta),
\end{equation}
with Green's function 
\begin{equation}
G_{\sigma\sigma^{\prime}}(\mathbf{r},\mathbf{r}^{\prime}\;\tau,\tau^{\prime})= - \bra T_\tau \psi_{\sigma}(\mathbf{r},\tau)\psi_{\sigma^{\prime}}^{\dagger}(\mathbf{r}^{\prime},\tau^{\prime})\ket .
\end{equation}

We need to calculate $G_{\sigma\sigma^{\prime}}(\mathbf{r},\mathbf{r}^{\prime}\;\tau,\tau^{\prime})$ first.

Hamiltonian as shown is Eq.(\ref{H}) can be written as
\begin{equation}
H=\int d \mathbf{r}\psi^{\dagger}_{\sigma}\hat{h}_{\sigma\sigma^{\prime}}\psi_{\sigma^{\prime}},
\end{equation}
\begin{equation}
\hat{h}:=\hat{h}_0+\hat{U}=\hat{h}_0-\frac{e}{2mc}(\mathbf{p}\cdot\mathbf{A}+\mathbf{A}\cdot\mathbf{p}).
\end{equation}
With Matsubara $\tau=it$, we have 
\begin{equation}
(-\frac{\partial}{\partial \tau}-\hat{h})G(\mathbf{r},\mathbf{r}^{\prime}\;\tau,\tau^{\prime})=\delta(\mathbf{r}-\mathbf{r}^{\prime})\delta(\tau-\tau^{\prime}).
\end{equation}
Also, for free electron,
\begin{equation}
(-\frac{\partial}{\partial \tau}-\hat{h}_{0})G_{0}=1.
\end{equation}
Therefore we have
\begin{equation}
(G^{-1}_{0}-U)G(\mathbf{r},\mathbf{r}^{\prime}\;\tau,\tau^{\prime})=\delta(\mathbf{r}-\mathbf{r}^{\prime})\delta(\tau-\tau^{\prime}).
\end{equation}
Expand $G$ to the first order:
\begin{equation}
 G=G_0+G_0 U G_0.
\end{equation}
with
\begin{equation}
G_0=\sum_{n\mathbf{k}}\frac{|n\mathbf{k}\ket \bra n\mathbf{k}|}{i\varepsilon-\xi_{n\mathbf{k}}},\quad \xi_{n\mathbf{k}}:=\varepsilon_{n\mathbf{k}}-\mu.
\end{equation}
Now we  can write
\begin{equation}
\mathbf{j}_{grad}(\mathbf{r},\tau)=\frac{e\hbar}{2mi}\lim_{\mathbf{r}^{\prime}\to \mathbf{r}}(\partial_{\mathbf{r}}-\partial_{\mathbf{r}^{\prime}})\sum_{\sigma}\int d\mathbf{r}_{1}d\tau_1 G_{\sigma\sigma^{\prime}}(\mathbf{r},\mathbf{r}_1;\tau,\tau_1)U(\mathbf{r}_1,\tau_1)G_{\sigma^{\prime}\sigma}(\mathbf{r}_1,\mathbf{r}^{\prime};\tau_1,\tau+\delta),\label{34}
\end{equation}
with
\begin{equation}
G_{\sigma\sigma^{\prime}}(\mathbf{r},\mathbf{r}^{\prime}\;\tau,\tau^{\prime})=\sum_{n\mathbf{k}}\frac{\psi_{n\mathbf{k}\sigma}(\mathbf{r},\tau)\psi^{*}_{n\mathbf{k}\sigma^{\prime}}(\mathbf{r}^{\prime},\tau^{\prime})}{i\varepsilon-\xi_{n\mathbf{k}}},
\end{equation}
and
\begin{equation}
U=-\frac{e}{2mc}(\mathbf{p}\cdot\mathbf{A}_{\tau}+\mathbf{A}_{\tau}\cdot\mathbf{p}).
\end{equation}
Fourier Transformation for $\mathbf{A}$:
\begin{equation}
\mathbf{A}_{\tau}(\mathbf{r})=\frac{1}{V}\sum_{\mathbf{q},\Omega}\mathbf{A}_{\mathbf{q},\Omega} \e^{i\mathbf{q}\mathbf{r}-i\Omega\tau}.
\end{equation}
Green's function only depends on the difference:
\begin{equation}
G_{\sigma\sigma^{\prime}}(\mathbf{r},\mathbf{r}^{\prime}\;\tau,\tau^{\prime})=\sum_{n{k}}\frac{\psi_{n\mathbf{k}\sigma}(\mathbf{r})\psi^{*}_{n\mathbf{k}\sigma^{\prime}}(\mathbf{r}^{\prime})}{i\varepsilon-\xi_{n\mathbf{k}}}\e^{-i\varepsilon(\tau-\tau^{\prime})}.
\end{equation}

Taking all these into Eq.(\ref{34}), we can get
\begin{align}
\mathbf{j}_{grad}(\mathbf{r},\tau)=&\frac{e\hbar}{2mi}\lim_{\mathbf{r}^{\prime}\to \mathbf{r}}(\partial_{\mathbf{r}}-\partial_{\mathbf{r}^{\prime}})\sum_{\substack{\Omega,\sigma,n,\mathbf{k}\\ \sigma^{\prime},n^{\prime},\mathbf{k}^{\prime}}}\int d\mathbf{r}_{1}d\tau_1 T\sum_{\varepsilon}\frac{\psi_{n\mathbf{k}\sigma}(\mathbf{r})\psi^{*}_{n\mathbf{k}\sigma^{\prime}}(\mathbf{r}_1)}{i\varepsilon-\xi_{n\mathbf{k}}}\e^{-i\varepsilon(\tau-\tau_1)}\\
&\cdot[-\frac{e}{2mc}(\mathbf{p}\cdot\mathbf{A}_{\Omega}(\mathbf{r}_1)+\mathbf{A}_{\Omega}(\mathbf{r}_1)\cdot\mathbf{p})\e^{-i\Omega\tau_1}]T\sum_{\varepsilon^{\prime}}\frac{\psi_{n^{\prime}\mathbf{k}^{\prime}\sigma^{\prime}}(\mathbf{r}_1)\psi^{*}_{n^{\prime}\mathbf{k}^{\prime}\sigma}(\mathbf{r}^{\prime})}{i\varepsilon^{\prime}-\xi_{n^{\prime}\mathbf{k}^{\prime}}}\e^{-i\varepsilon^{\prime}(\tau_1-\tau-\delta)}\\
=&\frac{e\hbar}{2mi}\lim_{\mathbf{r}^{\prime}\to \mathbf{r}}(\partial_{\mathbf{r}}-\partial_{\mathbf{r}^{\prime}})\sum_{\substack{\Omega,\sigma,n,\mathbf{k}\\\sigma^{\prime},n^{\prime},\mathbf{k}^{\prime}}}\int d\mathbf{r}_{1} T\sum_{\varepsilon}\frac{\psi_{n\mathbf{k}\sigma}(\mathbf{r})\psi^{*}_{n\mathbf{k}\sigma^{\prime}}(\mathbf{r}_1)}{i\varepsilon-\xi_{n\mathbf{k}}}\e^{-i\varepsilon \tau}\\
&\cdot[-\frac{e}{2mc}(\mathbf{p}\cdot\mathbf{A}_{\Omega}(\mathbf{r}_1)+\mathbf{A}_{\Omega}(\mathbf{r}_1)\cdot\mathbf{p})]T\sum_{\varepsilon^{\prime}}\frac{\psi_{n^{\prime}\mathbf{k}^{\prime}\sigma^{\prime}}(\mathbf{r}_1)\psi^{*}_{n^{\prime}\mathbf{k}^{\prime}\sigma}(\mathbf{r}^{\prime})}{i\varepsilon^{\prime}-\xi_{n^{\prime}\mathbf{k}^{\prime}}}\e^{i\varepsilon^{\prime}(\tau+\delta)}\frac{1}{T}\delta_{\varepsilon-\Omega-\varepsilon^{\prime},0}\\
=&\frac{e\hbar}{2mi}\lim_{\mathbf{r}^{\prime}\to \mathbf{r}}(\partial_{\mathbf{r}}-\partial_{\mathbf{r}^{\prime}})\sum_{\substack{\Omega,\sigma,n,\mathbf{k}\\\sigma^{\prime},n^{\prime},\mathbf{k}^{\prime}}}\int d\mathbf{r}_{1} T\sum_{\varepsilon}\frac{\psi_{n\mathbf{k}\sigma}(\mathbf{r})\psi^{*}_{n\mathbf{k}\sigma^{\prime}}(\mathbf{r}_1)}{i\varepsilon-\xi_{n\mathbf{k}}}\e^{-i\varepsilon \tau}\\
&\cdot[-\frac{e}{2mc}(\mathbf{p}\cdot\mathbf{A}_{\Omega}(\mathbf{r}_1)+\mathbf{A}_{\Omega}(\mathbf{r}_1)\cdot\mathbf{p})]\frac{\psi_{n^{\prime}\mathbf{k}^{\prime}\sigma^{\prime}}(\mathbf{r}_1)\psi^{*}_{n^{\prime}\mathbf{k}^{\prime}\sigma}(\mathbf{r}^{\prime})}{i(\varepsilon-\Omega)-\xi_{n^{\prime}\mathbf{k}^{\prime}}}\e^{i(\varepsilon-\Omega)(\tau+\delta)}\\
=&\frac{e\hbar}{2mi}\lim_{\mathbf{r}^{\prime}\to \mathbf{r}}(\partial_{\mathbf{r}}-\partial_{\mathbf{r}^{\prime}})\sum_{\substack{\Omega,\sigma,n,\mathbf{k}\\\sigma^{\prime},n^{\prime},\mathbf{k}^{\prime}}}\int d\mathbf{r}_{1} T\sum_{\varepsilon}\frac{\psi_{n\mathbf{k}\sigma}(\mathbf{r})\psi^{*}_{n\mathbf{k}\sigma^{\prime}}(\mathbf{r}_1)}{i\varepsilon-\xi_{n\mathbf{k}}}\\
&\cdot[-\frac{e}{2mc}(\mathbf{p}\cdot\mathbf{A}_{\Omega}(\mathbf{r}_1)+\mathbf{A}_{\Omega}(\mathbf{r}_1)\cdot\mathbf{p})]\frac{\psi_{n^{\prime}\mathbf{k}^{\prime}\sigma^{\prime}}(\mathbf{r}_1)\psi^{*}_{n^{\prime}\mathbf{k}^{\prime}\sigma}(\mathbf{r}^{\prime})}{i(\varepsilon-\Omega)-\xi_{n\mathbf{k}}}\e^{-i\Omega\tau}\\
=&\sum_{\Omega}\mathbf{j}_{\Omega}(\mathbf{r})\e^{-i\Omega\tau}.
\end{align}
Since
\begin{equation}
T\cdot\sum_{\varepsilon}\frac{1}{i\varepsilon-\xi_{n\mathbf{k}}}\cdot\frac{1}{i(\varepsilon-\Omega)-\xi_{n\mathbf{k}}}=\frac{f_{n^{\prime}\mathbf{k}^{\prime}}-f_{n\mathbf{k}}}{i\Omega+\xi_{n^{\prime}\mathbf{k}^{\prime}}-\xi_{n\mathbf{k}}},
\end{equation}
\begin{align}
\mathbf{j}_{\Omega}(\mathbf{r})=&\sum_{\substack{\sigma,n,\mathbf{k}\\\sigma^{\prime},n^{\prime},\mathbf{k}^{\prime}}}\frac{f_{n^{\prime}\mathbf{k}^{\prime}}-f_{n\mathbf{k}}}{i\Omega+\xi_{n^{\prime}\mathbf{k}^{\prime}}-\xi_{n\mathbf{k}}}\frac{e\hbar}{2mi}\lim_{\mathbf{r}^{\prime}\to \mathbf{r}}(\partial_{\mathbf{r}}-\partial_{\mathbf{r}^{\prime}})\int d\mathbf{r}_{1}\psi_{n\mathbf{k}\sigma}(\mathbf{r})\psi^{*}_{n\mathbf{k}\sigma^{\prime}}(\mathbf{r}_1)\\
&\cdot[-\frac{e}{2mc}(\frac{\hbar}{i}\partial_{\mathbf{r}_1}\cdot\mathbf{A}_{\Omega}(\mathbf{r}_1)+\mathbf{A}_{\Omega}(\mathbf{r}_1)\cdot\frac{\hbar}{i}\partial_{\mathbf{r}_1})]\psi_{n^{\prime}\mathbf{k}^{\prime}\sigma^{\prime}}(\mathbf{r}_1)\psi^{*}_{n^{\prime}\mathbf{k}^{\prime}\sigma}(\mathbf{r}^{\prime})\\
=&\sum_{\substack{\sigma,n,\mathbf{k}\\\sigma^{\prime},n^{\prime},\mathbf{k}^{\prime}}}\frac{f_{n^{\prime}\mathbf{k}^{\prime}}-f_{n\mathbf{k}}}{i\Omega+\xi_{n^{\prime}\mathbf{k}^{\prime}}-\xi_{n\mathbf{k}}}\frac{e\hbar}{2mi}(-\frac{e}{2mc})\int d\mathbf{r}_{1}\{\\
&\partial_{\mathbf{r}}\psi_{n\mathbf{k}\sigma}(\mathbf{r})\psi^{*}_{n\mathbf{k}\sigma^{\prime}}(\mathbf{r}_1)(\frac{\hbar}{i}\partial_{\mathbf{r}_1}\cdot\mathbf{A}_{\Omega}(\mathbf{r}_1)+\mathbf{A}_{\Omega}(\mathbf{r}_1)\cdot\frac{\hbar}{i}\partial_{\mathbf{r}_1})\psi_{n^{\prime}\mathbf{k}^{\prime}\sigma^{\prime}}(\mathbf{r}_1)\psi^{*}_{n^{\prime}\mathbf{k}^{\prime}\sigma}(\mathbf{r})\\
&-\psi_{n\mathbf{k}\sigma}(\mathbf{r})\psi^{*}_{n\mathbf{k}\sigma^{\prime}}(\mathbf{r}_1)(\frac{\hbar}{i}\partial_{\mathbf{r}_1}\cdot\mathbf{A}_{\Omega}(\mathbf{r}_1)+\mathbf{A}_{\Omega}(\mathbf{r}_1)\cdot\frac{\hbar}{i}\partial_{\mathbf{r}_1})\psi_{n^{\prime}\mathbf{k}^{\prime}\sigma^{\prime}}(\mathbf{r}_1)\partial_{\mathbf{r}}\psi^{*}_{n^{\prime}\mathbf{k}^{\prime}\sigma}(\mathbf{r})\}\\
=&-\frac{e^2}{4m^2cV}\sum_{\substack{\mathbf{q}, \sigma,n,\mathbf{k}\\\sigma^{\prime},n^{\prime},\mathbf{k}^{\prime}}}\frac{f_{n^{\prime}\mathbf{k}^{\prime}}-f_{n\mathbf{k}}}{i\Omega+\xi_{n^{\prime}\mathbf{k}^{\prime}}-\xi_{n\mathbf{k}}}\int d\mathbf{r}_{1}\{\\
&\frac{\hbar}{i}\partial_{\mathbf{r}}\psi_{n\mathbf{k}\sigma}(\mathbf{r})\psi^{*}_{n\mathbf{k}\sigma^{\prime}}(\mathbf{r}_1)[\mathbf{A}_{\mathbf{q},\Omega}\e^{i\mathbf{q}\cdot\mathbf{r}_1}\cdot(2\frac{\hbar}{i}\partial_{\mathbf{r}_1}+\hbar\mathbf{q})\psi_{n^{\prime}\mathbf{k}^{\prime}\sigma^{\prime}}(\mathbf{r}_1)\psi^{*}_{n^{\prime}\mathbf{k}^{\prime}\sigma}(\mathbf{r})]\\
&-\psi_{n\mathbf{k}\sigma}(\mathbf{r})\psi^{*}_{n\mathbf{k}\sigma^{\prime}}(\mathbf{r}_1)[\mathbf{A}_{\mathbf{q},\Omega}\e^{i\mathbf{q}\cdot\mathbf{r}_1}\cdot(2\frac{\hbar}{i}\partial_{\mathbf{r}_1}+\hbar\mathbf{q})\psi_{n^{\prime}\mathbf{k}^{\prime}\sigma^{\prime}}(\mathbf{r}_1)\frac{\hbar}{i}\partial_{\mathbf{r}}\psi^{*}_{n^{\prime}\mathbf{k}^{\prime}\sigma}(\mathbf{r})]\}\\
=&-\frac{e^2}{4m^2cV}\sum_{\substack{\mathbf{q}, \sigma,n,\mathbf{k}\\\sigma^{\prime},n^{\prime},\mathbf{k}^{\prime}}}\frac{f_{n^{\prime}\mathbf{k}^{\prime}}-f_{n\mathbf{k}}}{i\Omega+\xi_{n^{\prime}\mathbf{k}^{\prime}}-\xi_{n\mathbf{k}}}\int d\mathbf{r}_{1}\{\\
&\psi^{*}_{n^{\prime}\mathbf{k}^{\prime}\sigma}(\mathbf{r})\frac{\hbar}{i}\partial_{\mathbf{r}}\psi_{n\mathbf{k}\sigma}(\mathbf{r})[\psi^{*}_{n\mathbf{k}\sigma^{\prime}}(\mathbf{r}_1)\e^{i\mathbf{q}\cdot\mathbf{r}_1}(2\frac{\hbar}{i}\partial_{\mathbf{r}_1}+\hbar\mathbf{q})\psi_{n^{\prime}\mathbf{k}^{\prime}\sigma^{\prime}}(\mathbf{r}_1)\cdot\mathbf{A}_{\mathbf{q},\Omega}]\\
&-\psi_{n\mathbf{k}\sigma}(\mathbf{r})\frac{\hbar}{i}\partial_{\mathbf{r}}\psi^{*}_{n^{\prime}\mathbf{k}^{\prime}\sigma}(\mathbf{r})[\psi^{*}_{n\mathbf{k}\sigma^{\prime}}(\mathbf{r}_1)\e^{i\mathbf{q}\cdot\mathbf{r}_1}(2\frac{\hbar}{i}\partial_{\mathbf{r}_1}+\hbar\mathbf{q})\psi_{n^{\prime}\mathbf{k}^{\prime}\sigma^{\prime}}(\mathbf{r}_1)\cdot\mathbf{A}_{\mathbf{q},\Omega}]\}\\
=&-\frac{e^2}{4m^2cV^2}\sum_{\substack{\mathbf{q}, \sigma,n,\mathbf{k}\\\sigma^{\prime},n^{\prime},\mathbf{k}^{\prime}}}\frac{f_{n^{\prime}\mathbf{k}^{\prime}}-f_{n\mathbf{k}}}{i\Omega+\xi_{n^{\prime}\mathbf{k}^{\prime}}-\xi_{n\mathbf{k}}}\int d\mathbf{r}_{1}\e^{i(\mathbf{q}-\mathbf{k}+\mathbf{k}^{\prime})\cdot\mathbf{r}_1}\{\\
&\psi^{*}_{n^{\prime}\mathbf{k}^{\prime}\sigma}(\mathbf{r})\frac{\hbar}{i}\partial_{\mathbf{r}}\psi_{n\mathbf{k}\sigma}(\mathbf{r})[u^{*}_{n\mathbf{k}\sigma^{\prime}}(\mathbf{r}_1)(2\frac{\hbar}{i}\partial_{\mathbf{r}_1}+2\hbar\mathbf{k}^{\prime}+\hbar\mathbf{q})u_{n^{\prime}\mathbf{k}^{\prime}\sigma^{\prime}}(\mathbf{r}_1)\cdot\mathbf{A}_{\mathbf{q},\Omega}]\\
&-\psi_{n\mathbf{k}\sigma}(\mathbf{r})\frac{\hbar}{i}\partial_{\mathbf{r}}\psi^{*}_{n^{\prime}\mathbf{k}^{\prime}\sigma}(\mathbf{r})[u^{*}_{n\mathbf{k}\sigma^{\prime}}(\mathbf{r}_1)(2\frac{\hbar}{i}\partial_{\mathbf{r}_1}+2\hbar\mathbf{k}^{\prime}+\hbar\mathbf{q})u_{n^{\prime}\mathbf{k}^{\prime}\sigma^{\prime}}(\mathbf{r}_1)\cdot\mathbf{A}_{\mathbf{q},\Omega}]\}\\
\end{align}
Here we used Bloch wave function
\begin{equation}
\psi_{n\mathbf{k}\sigma}(\mathbf{r})=\frac{1}{\sqrt{V}}\e^{i\mathbf{k}\cdot\mathbf{r}}u_{n\mathbf{k}\sigma}(\mathbf{r}).
\end{equation}
 
Since $u_{n\mathbf{k}\sigma}(\mathbf{r})$ is a periodic function:
\begin{equation}
u_{n\mathbf{k}\sigma}(\mathbf{r})=u_{n\mathbf{k}\sigma}(\mathbf{r}+\mathbf{R}).
\end{equation}
\begin{align}
&\int_{-\infty}^{\infty} d\mathbf{r}_{1}\e^{i(\mathbf{q}-\mathbf{k}+\mathbf{k}^{\prime})\cdot\mathbf{r}_1}u^{*}_{n\mathbf{k}\sigma^{\prime}}(\mathbf{r}_1)(2\frac{\hbar}{i}\partial_{\mathbf{r}_1}+2\hbar\mathbf{k}^{\prime}+\hbar\mathbf{q})u_{n^{\prime}\mathbf{k}^{\prime}\sigma^{\prime}}(\mathbf{r}_1)\\
=&\int_{-\infty}^{\infty} d\mathbf{r}_{1}\e^{i(\mathbf{q}-\mathbf{k}+\mathbf{k}^{\prime})\cdot(\mathbf{r}_1+\mathbf{R})}u^{*}_{n\mathbf{k}\sigma^{\prime}}(\mathbf{r}_1+\mathbf{R})(2\frac{\hbar}{i}\partial_{\mathbf{r}_1}+2\hbar\mathbf{k}^{\prime}+\hbar\mathbf{q})u_{n^{\prime}\mathbf{k}^{\prime}\sigma^{\prime}}(\mathbf{r}_1+\mathbf{R})\\
=&\int_{-\infty}^{\infty} d\mathbf{r}_{1}\e^{i(\mathbf{q}-\mathbf{k}+\mathbf{k}^{\prime})\cdot(\mathbf{r}_1+\mathbf{R})}u^{*}_{n\mathbf{k}\sigma^{\prime}}(\mathbf{r}_1)(2\frac{\hbar}{i}\partial_{\mathbf{r}_1}+2\hbar\mathbf{k}^{\prime}+\hbar\mathbf{q})u_{n^{\prime}\mathbf{k}^{\prime}\sigma^{\prime}}(\mathbf{r}_1)
\end{align}
We have
\begin{equation}
(\mathbf{q}-\mathbf{k}+\mathbf{k}^{\prime})\cdot\mathbf{R}=2\pi n, \qquad n=0, \pm 1, \pm 2, \cdots .\label{69}
\end{equation}
As all these vectors $\mathbf{q}, \mathbf{k}, \mathbf{k}^{\prime}$ are within Brillouin Zone, our $(\mathbf{q}-\mathbf{k}+\mathbf{k}^{\prime})\cdot\mathbf{R}$ can only take the value $0, \pm 2\pi n$. But we are concerned with the $\mathbf{q}\rightarrow 0$ result at last. When $\mathbf{q}\rightarrow 0$, $(\mathbf{q}-\mathbf{k}+\mathbf{k}^{\prime})\cdot\mathbf{R}$ can only take the value $0$. Therefore, we neglect $\pm 2\pi n$ terms in our calculation. However, as a reminder, if we want to get $\sigma(\mathbf{q},\omega)$ for a larger $\mathbf{q}$, we should include $\pm 2\pi n$ terms.
\begin{align}
&\int_{-\infty}^{\infty} d\mathbf{r}_{1}\e^{i(\mathbf{q}-\mathbf{k}+\mathbf{k}^{\prime})\cdot\mathbf{r}_1}u^{*}_{n\mathbf{k}\sigma^{\prime}}(\mathbf{r}_1)(2\frac{\hbar}{i}\partial_{\mathbf{r}_1}+2\hbar\mathbf{k}^{\prime}+\hbar\mathbf{q})u_{n^{\prime}\mathbf{k}^{\prime}\sigma^{\prime}}(\mathbf{r}_1)\\
=&\sum_{R}\int_{v_0} d\bm{\rho}\; u^{*}_{n\mathbf{k}\sigma^{\prime}}(\bm{\rho})(2\frac{\hbar}{i}\partial_{\bm{\rho}}+2\hbar\mathbf{k}^{\prime}+\hbar\mathbf{q})u_{n^{\prime}\mathbf{k}^{\prime}\sigma^{\prime}}(\bm{\rho})\delta_{\mathbf{q}-\mathbf{k}+\mathbf{k}^{\prime},0}\\
=&N\bra u_{n\mathbf{k}\sigma^{\prime}}|(2\frac{\hbar}{i}\partial_{\bm{\rho}}+2\hbar\mathbf{k}^{\prime}+\hbar\mathbf{q})|u_{n^{\prime}\mathbf{k}^{\prime}\sigma^{\prime}}\ket \delta_{\mathbf{q}-\mathbf{k}+\mathbf{k}^{\prime},0}\label{72}
\end{align}
Therefore, we can get our $\mathbf{j}_\Omega (\mathbf{r})$ simplified as
\begin{align}
\mathbf{j}_\Omega (\mathbf{r})=&-\frac{e^2N}{4m^2cV^2}\sum_{\substack{\mathbf{q}, \sigma,\mathbf{k}\\n,n^{\prime}}}\frac{f_{n^{\prime}(\mathbf{k}-\mathbf{q})}-f_{n\mathbf{k}}}{i\Omega+\xi_{n^{\prime}(\mathbf{k}-\mathbf{q})}-\xi_{n\mathbf{k}}}\{\psi^{*}_{n^{\prime}(\mathbf{k}-\mathbf{q})\sigma}(\mathbf{r})\frac{\hbar}{i}\partial_{\mathbf{r}}\psi_{n\mathbf{k}\sigma}(\mathbf{r})\\
&-\psi_{n\mathbf{k}\sigma}(\mathbf{r})\frac{\hbar}{i}\partial_{\mathbf{r}}\psi^{*}_{n^{\prime}(\mathbf{k}-\mathbf{q})\sigma}(\mathbf{r})\}[\bra u_{n\mathbf{k}}|(2\frac{\hbar}{i}\partial_{\bm{\rho}}+2\hbar\mathbf{k}-\hbar\mathbf{q})|u_{n^{\prime}(\mathbf{k}-\mathbf{q})}\ket \cdot\mathbf{A}_{\mathbf{q},\Omega}]\\
=&-\frac{e^2N}{4m^2cV^3}\sum_{\substack{\mathbf{q}, \sigma,\mathbf{k}\\n,n^{\prime}}}\frac{f_{n^{\prime}(\mathbf{k}-\mathbf{q})}-f_{n\mathbf{k}}}{i\Omega+\xi_{n^{\prime}(\mathbf{k}-\mathbf{q})}-\xi_{n\mathbf{k}}}\{\e^{i\mathbf{q}\cdot\mathbf{r}}u^{*}_{n^{\prime}(\mathbf{k}-\mathbf{q})\sigma}(\mathbf{r})(\hbar\mathbf{k}+\frac{\hbar}{i}\partial_{\mathbf{r}})u_{n\mathbf{k}\sigma}(\mathbf{r})\\
&-\e^{i\mathbf{q}\cdot\mathbf{r}}u_{n\mathbf{k}\sigma}(\mathbf{r})(\hbar(\mathbf{q}-\mathbf{k})+\frac{\hbar}{i}\partial_{\mathbf{r}})u^{*}_{n^{\prime}(\mathbf{k}-\mathbf{q})\sigma}(\mathbf{r})\}[\bra u_{n\mathbf{k}}|(2\frac{\hbar}{i}\partial_{\bm{\rho}}+2\hbar\mathbf{k}-\hbar\mathbf{q})|u_{n^{\prime}(\mathbf{k}-\mathbf{q})}\ket \cdot\mathbf{A}_{\mathbf{q},\Omega}]\\
=&-\frac{e^2N}{4m^2cV^3}\sum_{\substack{\mathbf{q}, \sigma,\mathbf{k}\\n,n^{\prime}}}\frac{f_{n^{\prime}(\mathbf{k}-\mathbf{q})}-f_{n\mathbf{k}}}{i\Omega+\xi_{n^{\prime}(\mathbf{k}-\mathbf{q})}-\xi_{n\mathbf{k}}}\e^{i\mathbf{q}\cdot\mathbf{r}}\{u^{*}_{n^{\prime}(\mathbf{k}-\mathbf{q})\sigma}(\mathbf{r})(\hbar\mathbf{k}+\frac{\hbar}{i}\partial_{\mathbf{r}})u_{n\mathbf{k}\sigma}(\mathbf{r})\\
&-u_{n\mathbf{k}\sigma}(\mathbf{r})(\hbar(\mathbf{q}-\mathbf{k})+\frac{\hbar}{i}\partial_{\mathbf{r}})u^{*}_{n^{\prime}(\mathbf{k}-\mathbf{q})\sigma}(\mathbf{r})\}[\bra u_{n\mathbf{k}}|(2\frac{\hbar}{i}\partial_{\bm{\rho}}+2\hbar\mathbf{k}-\hbar\mathbf{q})|u_{n^{\prime}(\mathbf{k}-\mathbf{q})}\ket \cdot\mathbf{A}_{\mathbf{q},\Omega}].\label{78}
\end{align}
Do Fourier Transformation
\begin{equation}
\mathbf{j}_\Omega(\mathbf{q})=\int d\mathbf{r} \mathbf{j}_\Omega (\mathbf{r})\e^{-i\mathbf{q}\mathbf{r}}.\label{79}
\end{equation}
First, let us do partial integral to simplify it a little bit:
\begin{align}
&\int d\mathbf{r}\e^{i(\mathbf{q}^\prime-\mathbf{q})\cdot\mathbf{r}}\{u^{*}_{n^{\prime}(\mathbf{k}-\mathbf{q}^\prime)\sigma}(\mathbf{r})(\hbar\mathbf{k}+\frac{\hbar}{i}\partial_{\mathbf{r}})u_{n\mathbf{k}\sigma}(\mathbf{r})-u_{n\mathbf{k}\sigma}(\mathbf{r})(\hbar(\mathbf{q}^\prime-\mathbf{k})+\frac{\hbar}{i}\partial_{\mathbf{r}})u^{*}_{n^{\prime}(\mathbf{k}-\mathbf{q}^\prime)\sigma}(\mathbf{r})\}\\
=&\int d\mathbf{r}\e^{i(\mathbf{q}^\prime-\mathbf{q})\cdot\mathbf{r}}\{u^{*}_{n^{\prime}(\mathbf{k}-\mathbf{q}^\prime)\sigma}(\mathbf{r})(\hbar\mathbf{k}+\frac{\hbar}{i}\partial_{\mathbf{r}})u_{n\mathbf{k}\sigma}(\mathbf{r})-u^{*}_{n^{\prime}(\mathbf{k}-\mathbf{q}^\prime)\sigma}(\mathbf{r})(\hbar(\mathbf{q}-\mathbf{k})-\frac{\hbar}{i}\partial_{\mathbf{r}})u_{n\mathbf{k}\sigma}(\mathbf{r})\}\\
=&\int d\mathbf{r}\e^{i(\mathbf{q}^\prime-\mathbf{q})\cdot\mathbf{r}}u^{*}_{n^{\prime}(\mathbf{k}-\mathbf{q}^\prime)\sigma}(\mathbf{r})(2\hbar\mathbf{k}-\hbar\mathbf{q}+2\frac{\hbar}{i}\partial_{\mathbf{r}})u_{n\mathbf{k}\sigma}(\mathbf{r}).
\end{align}
Same analysis process as what we did from Eq.(\ref{69}) to Eq.(\ref{72}) give us similar result
\begin{align}
&\int d\mathbf{r}\e^{i(\mathbf{q}^\prime-\mathbf{q})\cdot\mathbf{r}}u^{*}_{n^{\prime}(\mathbf{k}-\mathbf{q}^\prime)\sigma}(\mathbf{r})(2\hbar\mathbf{k}-\hbar\mathbf{q}+2\frac{\hbar}{i}\partial_{\mathbf{r}})u_{n\mathbf{k}\sigma}(\mathbf{r})\\
=&N\bra u_{n^{\prime}(\mathbf{k}-\mathbf{q}^\prime)\sigma}|(2\frac{\hbar}{i}\partial_{\mathbf{r}}+2\hbar\mathbf{k}-\hbar\mathbf{q})|u_{n\mathbf{k}\sigma}\ket \delta_{\mathbf{q}^{\prime}-\mathbf{q},0}.
\end{align}
Substitute it into Eq.(\ref{78}) and Eq.(\ref{79}), we get
\begin{align}
\mathbf{j}_\Omega(\mathbf{q})=&-\frac{e^2N^2}{4m^2cV^3}\sum_{\mathbf{k},n,n^{\prime}}\frac{f_{n^{\prime}(\mathbf{k}-\mathbf{q})}-f_{n\mathbf{k}}}{i\Omega+\xi_{n^{\prime}(\mathbf{k}-\mathbf{q})}-\xi_{n\mathbf{k}}}\bra u_{n^{\prime}(\mathbf{k}-\mathbf{q})}|(2\frac{\hbar}{i}\partial_{\mathbf{r}}+2\hbar\mathbf{k}-\hbar\mathbf{q})|u_{n\mathbf{k}}\ket \\
&[\bra u_{n\mathbf{k}}|(2\frac{\hbar}{i}\partial_{\bm{\rho}}+2\hbar\mathbf{k}-\hbar\mathbf{q})|u_{n^{\prime}(\mathbf{k}-\mathbf{q})}\ket \cdot\mathbf{A}_{\mathbf{q},\Omega}]
\end{align}
\section{Prove that $\mathbf{j}_{grad}(\Omega\to 0,\mathbf{q}\to 0)+\mathbf{j}_{dia}=0$.}
Denote Eq.(\ref{H}) as
\begin{equation}
H=\int d\mathbf{r} \; \psi_{\sigma}^{\dagger}\hat{h}_{\sigma \sigma^{\prime} }\psi_{\sigma^{\prime}},
\end{equation}
\begin{equation}
\hat{h}=\frac{(\mathbf{p}-\frac{e}{c}\mathbf{A})^2}{2m}+\lambda\bm{\sigma} \cdot \EE \times (\mathbf{p}-\frac{e}{c}\mathbf{A})+U(r)-g\mu\BB \cdot\bm{\sigma}.\label{h}
\end{equation}
$\hat{h}$ have eigenstate $\psi_{n\mathbf{k}\sigma}(\mathbf{r})=\frac{1}{\sqrt{V}}\e^{i\mathbf{k}\cdot\mathbf{r}}u_{n\mathbf{k}\sigma}(\mathbf{r})$, with corresponding eigenvalue as $\varepsilon_{n\mathbf{k}}$:
\begin{equation}
\hat{h}_{\mathbf{k}}|u_{n\mathbf{k}}\ket =\varepsilon_{n\mathbf{k}}|u_{n\mathbf{k}}\ket \label{eigen}.
\end{equation}

Now we neglect the spin-orbit interaction as what we did before. We want to prove that when $\Omega\to 0,\mathbf{q}\to 0$, the gradient part of the current density 
\begin{align}
\mathbf{j}_{grad}^{i}(\Omega,\mathbf{q})=&-\frac{e^2N^2}{4m^2cV^3}\sum_{\mathbf{k},n,n^{\prime}}\frac{f_{n^{\prime}(\mathbf{k}-\mathbf{q})}-f_{n\mathbf{k}}}{i\Omega+\xi_{n^{\prime}(\mathbf{k}-\mathbf{q})}-\xi_{n\mathbf{k}}}\bra u_{n^{\prime}(\mathbf{k}-\mathbf{q})}|(2\hbar \hat{p}^i+2\hbar k^i-\hbar q^i)|u_{n\mathbf{k}}\ket \\
&\bra u_{n\mathbf{k}}|(2\hbar \hat{p}^j+2\hbar k^j-\hbar q^j)|u_{n^{\prime}(\mathbf{k}-\mathbf{q})}\ket A^j_{\mathbf{q},\Omega}\label{gradient}
\end{align}
cancels diamagnetic current density
\begin{equation}
\mathbf{j}^{i}_{dia}(\Omega,\mathbf{q})=-\frac{e^2}{mc} \frac{N}{V} A^i_{\Omega,\mathbf{q}}\; .
\end{equation}
Write
\begin{equation}
\mathbf{j}^i_{\mathbf{q}\Omega}=Q^{ij}_{\mathbf{q}\Omega}A^j_{\mathbf{q}\Omega},
\end{equation}
Now we try to prove that these two $Q$s are same with opposite sign.
\subsection{Diagonal contribution $n=n^{\prime}$ (intraband)}
From Eq.(\ref{gradient}), we have our intraband
\begin{equation}
\lim_{\substack{\Omega\to 0\\ \mathbf{q}\to 0}}Q^{ij}_{\mathbf{q}\Omega}=-\frac{e^2N^2}{4m^2cV^3}\sum_{\mathbf{k},n}\frac{-\frac{\partial f_{n\mathbf{k}}}{\partial \mathbf{k}}\cdot\mathbf{q}}{-\frac{\partial \xi_{n\mathbf{k}}}{\partial\mathbf{k}}\cdot{\mathbf{q}}}\bra u_{n\mathbf{k}}|(2\hbar \hat{p}^i+2\hbar k^i)|u_{n\mathbf{k}}\ket \bra u_{n\mathbf{k}}|(2\hbar \hat{p}^j+2\hbar k^j)|u_{n\mathbf{k}}\ket  .\label{94}
\end{equation}
Here
\begin{equation}
f_{n\mathbf{k}}=f_{\mathrm{th}}(\xi_{n\mathbf{k}}),
\end{equation}
$f_{\mathrm{th}}$ is the thermal distribution function.
\begin{equation}
\frac{\partial f_{n\mathbf{k}}}{\partial \mathbf{k}}=\frac{\partial f_{\mathrm{th}}(\xi_{n\mathbf{k}})}{\partial \xi_{n\mathbf{k}}}\cdot\frac{\partial \xi_{n\mathbf{k}}}{\partial\mathbf{k}}.
\end{equation}
Therefore, we have
\begin{equation}
\lim_{\substack{\Omega\to 0\\ \mathbf{q}\to 0}}Q^{ij}_{\mathbf{q}\Omega}=-\frac{e^2N^2}{4m^2cV^3}\sum_{\mathbf{k},n}\frac{\partial f_{\mathrm{th}}(\xi_{n\mathbf{k}})}{\partial \xi_{n\mathbf{k}}}\bra u_{n\mathbf{k}}|(2\hbar \hat{p}^i+2\hbar k^i)|u_{n\mathbf{k}}\ket \bra u_{n\mathbf{k}}|(2\hbar \hat{p}^j+2\hbar k^j)|u_{n\mathbf{k}}\ket  .\label{97}
\end{equation}
Take the derivative of $\mathbf{k}$ to each side of Eq.(\ref{eigen}):
\begin{equation}
\partial_{\mathbf{k}}\hat{h}_{\mathbf{k}}|u_{n\mathbf{k}}\ket +\hat{h}_{\mathbf{k}}\partial_{\mathbf{k}}|u_{n\mathbf{k}}\ket =\partial_{\mathbf{k}}\varepsilon_{n\mathbf{k}}|u_{n\mathbf{k}}\ket +\varepsilon_{n\mathbf{k}}\partial_{\mathbf{k}}|u_{n\mathbf{k}}\ket .\label{98eigen}
\end{equation}
Applying $\bra u_{n\mathbf{k}}|$ to the left side, we get
\begin{equation}
\bra u_{n\mathbf{k}}|\partial_{\mathbf{k}}\hat{h}_{\mathbf{k}}|u_{n\mathbf{k}}\ket =\partial_{\mathbf{k}}\varepsilon_{n\mathbf{k}}\bra u_{n\mathbf{k}}|u_{n\mathbf{k}}\ket =\frac{V}{N}\partial_{\mathbf{k}}\varepsilon_{n\mathbf{k}}=\frac{V}{N}\partial_{\mathbf{k}}\xi_{n\mathbf{k}}.
\end{equation}
From Eq.(\ref{h}), we see that, up to the lowest order
\begin{equation}
\partial_{\mathbf{k}}h_{\mathbf{k}}\approx \frac{\hbar\hat{p}+\hbar\mathbf{k}}{m}.
\end{equation}
Substitute it back to Eq.(\ref{97}), we have 
\begin{align}
\lim_{\substack{\Omega\to 0\\ \mathbf{q}\to 0}}Q^{ij}_{\mathbf{q}\Omega}=&-\frac{e^2}{cV}\sum_{\mathbf{k},n}\frac{\partial f_{\mathrm{th}}(\xi_{n\mathbf{k}})}{\partial \xi_{n\mathbf{k}}}\frac{\partial \xi_{n\mathbf{k}}}{\partial k_i}\frac{\partial \xi_{n\mathbf{k}}}{\partial k_j}\\
=&\frac{e^2}{cV}\sum_{\mathbf{k},n}\frac{\partial^2 \xi_{n\mathbf{k}}}{\partial k_i \partial k_j}f_{\mathrm{th}}(\xi_{n\mathbf{k}}).
\end{align}
By taking a second derivative to Eq.(\ref{eigen}) $\partial_{k_i}\partial_{k_j} (\hat{h}_{\mathbf{k}}|u_{n\mathbf{k}}\ket =\varepsilon_{n\mathbf{k}}|u_{n\mathbf{k}}\ket )$, we get
\begin{equation}
\partial_{k_i} (\frac{\hat{p}_j+k_j}{m}|u_{n\mathbf{k}}\ket +\hat{h}_{\mathbf{k}}\partial_{k_j}|u_{n\mathbf{k}}\ket )=\partial_{k_i} (\partial_{k_j}\varepsilon_{n\mathbf{k}}|u_{n\mathbf{k}}\ket +\varepsilon_{n\mathbf{k}}\partial_{k_j}|u_{n\mathbf{k}}\ket ).
\end{equation}
\begin{align}
\Rightarrow \quad &\frac{\hat{p}_j+k_j}{m}\partial_{k_i}|u_{n\mathbf{k}}\ket +\frac{1}{m}\delta_{ij}|u_{n\mathbf{k}}\ket + \frac{\hat{p}_i+k_i}{m}\partial_{k_j}|u_{n\mathbf{k}}\ket +\hat{h}_{\mathbf{k}}\partial_{k_i}\partial_{k_j}|u_{n\mathbf{k}}\ket \\
&=\partial_{k_i} \partial_{k_j}\varepsilon_{n\mathbf{k}}|u_{n\mathbf{k}}\ket +\partial_{k_j}\varepsilon_{n\mathbf{k}}\partial_{k_i}|u_{n\mathbf{k}}\ket +\partial_{k_i}\varepsilon_{n\mathbf{k}}\partial_{k_j}|u_{n\mathbf{k}}\ket +\varepsilon_{n\mathbf{k}}\partial_{k_i} \partial_{k_j}|u_{n\mathbf{k}}\ket .
\end{align}

applying $\bra u_{n\mathbf{k}}|$ to the left, we have
\begin{equation}
\frac{\partial^2 \varepsilon_{n\mathbf{k}}}{\partial k_i \partial k_j}=\frac{1}{m}\delta_{ij}+\frac{1}{v_o}\bra u_{n\mathbf{k}}|(\frac{\hat{p}_j+k_j}{m}-\partial_{k_j}\varepsilon_{n\mathbf{k}})\partial_{k_i}|u_{n\mathbf{k}}\ket +\frac{1}{v_o}\bra u_{n\mathbf{k}}|(\frac{\hat{p}_i+k_i}{m}-\partial_{k_i}\varepsilon_{n\mathbf{k}})\partial_{k_j}|u_{n\mathbf{k}}\ket .
\end{equation}
Finally, we get the intraband contribution:
\begin{align}
\lim_{\substack{\Omega\to 0\\ \mathbf{q}\to 0}}Q^{ij}_{\mathbf{q}\Omega}=&\frac{e^2 }{cV}\sum_{\mathbf{k},n}f_{\mathrm{th}}(\xi_{n\mathbf{k}})\{\frac{\delta_{ij}}{m}+\frac{1}{v_o}\bra u_{n\mathbf{k}}|(\frac{\hat{p}_j+k_j}{m}-\partial_{k_j}\varepsilon_{n\mathbf{k}})\partial_{k_i}|u_{n\mathbf{k}}\ket \\
&+\frac{1}{v_o}\bra u_{n\mathbf{k}}|(\frac{\hat{p}_i+k_i}{m}-\partial_{k_i}\varepsilon_{n\mathbf{k}})\partial_{k_j}|u_{n\mathbf{k}}\ket \}\\
=&\frac{e^2 N }{mcV}\delta_{ij}+\frac{e^2 }{cVv_0}\sum_{\mathbf{k},n}f_{\mathrm{th}}(\xi_{n\mathbf{k}})\bra u_{n\mathbf{k}}|(\frac{\hat{p}_j+k_j}{m}-\partial_{k_j}\varepsilon_{n\mathbf{k}})\partial_{k_i}|u_{n\mathbf{k}}\ket \\
&+\frac{e^2 }{cVv_0}\sum_{\mathbf{k},n}f_{\mathrm{th}}(\xi_{n\mathbf{k}})\bra u_{n\mathbf{k}}|(\frac{\hat{p}_i+k_i}{m}-\partial_{k_i}\varepsilon_{n\mathbf{k}})\partial_{k_j}|u_{n\mathbf{k}}\ket .
\end{align}
\subsection{Off-diagonal contribution $n\neq n^{\prime}$ (interband)}
From Eq.(\ref{gradient}), we have the interband
\begin{equation}
\lim_{\substack{\Omega\to 0\\ \mathbf{q}\to 0}}Q^{ij}_{\mathbf{q}\Omega}=-\frac{e^2N^2}{4m^2cV^3}\sum_{\mathbf{k},n\neq n^{\prime}}\frac{f_{n^{\prime}\mathbf{k}}-f_{n\mathbf{k}}}{\xi_{n^{\prime}\mathbf{k}}-\xi_{n\mathbf{k}}}\bra u_{n^{\prime}\mathbf{k}}|(2\hbar \hat{p}^i+2\hbar k^i)|u_{n\mathbf{k}}\ket \bra u_{n\mathbf{k}}|(2\hbar \hat{p}^j+2\hbar k^j)|u_{n^{\prime}\mathbf{k}}\ket  .\label{interband}
\end{equation}
In order to get $\bra u_{n^{\prime}\mathbf{k}}|(2\hbar \hat{p}^i+2\hbar k^i)|u_{n\mathbf{k}}\ket $, we left multiply $\bra u_{n^{\prime}\mathbf{k}}|$ to Eq.(\ref{98eigen}):
\begin{equation}
\bra u_{n^{\prime}\mathbf{k}}|\partial_{\mathbf{k}}\hat{h}_{\mathbf{k}}|u_{n\mathbf{k}}\ket +\varepsilon_{n^{\prime},\mathbf{k}}\bra u_{n^{\prime}\mathbf{k}}|\partial_{\mathbf{k}}|u_{n\mathbf{k}}\ket =\partial_{\mathbf{k}}\varepsilon_{n\mathbf{k}}\bra u_{n^{\prime}\mathbf{k}}|u_{n\mathbf{k}}\ket +\varepsilon_{n\mathbf{k}}\bra u_{n^{\prime}\mathbf{k}}|\partial_{\mathbf{k}}|u_{n\mathbf{k}}\ket .
\end{equation}
\begin{equation}
\Rightarrow \quad\bra u_{n^{\prime}\mathbf{k}}|\partial_{\mathbf{k}}\hat{h}_{\mathbf{k}}|u_{n\mathbf{k}}\ket =\bra u_{n^{\prime}\mathbf{k}}|\frac{\hbar\hat{p}+\hbar\mathbf{k}}{m}|u_{n\mathbf{k}}\ket =(\xi_{n\mathbf{k}}-\xi_{n^{\prime},\mathbf{k}})\bra u_{n^{\prime}\mathbf{k}}|\partial_{\mathbf{k}}|u_{n\mathbf{k}}\ket .
\end{equation}
Substituting it to Eq.(\ref{interband}), we get
\begin{align}
\lim_{\substack{\Omega\to 0\\ \mathbf{q}\to 0}}Q^{ij}_{\mathbf{q}\Omega}=&\frac{e^2N^2}{cV^3}\sum_{\mathbf{k},n\neq n^{\prime}}\frac{f_{n^{\prime}\mathbf{k}}-f_{n\mathbf{k}}}{\xi_{n^{\prime}\mathbf{k}}-\xi_{n\mathbf{k}}}(\xi_{n^{\prime}\mathbf{k}}-\xi_{n\mathbf{k}})^2\bra u_{n^{\prime}\mathbf{k}}|\partial_{k_i}|u_{n\mathbf{k}}\ket \bra u_{n \mathbf{k}}|\partial_{k_j}|u_{n^{\prime}\mathbf{k}}\ket \\
=&\frac{e^2N^2}{cV^3}\sum_{\mathbf{k},n, n^{\prime}}(f_{n^{\prime}\mathbf{k}}-f_{n\mathbf{k}})(\xi_{n^{\prime}\mathbf{k}}-\xi_{n\mathbf{k}})\bra u_{n^{\prime}\mathbf{k}}|\partial_{k_i}|u_{n\mathbf{k}}\ket \bra u_{n \mathbf{k}}|\partial_{k_j}|u_{n^{\prime}\mathbf{k}}\ket 
\end{align}
\subsection{Total contribution}
By combining the interband and intraband result we get in section 4.1 and section 4.2, we get the $Q(\Omega\to 0,\mathbf{q}\to 0)$ of gradient part of the current 
\begin{align}
\lim_{\substack{\Omega\to 0\\ \mathbf{q}\to 0}}Q^{ij}_{\mathbf{q}\Omega}=&\frac{e^2 N }{mcV}\delta_{ij}+\frac{e^2 }{cVv_0}\sum_{\mathbf{k},n}f_{n\mathbf{k}}\bra u_{n\mathbf{k}}|(\frac{\hat{p}_j+k_j}{m}-\partial_{k_j}\varepsilon_{n\mathbf{k}})\partial_{k_i}|u_{n\mathbf{k}}\ket \\
&+\frac{e^2 }{cVv_0}\sum_{\mathbf{k},n}f_{n\mathbf{k}}\bra u_{n\mathbf{k}}|(\frac{\hat{p}_i+k_i}{m}-\partial_{k_i}\varepsilon_{n\mathbf{k}})\partial_{k_j}|u_{n\mathbf{k}}\ket \\
&+\frac{e^2}{cVv_0^2}\sum_{\mathbf{k},n, n^{\prime}}(f_{n^{\prime}\mathbf{k}}-f_{n\mathbf{k}})(\xi_{n^{\prime}\mathbf{k}}-\xi_{n\mathbf{k}})\bra u_{n^{\prime}\mathbf{k}}|\partial_{k_i}|u_{n\mathbf{k}}\ket \bra u_{n \mathbf{k}}|\partial_{k_j}|u_{n^{\prime}\mathbf{k}}\ket \\
=&\frac{e^2 N }{mcV}\delta_{ij}+\frac{e^2 }{cVv_0}\sum_{\mathbf{k},n,n^{\prime}}f_{n\mathbf{k}}\bra u_{n\mathbf{k}}|(\frac{\hat{p}_j+k_j}{m}-\partial_{k_j}\varepsilon_{n\mathbf{k}})\frac{|u_{n^\prime \mathbf{k}}\ket \bra u_{n^\prime \mathbf{k}}|}{v_0}\partial_{k_i}|u_{n\mathbf{k}}\ket \\
&+\frac{e^2 }{cVv_0}\sum_{\mathbf{k},n,n^{\prime}}f_{n\mathbf{k}}\bra u_{n\mathbf{k}}|(\frac{\hat{p}_i+k_i}{m}-\partial_{k_i}\varepsilon_{n\mathbf{k}})\frac{|u_{n^\prime \mathbf{k}}\ket \bra u_{n^\prime \mathbf{k}}|}{v_0}\partial_{k_j}|u_{n\mathbf{k}}\ket \\
&+\frac{e^2}{cVv_0^2}\sum_{\mathbf{k},n, n^{\prime}}(f_{n^{\prime}\mathbf{k}}-f_{n\mathbf{k}})(\xi_{n^{\prime}\mathbf{k}}-\xi_{n\mathbf{k}})\bra u_{n^{\prime}\mathbf{k}}|\partial_{k_i}|u_{n\mathbf{k}}\ket \bra u_{n \mathbf{k}}|\partial_{k_j}|u_{n^{\prime}\mathbf{k}}\ket \\
=&\frac{e^2 N }{mcV}\delta_{ij}+\frac{e^2 }{cVv_0^2}\sum_{\mathbf{k},n\neq n^{\prime}}f_{n\mathbf{k}}\bra u_{n\mathbf{k}}|\frac{\hat{p}_j+k_j}{m}|u_{n^\prime \mathbf{k}}\ket \bra u_{n^\prime \mathbf{k}}|\partial_{k_i}|u_{n\mathbf{k}}\ket \\
&+\frac{e^2 }{cVv_0^2}\sum_{\mathbf{k},n\neq n^{\prime}}f_{n\mathbf{k}}\bra u_{n\mathbf{k}}|\frac{\hat{p}_i+k_i}{m}|u_{n^\prime \mathbf{k}}\ket \bra u_{n^\prime \mathbf{k}}|\partial_{k_j}|u_{n\mathbf{k}}\ket \\
&+\frac{e^2}{cVv_0^2}\sum_{\mathbf{k},n\neq n^{\prime}}(f_{n^{\prime}\mathbf{k}}-f_{n\mathbf{k}})(\xi_{n^{\prime}\mathbf{k}}-\xi_{n\mathbf{k}})\bra u_{n^{\prime}\mathbf{k}}|\partial_{k_i}|u_{n\mathbf{k}}\ket \bra u_{n \mathbf{k}}|\partial_{k_j}|u_{n^{\prime}\mathbf{k}}\ket \\
=&\frac{e^2 N }{mcV}\delta_{ij}+\frac{e^2 }{cVv_0^2}\sum_{\mathbf{k},n\neq n^{\prime}}f_{n\mathbf{k}}(\xi_{n^{\prime}\mathbf{k}}-\xi_{n\mathbf{k}})\bra u_{n\mathbf{k}}|\partial_{k_j}|u_{n^\prime \mathbf{k}}\ket \bra u_{n^\prime \mathbf{k}}|\partial_{k_i}|u_{n\mathbf{k}}\ket \\
&+\frac{e^2 }{cVv_0^2}\sum_{\mathbf{k},n\neq n^{\prime}}f_{n\mathbf{k}}(\xi_{n^{\prime}\mathbf{k}}-\xi_{n\mathbf{k}})\bra u_{n\mathbf{k}}|\partial_{k_i}|u_{n^\prime \mathbf{k}}\ket \bra u_{n^\prime \mathbf{k}}|\partial_{k_j}|u_{n\mathbf{k}}\ket \\
&+\frac{e^2}{cVv_0^2}\sum_{\mathbf{k},n\neq n^{\prime}}(f_{n^{\prime}\mathbf{k}}-f_{n\mathbf{k}})(\xi_{n^{\prime}\mathbf{k}}-\xi_{n\mathbf{k}})\bra u_{n\mathbf{k}}|\partial_{k_i}|u_{n^{\prime}\mathbf{k}}\ket \bra u_{n^{\prime} \mathbf{k}}|\partial_{k_j}|u_{n\mathbf{k}}\ket \\
=&\frac{e^2 N }{mcV}\delta_{ij}.
\end{align}
Finally, we reach the result
\begin{equation}
\mathbf{j}^{i}_{grad}(\Omega\to 0,\mathbf{q}\to 0)=\frac{e^2}{mc} \frac{N}{V}\delta_{ij} A^j_{\Omega,\mathbf{q}}=\frac{e^2}{mc} \frac{N}{V} A^i_{\Omega,\mathbf{q}}\; .
\end{equation}
Obviously, it cancels the diamagnet current
\begin{equation}
\mathbf{j}^{i}_{dia}(\Omega,\mathbf{q})=-\frac{e^2}{mc} \frac{N}{V} A^i_{\Omega,\mathbf{q}}
\end{equation}
with $\Omega \to 0, \mathbf{q}\to 0$. We have no net current when $\Omega \to 0, \mathbf{q}\to 0$. QED.
\section{Conclusion}
A general Hamiltonian 
\begin{equation}
H=\int d\mathbf{r} \; \psi_{\sigma}^{\dagger}[\frac{(\mathbf{p}-\frac{e}{c}\mathbf{A})^2}{2m}+\lambda\bm{\sigma} \cdot \EE \times (\mathbf{p}-\frac{e}{c}\mathbf{A})+U(r)-g\mu\BB \cdot\bm{\sigma}]_{\sigma \sigma^{\prime} }\psi_{\sigma^{\prime}}.
\end{equation}
give a current containing three part
\begin{equation}
\mathbf{j}=\mathbf{j}_{grad}+\mathbf{j}_{dia}+\mathbf{j}_s.
\end{equation}

Magnetic current from spin density is
\begin{equation}
\mathbf{j}_{spin}=\mu_{B}gc \vec{\nabla}\times (\psi^{\dagger}\bm{\sigma}\psi).
\end{equation}

Diamagnetic current
\begin{equation}
\mathbf{j}_{dia}=-\frac{e^2}{mc}\bra \psi^{\dagger}_{\sigma}(\mathbf{r})\psi_{\sigma}(\mathbf{r})\ket \mathbf{A}(\mathbf{r},\tau).
\end{equation}
Averaging over a unit cell, the diamagnetic current become
\begin{equation}
\bar{\mathbf{j}}_{dia}=-\frac{e^2}{mc}\cdot\frac{N}{V}\cdot \mathbf{A}.
\end{equation}

Gradient part of the current with the form
\begin{equation}
\mathbf{j}_{grad}=\frac{e\hbar}{2mi}(\psi_{\sigma}^{\dagger}\nabla \psi_{\sigma}-\nabla \psi_{\sigma}^{\dagger}\cdot\psi_{\sigma})
\end{equation}
result in 
\begin{align}
\mathbf{j}_{grad}(\Omega, \mathbf{q})=&-\frac{e^2N^2}{4m^2cV^3}\sum_{\mathbf{k},n,n^{\prime}}\frac{f_{n^{\prime}(\mathbf{k}-\mathbf{q})}-f_{n\mathbf{k}}}{i\Omega+\xi_{n^{\prime}(\mathbf{k}-\mathbf{q})}-\xi_{n\mathbf{k}}}\bra u_{n^{\prime}(\mathbf{k}-\mathbf{q})}|(2\frac{\hbar}{i}\partial_{\mathbf{r}}+2\hbar\mathbf{k}-\hbar\mathbf{q})|u_{n\mathbf{k}}\ket \\
&[\bra u_{n\mathbf{k}}|(2\frac{\hbar}{i}\partial_{\bm{\rho}}+2\hbar\mathbf{k}-\hbar\mathbf{q})|u_{n^{\prime}(\mathbf{k}-\mathbf{q})}\ket \cdot\mathbf{A}_{\mathbf{q},\Omega}].
\end{align}

We have proven that, ignoring the current from spin density, the net current vanish when $\Omega\to 0, \mathbf{q} \to 0$.

%% file: appc.tex
\chapter{Detailed derivation of Optical Conductivity in Tight Binding Model}

For Tight Binding model, the Hamiltonian on the presence of $\mathbf{A}$ should be written as 
\begin{equation}
H=\sum_{ij} h_{ij}^{\alpha\beta} \e^{\frac{ie}{\hbar c}\int_{R_j}^{R_i} d\mathbf{r} \mathbf{A}(\mathbf{r})}a_{i\alpha}^{\dagger}a_{j\beta}.
\end{equation}
After some careful calculation we can get
\begin{align}
\mathbf{j}&=\mathbf{j}_{grad}+\mathbf{j}_{dia},\\
\mathbf{j}_{grad}&=\frac{e}{\hbar}\sum_{\kk }a^{\dagger}_{\kk -\qq /2,\alpha}\partial_{\kk }h_{\kk }^{\alpha\beta}a_{\kk +\qq /2,\beta},\label{jg}\\
\mathbf{j}_{dia}&=-\frac{e^2}{\hbar^2 c}\sum_{\kk }a^{\dagger}_{\kk -\qq /2,\alpha}\partial_{\kk }(\mathbf{A}\partial_{\kk }h_{\kk }^{\alpha\beta})a_{\kk +\qq /2,\beta}.
\end{align}
Write $\mathbf{j}(\kk ,\omega)=Q_{ij}(\kk ,\omega)\mathbf{A}_j(\kk ,\omega)$, we can calculate the response kernel. With the help of Green's Function, we can get the result for gradient part of current density(in Matsubara representation):
\begin{align}
Q_{ij}(\qq ,\Omega)=&-\frac{e^2}{\hbar^2c\beta}\sum_{m,\kk }\mathrm{Tr}[\frac{\partial h}{\partial k_i}G^M(\kk +\qq /2,\epsilon_m+\Omega)\frac{\partial h}{\partial k_j}G^M(\kk -\qq /2,\epsilon_m)],\\
=&-\frac{e^2}{\hbar^2 c}\sum_{\kk ,n,n^{\prime}}\frac{f_{n^{\prime},\kk -\qq /2}-f_{n,\kk +\qq /2}}{i\Omega+\xi_{n^{\prime},\kk -\qq /2}-\xi_{n,\kk +\qq /2}}\bra n^\prime,\kk -\frac{\qq }{2}|\frac{\partial h}{\partial k_i}|n,\kk +\frac{\qq }{2}\ket \bra n,\kk +\frac{\qq }{2}|\frac{\partial h}{\partial k_j}|n^\prime,\kk -\frac{\qq }{2}\ket .
\end{align}
Detailed calculation has been attached below.
\section{The expression of the current density}
The Hamiltonian for tight binding model can be written as
\begin{equation}
H=\sum_{ij} h_{ij}^{\alpha\beta}(\mathbf{A}) a_{i\alpha}^{\dagger}a_{j\beta},
\end{equation}
where
\begin{align}
h_{ij}^{\alpha\beta}(\mathbf{A})=&h_{ij}^{\alpha\beta}\e^{\frac{ie}{\hbar c}\int_{\mathbf{R}_j}^{\mathbf{R}_i} d\mathbf{r} \mathbf{A}(\mathbf{r})}\\
\approx&h_{ij}^{\alpha\beta}(\mathbf{R}_i-\mathbf{R}_j)[1+\frac{ie}{\hbar c}(\mathbf{R}_i-\mathbf{R}_j)\mathbf{A}(\frac{\mathbf{R}_i+\mathbf{R}_j}{2})].
\end{align}
Do Fourier transformation
\begin{align}
h_{\kk }^{\alpha\beta}(\mathbf{A})=&\sum_{\mathbf{R}} \e^{-i\kk \cdot\mathbf{R}}h_{ij}^{\alpha\beta}(\mathbf{R}_i-\mathbf{R}_j)[1+\frac{ie}{\hbar c}(\mathbf{R}_i-\mathbf{R}_j)\mathbf{A}(\frac{\mathbf{R}_i+\mathbf{R}_j}{2})]\\
=&h^{\alpha\beta}(\kk )+\frac{ie}{\hbar c}\sum_{\mathbf{R}}\e^{-i\kk \cdot\mathbf{R}}(\mathbf{R}\cdot\mathbf{A})h^{\alpha\beta}(\mathbf{R})\\
=&h^{\alpha\beta}(\kk )+\frac{ie}{\hbar c}\sum_{\mathbf{R}}\frac{-1}{i}\mathbf{A}\cdot(\partial_{\kk }\e^{-i\kk \cdot\mathbf{R}})h^{\alpha\beta}(\mathbf{R})\\
=&h^{\alpha\beta}(\kk )-\frac{e}{\hbar c}\mathbf{A}\cdot\partial_{\kk }h^{\alpha\beta}_{\kk }.\label{sep}
\end{align}

Now we need to calculate the current density. We are going to use the continuity equation $\dot{\rho}_{\qq }+i\qq \cdot\mathbf{j}_{\qq }=0$ together with the equation of motion function $\dot{\rho}_{\qq }=\frac{i}{\hbar}[\hat{H},\rho_{\qq }]$ (remember $\rho_i=e a^{\dagger}_{i\alpha} a_{i\alpha}$) to reach $\mathbf{j}_{\qq }$.

Fourier transformation can give us
\begin{equation}
H=\sum_{ij} h_{ij}^{\alpha\beta}a_{i\alpha}^{\dagger}a_{j\beta}=\sum_{\kk }h^{\alpha\beta}(\kk )a_{\kk \alpha}^{\dagger}a_{\kk \beta};\qquad \rho_{\qq }=e\sum_{\kk }a^{\dagger}_{\kk \alpha}a_{\kk +\qq ,\alpha}.
\end{equation}
Therefore,
\begin{align}
-i\qq \cdot\mathbf{j}_{\qq }=\dot{\rho}_{\qq }=&\frac{ie}{\hbar}\sum_{\kk ,\mathbf{p}}[a^\dagger_{\kk } h_{\kk } a_{\kk }, a^\dagger_{\mathbf{p}}a_{\mathbf{p}+\qq }]\\
=&\frac{ie}{\hbar}\sum_{\kk ,\mathbf{p}}a^\dagger_{\kk } h_{\kk }[ a_{\kk }, a^\dagger_{\mathbf{p}}a_{\mathbf{p}+\qq }]+[a^\dagger_{\kk } h_{\kk } , a^\dagger_{\mathbf{p}}a_{\mathbf{p}+\qq }]a_{\kk }\\
=&\frac{ie}{\hbar}\sum_{\kk ,\mathbf{p}}a^\dagger_{\kk } h_{\kk }a_{\kk +\qq }-a^\dagger_{\kk } h_{\kk +\qq }a_{\kk +\qq }\\
=&\frac{ie}{\hbar}\sum_{\kk ,\mathbf{p}}a^\dagger_{\kk } (h_{\kk }-h_{\kk +\qq })a_{\kk +\qq }
\end{align}
For small $\qq $ 's we get
\begin{equation}
\mathbf{j}_{grad}=\frac{e}{\hbar}\sum_{\kk }a^{\dagger}_{\kk -\qq /2,\alpha}\partial_{\kk }h_{\kk }^{\alpha\beta}(\mathbf{A})a_{\kk +\qq /2,\beta},
\end{equation}
and $h_{\kk }^{\alpha\beta}(\mathbf{A})$ is given by Eq.(\ref{sep}). Therefore, we reach our final result
\begin{align}
\mathbf{j}&=\mathbf{j}_{grad}+\mathbf{j}_{dia},\\
\mathbf{j}_{grad}&=\frac{e}{\hbar}\sum_{\kk }a^{\dagger}_{\kk -\qq /2,\alpha}\partial_{\kk }h_{\kk }^{\alpha\beta}a_{\kk +\qq /2,\beta},\\
\mathbf{j}_{dia}&=-\frac{e^2}{\hbar^2 c}\sum_{\kk }a^{\dagger}_{\kk -\qq /2,\alpha}\partial_{\kk }(\mathbf{A}\partial_{\kk }h_{\kk }^{\alpha\beta})a_{\kk +\qq /2,\beta}.
\end{align}
\section{Calculation of the response kernel}
\subsection{linear response}
\begin{equation}
\hat{H}=\hat{H}_0+\delta\hat{H},\qquad \delta\hat{H}=-\frac{1}{c}\int d\mathbf{r} \;\mathbf{j}(\mathbf{r})\mathbf{A}(\mathbf{r})=-\frac{1}{cV}\sum_{\qq } \mathbf{j}_{-\qq }\mathbf{A}_{\qq }
\end{equation}
\begin{equation}
i\frac{\partial}{\partial t}|En(t)\ket =\hat{H}|En(t)\ket .\label{eq8}
\end{equation}
Write the eigenstate $|En(t)\ket $ as the evolution of the state at $t=0$, or $|En(t)\ket =\e^{-i\hat{H}_0 t}\hat{U}(t)|En\ket $. Then Eq.(\ref{eq8}) becomes
\begin{align}
i\frac{\partial}{\partial t}|En(t)\ket =&\hat{H}_0 \e^{-i\hat{H}_0 t}\hat{U}(t)|En\ket +i\e^{-i\hat{H}_0 t}\frac{\partial \hat{U}(t)}{\partial t}|En\ket ;\\
=&\hat{H}_0 \e^{-i\hat{H}_0 t}\hat{U}(t)|En\ket +\delta \hat{H} \e^{-i\hat{H}_0 t} \hat{U}(t)|En\ket .\\
\Rightarrow i\frac{\partial \hat{U}(t)}{\partial t}|En\ket =&\e^{i\hat{H}_0 t}\delta \hat{H} \e^{-i\hat{H}_0 t} \hat{U}(t)|En\ket =\delta \hat{H} (t) \hat{U}(t)|En\ket .\\
\Rightarrow \hat{U}(t)=& 1-i\int_{-\infty}^{t}\delta \hat{H}(t^{\prime})\hat{U}(t^\prime)dt^\prime.
\end{align}
To the first order of $\delta \hat{H}$:
\begin{equation}
\hat{U}(t)=1-i\int_{-\infty}^{t}\delta \hat{H}(t^{\prime})dt^\prime.
\end{equation}

Now we can calculate the expectation value of the current density 
\begin{align}
\mathbf{j}(\mathbf{r},t)=&\bra En(t)|\hat{J}(\mathbf{r},t)|En(t)\ket =\bra En(t)|\mathbf{j}(\mathbf{r})|En(t)\ket -ne\mathbf{A}(\mathbf{r},t)\\
=&\bra En|(1+i\int_{-\infty}^{t}\delta \hat{H}(t^{\prime})dt^\prime)\e^{i\hat{H}_0 t}\mathbf{j}(\mathbf{r}) \e^{-i\hat{H}_0 t}(1-i\int_{-\infty}^{t}\delta \hat{H}(t^{\prime})dt^\prime)|En\ket -ne\mathbf{A}(\mathbf{r},t)\\
=&\bra En|\mathbf{j}(\mathbf{r})|En\ket +i\int_{-\infty}^{t}dt^\prime \bra En|[\delta H(t^\prime),\mathbf{j}(\mathbf{r},t)]|En\ket -ne\mathbf{A}(\mathbf{r},t)\\
=&i\int_{-\infty}^{t}dt^\prime \bra En|[\delta H(t^\prime),\mathbf{j}(\mathbf{r},t)]|En\ket -ne\mathbf{A}(\mathbf{r},t).\\
j_i(\mathbf{r},t)=&\int_{-\infty}^{+\infty}dt^\prime\int d\mathbf{r}^\prime\sum_j Q_{ij}(\mathbf{r}-\mathbf{r}',t-t^\prime)A_j(\mathbf{r}^\prime,t^\prime),\\
Q_{ij}(\mathbf{r}-\mathbf{r}',t-t^\prime)=&-i\theta(t-t^\prime)\frac{1}{c}\bra En|[\mathbf{j}_i(\mathbf{r},t),\mathbf{j}_j(\mathbf{r}^\prime,t^\prime)]|En\ket -ne\delta_{ij}\delta(\mathbf{r}-\mathbf{r}^\prime)\delta(t-t^\prime).
\end{align}

\subsection{Do Fourier transformation and Wick's contraction}
\begin{align}
\mathbf{j}(\kk ,\omega)=&\int d\mathbf{r}dt\mathbf{j}(\mathbf{r},t)\e^{-i\kk \cdot \mathbf{r}+i\omega t}\\
=&\int d\mathbf{r}dt\int dt^\prime d\mathbf{r}^\prime\sum_j Q_{ij}(\mathbf{r}-\mathbf{r}',t-t^\prime)A_j(\mathbf{r}^\prime,t^\prime)\e^{-i\kk \cdot \mathbf{r}+i\omega t}\\
=&\int d\mathbf{r}dt\int dt^\prime d\mathbf{r}^\prime\int(d\qq )d\Omega Q_{ij}(\qq ,\Omega)\e^{i\qq (\mathbf{r}-\mathbf{r}^\prime)-i\Omega(t-t^\prime)}\int(d\qq \prime)d \Omega^\prime A_j(\qq ^\prime,\Omega^\prime)\e^{i\qq \cdot\mathbf{r}^\prime-i\Omega t^\prime}\e^{-i\kk \cdot \mathbf{r}+i\omega t}\\
=&\int(d\qq )d\Omega(d\qq ^\prime)d\Omega^\prime\delta_{\qq ,\kk }\delta_{\omega,\Omega}\delta_{\qq ,\qq ^\prime}\delta_{\Omega,\Omega^\prime}Q_{ij}(\qq ,\vec{\Omega})A_j(\qq ^\prime,\Omega^\prime)\\
=&Q_{ij}(\kk ,\omega)A_j(\kk ,\omega).
\end{align}
Ignore diamagnetic current. 
\begin{align}
Q_{ij}(\kk ,t-t^\prime)=&\int d\mathbf{r} d\mathbf{r}^\prime Q_{ij}(\mathbf{r}-\mathbf{r}^\prime,t-t^\prime)\e^{-i\kk \cdot(\mathbf{r}-\mathbf{r}^\prime)}\\
=&\int d\mathbf{r}d\mathbf{r}^\prime [-i\theta(t-t^\prime)\frac{1}{c}\bra En|[\mathbf{j}_i(\mathbf{r},t),\mathbf{j}_j(\mathbf{r}^\prime,t^\prime)]|En\ket \e^{-i\kk \cdot(\mathbf{r}-\mathbf{r}^\prime)}]\\
=&\int d\mathbf{r}d\mathbf{r}^\prime(d\qq )(d\qq ^\prime) [-i\theta(t-t^\prime)\frac{1}{c}\bra En|[\mathbf{j}_i(\qq ,t),\mathbf{j}_j(\qq ^\prime,t^\prime)]|En\ket \e^{i\qq \cdot\mathbf{r}+i\qq ^\prime\cdot\mathbf{r}^\prime-i\kk \cdot(\mathbf{r}-\mathbf{r}^\prime)}]\\
=&\int (d\qq )(d\qq ^\prime) [-i\theta(t-t^\prime)\frac{1}{c}\bra En|[\mathbf{j}_i(\qq ,t),\mathbf{j}_j(\qq ^\prime,t^\prime)]|En\ket \delta_{\qq ,\kk }\delta_{\qq ^\prime,-\kk }]\\
=&-i\theta(t-t^\prime)\frac{1}{c}\bra En|[\mathbf{j}_i(\kk ,t),\mathbf{j}_j(-\kk ,t^\prime)]|En\ket .\\
\Rightarrow \qquad &\quad Q_{ij}(\kk ,\tau-\tau^\prime)=\frac{1}{c}\bra T_\tau[\mathbf{j}_i(\kk ,\tau),\mathbf{j}_j(-\kk ,\tau^\prime)]\ket .
\end{align}
Take Eq.(\ref{jg}) in to the equation above, and set $\tau^\prime=0$, we can get
\begin{equation}
Q_{ij}(\qq ,\tau)=\frac{e^2}{\hbar^2 c}\sum_{\kk ,\mathbf{p}}\frac{\partial h^{\alpha\beta}}{\partial k_i}\frac{\partial h^{\gamma\delta}}{\partial p_j}\bra T_\tau[a^\dagger_{\kk -\qq /2,\alpha}(\tau)a_{\kk +\qq /2,\beta}(\tau)a^\dagger_{\kk +\qq /2,\gamma}a_{\kk -\qq /2,\delta}]\ket .
\end{equation}
Wick's contraction gives 
\begin{align}
Q_{ij}(\qq ,\tau)=-&\frac{e^2}{\hbar^2 c}\sum_{\kk ,\mathbf{p}}\frac{\partial h^{\alpha\beta}}{\partial k_i}\frac{\partial h^{\gamma\delta}}{\partial p_j}\bra T_\tau a_{\kk -\qq /2,\delta}a^\dagger_{\kk -\qq /2,\alpha}(\tau)\ket \bra T_\tau a_{\kk +\qq /2,\beta}(\tau)a^\dagger_{\kk +\qq /2,\gamma}\ket \\
=-&\frac{e^2}{\hbar^2 c}\sum_{\kk }\frac{\partial h^{\alpha\beta}}{\partial k_i}\frac{\partial h^{\gamma\delta}}{\partial k_j}G^{M}_{\delta\alpha}(\kk -\qq /2,-\tau)G^{M}_{\beta\gamma}(\kk +\qq /2,\tau).
\end{align}
Do Fourier transformation for the response kernel, with $G^{M}(\mathbf{p},\tau)=\frac{1}{\beta}\sum_n \e^{-i\epsilon_n \tau}G^M(\mathbf{p},\epsilon_n)$, we obtain
\begin{align}
Q_{ij}(\qq ,\Omega)=&\int d\tau\e^{i\Omega\tau}Q_{ij}(\qq ,\tau)\\
=-&\frac{e^2}{\hbar^2 c}\sum_{\kk }\frac{\partial h^{\alpha\beta}}{\partial k_i}\frac{\partial h^{\gamma\delta}}{\partial k_j}\int d\tau \frac{1}{\beta^2}\sum_{m,n}\e^{i\epsilon_m \tau}G^{M}_{\delta\alpha}(\kk -\qq /2,\epsilon_m)\e^{-i\epsilon_n \tau}G^{M}_{\beta\gamma}(\kk +\qq /2,\epsilon_n)\e^{i\Omega\tau}\\
=-&\frac{e^2}{\hbar^2 c\beta}\sum_{m,\kk }\frac{\partial h^{\alpha\beta}}{\partial k_i}\frac{\partial h^{\gamma\delta}}{\partial k_j}G^{M}_{\delta\alpha}(\kk -\qq /2,\epsilon_m)G^{M}_{\beta\gamma}(\kk +\qq /2,\epsilon_m+\Omega)\\
=-&\frac{e^2}{\hbar^2 c\beta}\sum_{m,\kk }\mathrm{Tr}[\frac{\partial h}{\partial k_i}G^{M}(\kk +\qq /2,\epsilon_m+\Omega)\frac{\partial h}{\partial k_j}G^{M}(\kk -\qq /2,\epsilon_m)]
\end{align}

Input the general express of the Masubara Green's Function, we finally obtain
\begin{align}
Q_{ij}(\qq ,\Omega)=&-\frac{e^2}{\hbar^2 c\beta}\sum_{m,\kk ,n,n^\prime}\mathrm{Tr}[\frac{\partial h}{\partial k_i}\frac{|n,\kk +\frac{\qq }{2}\ket \bra n,\kk +\frac{\qq }{2}|}{i(\epsilon_m+\Omega)-\epsilon_{n,\kk +\qq /2}}\frac{\partial h}{\partial k_j}\frac{|n^\prime,\kk -\frac{\qq }{2}\ket \bra n^\prime,\kk -\frac{\qq }{2}|}{i\epsilon_m-\epsilon_{n^\prime,\kk -\qq /2}}]\\
=&-\frac{e^2}{\hbar^2 c\beta}\sum_{m,\kk ,n,n^\prime}\frac{1}{i(\epsilon_m+\Omega)-\epsilon_{n,\kk +\qq /2}}\frac{1}{i\epsilon_m-\epsilon_{n^\prime,\kk -\qq /2}}\nonumber\\
&*\bra n^\prime,\kk -\frac{\qq }{2}|\frac{\partial h}{\partial k_i}|n,\kk +\frac{\qq }{2}\ket \bra n,\kk +\frac{\qq }{2}|\frac{\partial h}{\partial k_j}|n^\prime,\kk -\frac{\qq }{2}\ket \\
=&-\frac{e^2}{\hbar^2 c}\sum_{\kk ,n,n^{\prime}}\frac{f_{n^{\prime},\kk -\qq /2}-f_{n,\kk +\qq /2}}{i\Omega+\xi_{n^{\prime},\kk -\qq /2}-\xi_{n,\kk +\qq /2}}\bra n^\prime,\kk -\frac{\qq }{2}|\frac{\partial h}{\partial k_i}|n,\kk +\frac{\qq }{2}\ket \bra n,\kk +\frac{\qq }{2}|\frac{\partial h}{\partial k_j}|n^\prime,\kk -\frac{\qq }{2}\ket .
\end{align}